\documentclass[MSNbibl,number,citesort,dvips]{arxstspdf}
\usepackage{flushend}
\usepackage{stfloats}
\usepackage{graphicx}

\volume{26}
\issue{4}
\pubyear{2011}
\firstpage{613}
\lastpage{646}
\doi{10.1214/11-STS367}

\makeatletter
\newcommand{\eqref}[1]{(\ref{#1})}

\newcommand{\modelJapPoi}{\model H}
\newcommand{\modelJapSof}{\model I}
\newcommand{\modelJapOgT}{\model J}

\newcommand{\modelTwoCSR}{\model E}
\newcommand{\modelTwoAre}{\model F}
\newcommand{\modelTwoGey}{\model G}

\newcommand{\model}[1]{\textsf{({#1})}}
\newcommand{\modelOneCSR}{\model A}
\newcommand{\modelOneInh}{\model B}
\newcommand{\modelOneStr}{\model C}
\newcommand{\modelOneInS}{\model D}

\newcommand{\transpose}[1]{{#1}^{\top}}

\newcommand{\outerprod}[2]{{#1} \, \transpose{#2}}
\newcommand{\outersquare}[1]{\outerprod{#1}{#1}}

\newcommand{\ddt}{\frac{\partial}{\partial\theta}}

\newcommand{\PL}{\mathsf{PL}}
\newcommand{\PU}{\mathsf{PU}}

\newcommand{\bx}{\mathbf{x}}
\newcommand{\by}{\mathbf{y}}
\newcommand{\bX}{\mathbf{X}}
\newcommand{\bY}{\mathbf{Y}}

\newcommand{\probab}{\mathbb{P}}

\newcommand{\expected}{\mathbb E}
\newcommand{\Var}{\mathbb{V}\mathrm{ar}}
\newcommand{\real}{\mathbb{R}}

\newcommand{\Khat}{\hat{K}}
\newcommand{\Fhat}{\hat{F}}
\newcommand{\Ghat}{\hat{G}}

\newcommand{\indicate}[1]{{\mathbb I}\{ {#1} \}}
\newcommand{\dee}[1]{\,\mathrm{d}{#1}}

\newcommand{\compen}{\mathcal{C}}
\newcommand{\resid}{\mathcal{R}}
\newcommand{\stdres}{\mathcal{T}}
\newcommand{\compvar}{\mathcal{C}^2}
\newcommand{\innov}{\mathcal{I}}

\newcommand{\psum}{\Sigma\Delta}
\newcommand{\pcom}{\mathcal{C}\Delta}
\newcommand{\pres}{\mathcal{R}\Delta}
\newcommand{\pstd}{\mathcal{T} \Delta}
\newcommand{\pvar}{\mathcal{C}^2 \Delta}

\newcommand{\free}[1]{{#1}^\circ}
\newcommand{\fixed}[1]{{#1}^{+}}
\newcommand{\bxfree}{\free\bx}
\newcommand{\bxfixed}{\fixed\bx}
\newcommand{\bXfree}{\free\bX}
\newcommand{\bXfixed}{\fixed\bX}

\newcommand{\Wfree}{\free W}
\newcommand{\Wfixed}{\fixed W}

\newcommand{\augm}[2]{{#1} \cup \{ {#2} \}}
\newcommand{\del}[2]{{#1} \setminus \{ {#2} \}}
\newcommand{\bxplus}[1]{\augm{\bx}{#1}}
\newcommand{\bxminus}[1]{\bx_{-{#1}}}
\newcommand{\bxfreeplus}[1]{\augm{\bxfree}{#1}}
\newcommand{\bxfreeminus}[1]{\bxfree_{-{#1}}}

\newcommand{\bXminus}[1]{\bX_{-{#1}}}

\newcommand{\PoisProc}[2]{\operatorname{Poisson}({#1},{#2})}

\renewcommand{\emptyset}{\varnothing}
\renewcommand{\epsilon}{\varepsilon}
\makeatother

\begin{document}
\begin{frontmatter}

\title{Score, Pseudo-Score and Residual Diagnostics for Spatial Point Process~Models}
\runtitle{Diagnostics for Spatial Point Processes}

\begin{aug}
\author{\fnms{Adrian} \snm{Baddeley}\ead[label=e1]{Adrian.Baddeley@csiro.au}},
\author{\fnms{Ege} \snm{Rubak}\corref{}\ead[label=e2]{rubak@math.aau.dk}}
\and
\author{\fnms{Jesper} \snm{M{\o}ller}\ead[label=e3]{jm@math.aau.dk}}

\runauthor{A. Baddeley, E. Rubak and J. M{\o}ller}

\affiliation{CSIRO, University of Western Australia, Aalborg University and Aalborg University}

\address{Adrian Baddeley is Research Scientist, CSIRO Mathematics, Informatics and Statistics, Private Bag 5,
  Wembley WA 6913, Australia and Adjunct
Professor, University of Western Australia \printead{e1}. Ege Rubak is Postdoctoral Scholar, Department of Mathematical Sciences, Aalborg University,
  Fredrik Bajers Vej 7G, DK-9220 Aalborg \O, Denmark \printead{e2}. Jesper M{\o}ller is Professor of Statistics, Department of Mathematical Sciences, Aalborg University,
  Fredrik Bajers Vej 7G, DK-9220 Aalborg \O, Denmark \printead{e3}.}

\end{aug}

\begin{abstract}
We develop new tools for formal inference and informal model valid\-ation
in the analysis of spatial point pattern data.
The score test is generalized to a
``pseudo-score'' test derived from Besag's pseudo-likelihood,
and to a class of diagnostics based on point process residuals.
The results lend theoretical support to the established practice
of using functional summary statistics, such as Ripley's $K$-function,
when testing for complete spatial randomness; and they provide
new tools such as the \textit{compensator} of the $K$-function
for testing other fitted models. The results also support localization methods
such as the scan statistic and smoothed residual plots.
Software for computing the diagnostics is provided.
\end{abstract}


\vspace*{-6pt}
\end{frontmatter}

\section{Introduction}
\label{S:intro}

This paper develops new tools for formal inference and informal model
valid\-ation in the analysis of spatial point pattern data. The score
test statistic, based on the point process likelihood, is generalized
to a ``pseudo-score'' test statistic derived from Besag's
pseudo-likelihood. The score and pseudo-score can be viewed as
residuals, and further generalized to a class of residual
diagnostics.\vadjust{\goodbreak}

The likelihood score and the score test \cite{rao48,wald41},
\cite{coxhink74}, pages 315 and 324, are used frequently in applied
statistics to provide diagnostics for model selection and model
validation \cite{atki82,cookweis83,preg82,chen83,wang85}.
In spatial statistics, the score test has been used mainly to support
formal inference about covariate effects
\cite{berm86,laws93a,walletal92} assuming the underlying point process
is Poisson under both the null and alternative hypotheses. Our approach
extends this to a much wider class of point processes, making it
possible (for example) to check for covariate effects or localized
hot-spots in a clustered point pattern.

\begin{figure*}
\centering
\begin{tabular}{@{}ccc@{}}

\includegraphics{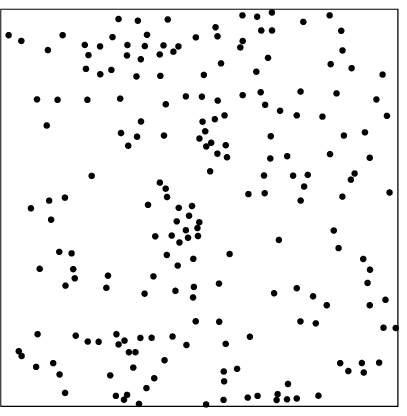}&
\includegraphics{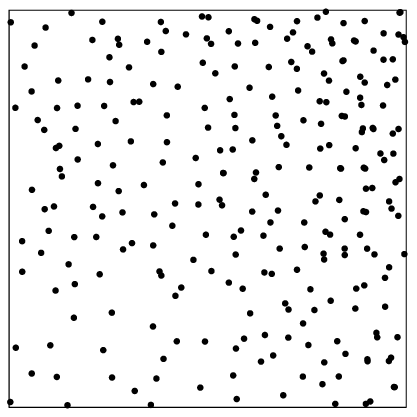}&
\includegraphics{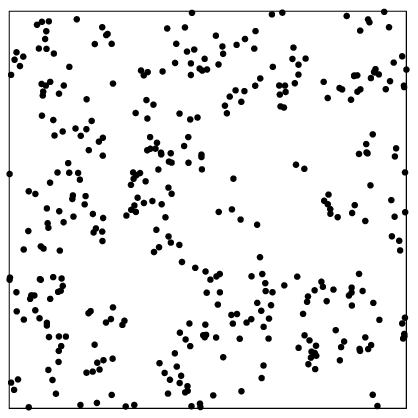}\\
\scriptsize{(a)}&\scriptsize{(b)}&\scriptsize{(c)}
\end{tabular}
\vspace*{-3pt}
\caption{Point pattern data sets. \textup{(a)} Japanese black pine seedlings and saplings
    in a $10 \times 10$ metre quadrat \protect\cite{numa61,numa64}.
    Reprinted by kind permission of Professors M. Numata and Y. Ogata.
    \textup{(b)} Simulated realization
    of inhomogeneous Strauss process
    showing strong inhibition and spatial trend \protect\cite{baddetal05}, Figure 4\textup{b}.
    \textup{(c)} Simulated realization of homogeneous Geyer
    saturation process showing moderately strong clustering
    without spatial trend \protect\cite{baddetal05}, Figure 4\textup{c}.}\label{fig:data}
\end{figure*}

Figure \ref{fig:data} shows three example data sets studied in the
paper. Our techniques make it possible to check separately for
``inhomogeneity'' (spatial variation in abundance of points) and
``interaction'' (localized dependence between points) in these data.

Our approach also provides theoretical support for the established
practice of using functional summary statistics such as Ripley's
$K$-function \cite{ripl76,ripl77} to study clustering and inhibition
between points. In one class of models, the score test statistic is
equivalent to the empirical $K$-function, and the score test procedure
is closely related to the customary goodness-of-fit procedure based on
comparing the empirical $K$-function with its null expected value.
Similar statements apply to the nearest neighbor distance distribution
function $G$ and the empty space function $F$.

For computational efficiency, especially in large data sets, the point
process likelihood is often replaced by Besag's \cite{besa78}
pseudo-likelihood. The resulting ``pseudo-score'' is a possible surrogate
for the likelihood score in the score test. In one model, this
pseudo-score test statistic is equivalent to a \textit{residual} version
of the empirical $K$-function, yielding a~new, efficient diagnostic for
model fit. However, in general, the interpretation of the pseudo-score
test statistic is conceptually more complicated than that of the
likelihood score
 test statistic, and hence
difficult to employ as a diagnostic.

In classical settings the score test statistic is\break a~weighted sum of
residuals. For point processes the pseudo-score test statistic is
a~weighted point process residual in the sense of
\cite{baddetal05,baddmollpake08}. This suggests a~simplification, in
which the pseudo-score test statistic is replaced by another residual
diagnostic that is easier to interpret and to compute.

\begin{figure}[b]
\vspace*{-3pt}
\includegraphics{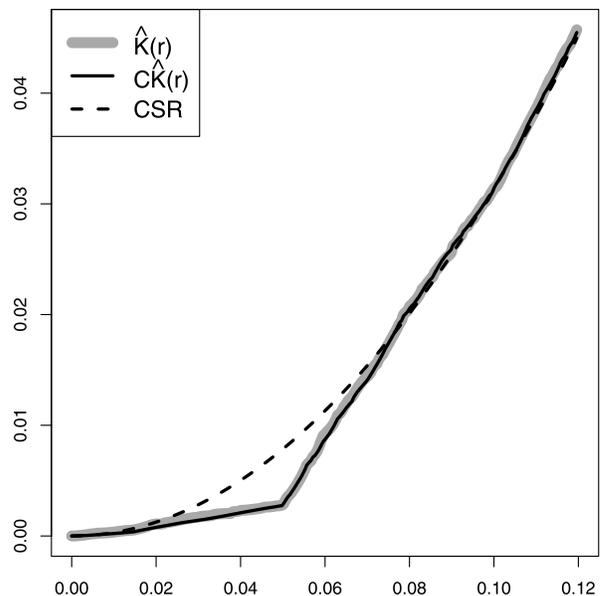}
  \caption{Empirical $K$-function (thick grey line) for the
    point pattern data in Figure \protect\ref{fig:data}\textup{(b)},
    compensator of the $K$-function (solid black line)
    for a model of the correct form, and
    expected $K$-function for a homogeneous Poisson process (dashed line).}
  \label{F:KrightKcom}
\end{figure}

In special cases this diagnostic is a residual version of one of the
classical functional summary statistics~$K$, $G$ or $F$ obtained by
subtracting a ``\textit{compensator}'' from the functional summary
statistic.
The compensator depends on the fitted model, and may also depend on the
observed data. For example, suppose the fitted model is the homogeneous
Poisson process. Then (ignoring some details) the compensator of the
empirical $K$-function $\hat K(r)$ is its expectation $K_0(r) = \pi
r^2$ under the model, while the compensator of the empirical nearest
neighbor function~$\hat G(r)$ is the empirical empty space function
$\hat F(r)$ for the same data.\vadjust{\goodbreak}
This approach provides a~new class of residual summary statistics that
can be used as informal diagnostics for
model fit, for a~wide range of point process models, in close analogy
with current practice. The diagnostics apply under very general
conditions, including the case of inhomogeneous point process models,
where exploratory methods are underdeveloped or inapplicable. For
instance, Figure \ref{F:KrightKcom} shows the compensator of $K(r)$ for
an inhomogeneous Strauss process.

Section \ref{S:assumptions} introduces basic definitions and
assumptions. Section \ref{S:scoretest} describes the score test for a
general point\vadjust{\goodbreak} process model, and Section \ref{S:scorePoisson} develops
the important case of Poisson point process models.
Section~\ref{S:nonPois} gives examples and technical tools for
non-Poisson point process models. Section~\ref{S:theory} develops the
general theory for our diagnostic tools. Section \ref{S:firstorder}
applies these tools to tests for first order trend and hotspots.
Sections \ref{S:interaction}--\ref{S:area} develop diagnostics for
interaction between points, based on pairwise distances, nearest
neighbor distances and empty space distances, respectively. The tools
are demonstrated on data in
Sections \ref{S:TrendInhib}--\ref{S:summary}.
  Further examples of diagnostics are given in
  Appendix \ref{App:further}.
Appendices \ref{App:var}--\ref{App:G} provide technical details.

\section{Assumptions}
\label{S:assumptions}

\subsection{Fundamentals}

A spatial point pattern data set is a finite set $\bx =
\{x_1,\ldots,x_n\}$ of points $x_i \in W$, where the number of points
$n(\bx)=n \ge 0$ is not fixed in advance, and the domain of observation
$W \subset\real^d$ is a fixed, known region of $d$-dimensional space
with finite positive volume $|W|$. We take $d=2$, but the results
generalize easily to all dimensions.

A point process model assumes that $\bx$ is a realization of a finite
point process $\bX$ in $W$ without multiple points. We can equivalently
view $\bX$ as a~random finite subset of $W$. Much of the literature on
spatial statistics assumes that $\bX$ is the restriction $\bX = \bY
\cap W$ of a~stationary point process $\bY$ on the entire space
$\real^2$. We do not assume this; there is no assumption of
stationarity, and some of the models considered here are intrinsically
confined to the domain $W$. For further background material including
measure theoretical details, see, for example, \cite{mollwaag04}, Appendix
B.

Write $\bX \sim \PoisProc W \rho$ if $\bX$ follows the Poisson process
on $W$ with intensity function $\rho$, where we assume
$\nu=\int_W\rho(u)\,\mathrm{d}u$ is finite. Then $n(\bX)$ is Poisson
distributed with mean $\nu$, and, conditional on~$n(\bX)$, the points
in $\bX$ are i.i.d.
 with density $\rho(u)/\nu$.

Every point process model considered here is assumed to have a
probability density with respect to $\PoisProc W 1$, the unit rate
Poisson process, under one of the following scenarios.

\subsection{Unconditional Case}
\label{S:assump:uncon}

In the \textit{unconditional case} we assume $\bX$ has a density $f$ with
respect to $\PoisProc W 1$. Then the density is
characterized by the property
%
\begin{equation}\label{e:f}
  \expected[ h(\bX) ] = \expected[ h(\bY) f(\bY) ]
\end{equation}
for all nonnegative measurable 
functionals $h$, where $\bY \sim \PoisProc W 1$.\vadjust{\goodbreak} In particular, the
density of\break $\PoisProc W \rho$ is
\begin{equation}\label{e:poisson}
  f(\bx) = \exp\biggl( \int_W \bigl(1 - \rho(u)\bigr) \dee u \biggr)\prod_{i} \rho(x_i).
\end{equation}
We assume that $f$ is hereditary, that is, $f(\bx)>0$ implies
$f(\by)>0$ for all finite $\by\subset\bx\subset W$.
%
%
Processes satisfying these assumptions include (under integrability
conditions) inhomogeneous Poisson processes with an intensity function,
finite Gibbs processes contained in~$W$, and Cox processes driven by
random fields. See \cite{illietal08}, Chapter 3, for an overview of
finite point processes including these examples. In practice, our
methods require the density to have a~tractable form, and are only
developed for Poisson and Gibbs processes.

\subsection{Conditional Case}
\label{S:assump:cond}

In the \textit{conditional case}, we assume $\bX = \bY \cap W$ where
$\bY$ is a point process. Thus, $\bX$ may depend on unobserved points
of $\bY$ lying outside $W$. The density of $\bX$ may be unknown or
intractable. Under suitable conditions (explained in
Section~\ref{S:Markov}) modeling and inference can be based on the
conditional distribution of $\bXfree = \bX \cap \Wfree$ given $\bXfixed
= \bX \cap \Wfixed = \bxfixed$, where $\Wfixed \subset W$ is a
subregion, typically a region near the boundary of~$W$, and only the
points in $\Wfree = W \setminus \Wfixed$ are treated as random. We
assume that the conditional distribution of $\bXfree = \bX \cap
\Wfree$ given $\bXfixed = \bX \cap \Wfixed = \bxfixed$ has an
hereditary density $f(\bxfree \vert \bxfixed)$ with respect to
$\PoisProc \Wfree 1$.
Processes satisfying these assumptions include Markov point processes
\cite{lies00}, \cite{mollwaag04}, Section 6.4,
together with all processes covered by the unconditional case. Our
methods are only developed for Poisson and Markov point processes.

For ease of exposition, we focus mainly on the unconditional case, with
occasional comments on the conditional case. For Poisson point process
models, we always take $W=\Wfree$ so that the two cases agree.

\section{Score Test for Point Processes}
\label{S:scoretest}

In principle, any technique for likelihood-based inference is
applicable to point process likelihoods. In practice, many likelihood
computations require extensive Monte Carlo simulation
\cite{geye99,mollwaag04,mollwaag07}. To minimize such difficulties,
when assessing the goodness of fit of a fitted point process model, it
is natural to choose the score test which only requires computations
for the null hypothesis \cite{wald41,rao48}.

Consider any parametric family of point process models for $\bX$ with
density $f_\theta$ indexed by a $k$-dimen\-sional\vadjust{\goodbreak} vector parameter
$\theta \in \Theta \subseteq \real^k$. For a \textit{simple} null
hypothesis $H_0\dvtx \theta = \theta_0$ where $\theta_0 \in \Theta$ is
fixed, the score test against any alternative $H_1\dvtx \theta \in
\Theta_1$, where $\Theta_1 \subseteq \Theta\setminus\{\theta_0\}$, is
based on the score test statistic (\cite{coxhink74}, page 315),
%
\begin{equation}\label{e:scoretest0}
T^2 = \transpose{U(\theta_0)} I(\theta_0)^{-1}U(\theta_0).
\end{equation}
Here $
  U(\theta) = \ddt \log f_\theta(\bx)
$ and $
  I(\theta) = \expected_\theta[ U(\theta)\transpose{U(\theta)} ]
$ are the score function and Fisher information, respectively, and the
expectation is with respect to $f_\theta$. Here and throughout, we
assume that the order of integration and differentation \textit{with
respect to $\theta$} can be interchanged. Under suitable conditions,
the null distribution of $T^2$ is $\chi^2$ with $k$ degrees of freedom.
In the case $k=1$ it may be informative to evaluate the signed square
root
%
\begin{equation}\label{e:scoretest:sqrt}
  T = {U(\theta_0)}/{\sqrt{I(\theta_0)}},
\end{equation}
which is asymptotically $N(0,1)$ distributed under the same conditions.

For a \textit{composite} null hypothesis $H_0\dvtx \theta \in \Theta_0$ where
$\Theta_0 \subset \Theta$ is an $m$-dimensional submanifold with $0 < m
< k$, the score test statistic is defined in \cite{coxhink74}, page~324. However, we shall not use this version of the score
test, as it assumes differentiability of the likelihood with respect to
nuisance parameters, which is not necessarily applicable here (as
exemplified in Section~\ref{S:pois:thresh}).

In the sequel we often consider models of the form
%
\begin{equation}\label{E:special.form}
f_{(\alpha,\beta)}(\bx)=c(\alpha,\beta)h_\alpha(\bx)\exp(\beta S(\bx)),
\end{equation}
where the parameter $\beta$ and the statistic $S(\bx)$ are one
dimensional, and the null hypothesis is \mbox{$H_0\dvtx\beta=0$}. For fixed
$\alpha$, this is a linear exponential family and~\eqref{e:scoretest:sqrt} becomes
\[
  T(\alpha)=\bigl(
    S(\bx)-\expected_{(\alpha,0)}[S(\bX)]
  \bigr)/\sqrt{\Var_{(\alpha,0)}[S(\bX)]}.
\]
In practice, when $\alpha$ is unknown,
 we replace $\alpha$ by its MLE under $H_0$ so that, with a
slight abuse of notation, the signed square root of the score test
statistic is approximated by
\begin{eqnarray}\label{e:T=std.S}
\hspace*{10pt}T&=&T(\hat\alpha)\nonumber\\ [-8pt]\\ [-8pt]
\hspace*{10pt}&=&\bigl(S(\bx)-\expected_{(\hat\alpha,0)}[S(\bX)]\bigr)/\sqrt{\Var_{(\hat\alpha,0)}[S(\bX)]}.\nonumber
\end{eqnarray}
%
Under suitable conditions,
 $T$ in \eqref{e:T=std.S} is asymptotically equivalent to
$T$ in \eqref{e:scoretest:sqrt}, and so a standard Normal approximation
may still apply.

\section{Score Test for Poisson Processes}
\label{S:scorePoisson}

Application of the score test to Poisson point process models appears
to originate with Cox \cite{cox72pp}. Consider a~parametric family of
Poisson processes,\break $\PoisProc W {\rho_\theta}$, where\vadjust{\goodbreak}
the intensity function is indexed by $\theta \in \Theta$. The score
test statistic is (\ref{e:scoretest0}), where
\begin{eqnarray*}
  U(\theta) &=& \sum_{i} \kappa_\theta(x_i)
                - \int_W \kappa_\theta(u) \rho_\theta(u) \dee u, \\
  I(\theta) &=& \int_W  \outersquare{\kappa_\theta(u)}
                       \rho_\theta(u) \dee u
\end{eqnarray*}
with $
    \kappa_\theta(u) = \ddt \log\rho_\theta(u).
$ Asymptotic results are giv\-en in \cite{kuto98,rathcres94b}.

\subsection{Log-Linear Alternative}
\label{S:pois:loglin}

The score test is commonly used in spatial epidemiology to assess
whether disease incidence depends on environmental exposure.
As a particular case of \eqref{E:special.form},   
suppose the Poisson model has a log-linear intensity function
%
\begin{equation}\label{e:rho=expZ}
\rho_{(\alpha,\beta)}(u) = \exp\bigl(\alpha + \beta Z(u)\bigr),
\end{equation}
where $Z(u), u \in W$, is a known, real-valued and nonconstant covariate
function, and $\alpha$ and $\beta$ are real parameters. Cox
\cite{cox72pp} noted that the uniformly most powerful test of $H_0\dvtx
\beta=0$ (the homogeneous Poisson process)  against $H_1\dvtx \beta > 0$ is
based on the statistic
%
\begin{equation}\label{e:S=sumZ}
S(\bx) = \sum_i Z(x_i).
\end{equation}
Recall that, for a point process $\bX$ on $W$ with intensity function
$\rho$, we have Campbell's Formula (\cite{dalevere03}, page~163),
%
\begin{equation}\label{e:meanPois}
\expected\biggl( \sum_{x_i\in\bX} h(x_i) \biggr) = \int_W h(u) \rho(u) \dee u
\end{equation}
for any Borel function $h$ such that the integral on the right-hand
side exists; and for the Poisson process $\PoisProc W \rho$,
%
\begin{equation}\label{e:varPois}
  \Var\biggl( \sum_{x_i\in\bX} h(x_i) \biggr) = \int_W h(u)^2 \rho(u) \dee u
\end{equation}
for any Borel function $h$ such that the integral on the right-hand
side exists. Hence, the standardized version of (\ref{e:S=sumZ}) is
%
\begin{equation}\label{e:T=std.sumZ}
\hspace*{20pt}T=\biggl(S(\bx)-\hat\kappa \int_W Z(u)\,\mathrm{d}u\biggr)\Big/
\sqrt{\hat\kappa \int_W Z(u)^2\,\mathrm{d}u},\hspace*{-10pt}
\end{equation}
where $\hat\kappa = n/|W|$ is the MLE of the intensity
$\kappa=\exp(\alpha)$ under the null hypothesis. This is a direct
application of the approximation \eqref{e:T=std.S} of the signed square
root of the score test statistic.

Berman \cite{berm86} proposed several tests and diagnostics for spatial
association between a point process~$\bX$ and a covariate function\vadjust{\goodbreak}
$Z(u)$. Berman's $Z_1$ test is equivalent to the Cox score test
described above. Waller et al. \cite{walletal92} and Lawson
\cite{laws93a} proposed tests for the dependence of disease incidence
on environmental exposure, based on data giving point locations of
disease cases. These are also applications of the score test. Berman
conditioned on the number of points when making inference. This is in
accordance with the observation that the statistic $n(\bx)$ is
S-ancillary for $\beta$, while $S(\bx)$ is S-sufficient for~$\beta$.

\subsection{Threshold Alternative and Nuisance~Parameters}
\label{S:pois:thresh}

Consider the Poisson process with an intensity function of
``threshold''
form,
\[
\rho_{z,\kappa,\phi}(u) =
\cases{
\kappa \exp(\phi) & if $Z(u) \le z$, \cr
\kappa         & if $Z(u) > z$,}
\]
where $z$ is the threshold level. If $z$ is fixed, this model is a
special case of \eqref{e:rho=expZ} with $Z(u)$ replaced by
$\indicate{Z(u) \le z}$, and so \eqref{e:S=sumZ} is replaced by
\[
 S(\bx) = S(\bx, z) = \sum_{i} \indicate{Z(x_i) \le z},
\]
where $\indicate{\cdot}$ denotes the indicator function. By
\eqref{e:T=std.sumZ} the (approximate) score test of $H_0\dvtx \phi=0$
against $H_1\dvtx \phi \neq 0$ is based on
\[
T = T(z) = \bigl({S(\bx,z) - \hat\kappa A(z)}\bigr)/{\sqrt{\hat\kappa A(z)}},
\]
where $A(z) = |\{ u \in W\dvtx Z(u) \le z\}|$ is the area of the
corresponding level set of $Z$.

If $z$ is not fixed, then it plays the role of a nuisance parameter in
the score test: the value of $z$ affects inference about the canonical
parameter $\phi$, which is the parameter of primary interest in the
score test. Note that the likelihood is not differentiable with respect
to $z$.

In most applications of the score test, a nuisance parameter would be
replaced by its MLE under the null hypothesis. However, in this
context, $z$ is not identifiable under the null hypothesis.
Several 
solutions have been proposed \cite{conn01,davi77,davi87,hans96,silv96}.
They include replacing $z$ by its MLE under the alternative~\cite{conn01},
maximizing $T(z)$ or $|T(z)|$ over $z$
\cite{davi77,davi87}, and finding the maximum $p$-value of $T(z)$ or
$|T(z)|$ over a confidence region for $z$ under the alternative~\cite{silv96}.

These approaches appear to be inapplicable to the current context.
While the null distribution of $T(z)$ is asymptotically $N(0,1)$ for
each fixed $z$ as $\kappa \to\infty$, this convergence is not uniform
in $z$. The null distribution of $S(\bx,z)$ is Poisson with parameter
$\kappa A(z)$; sample paths of $T(z)$ will be governed by Poisson
behavior where $A(z)$ is small.

In this paper, our approach is simply to plot the score test statistic
as a function of the nuisance parameter. This turns the score test into
a graphical exploratory tool,
following the approach adopted in many other areas
\cite{atki82,cookweis83,preg82,chen83,wang85}.
A second style of plot based on $S(x,z) - \hat\kappa A(z)$ against $z$
may be more appropriate visually. Such a plot is the lurking variable
plot of \cite{baddetal05}. Berman \cite{berm86} also proposed a plot of
$S(\bx,z)$ against $z$, together with a plot of~$\hat\kappa A(z)$
against $z$, as a diagnostic for dependence on~$Z$. This is related to
the Kolmogorov--Smirnov test since, under $H_0$, the values $Y_i =
Z(x_i)$ are i.i.d. with distribution function $\probab(Y \le y) =
A(y)/|W|.$

\subsection{Hot Spot Alternative}

Consider the Poisson process with intensity
%
\begin{equation}\label{e:hotspot}
  \rho_{\kappa,\phi,v}(u) = \kappa \exp\bigl(\phi k(u-v)\bigr),
\end{equation}
where $k$ is a kernel\vspace*{1pt} (a probability density on $\real^2$), $\kappa >
0$ and $\phi$ are real parameters, and $v \in \real^2$ is a~nuisance
parameter. This process has a ``hot spot'' of elevated intensity in the
vicinity of the location~$v$. By~\eqref{e:T=std.sumZ} and
\eqref{e:meanPois}--\eqref{e:varPois} the score test of $H_0\dvtx \phi=0$
against $H_1\dvtx \phi \neq 0$ is based on
\[
T = T(v) = \bigl(S(\bx,v) - \hat\kappa M_1(v)\bigr)/\sqrt{\hat\kappa M_2(v)},
\]
where
\[
S(\bx,v) = \sum_{i} k(x_i - v)
\]
is the usual nonparametric kernel estimate of point process intensity
\cite{digg85} evaluated at $v$ without edge correction, and
\[
M_i(v) = \int_W k(u-v)^i \dee u,\quad   i = 1,2.
\]
The numerator $S(\bx,v) - \hat\kappa M_1(v)$ is the \textit{smoothed
residual field} \cite{baddetal05} of the null model. In the special
case where $k(u) \propto \indicate{ \| u\| \le h}$ is the uniform
density on a disc of radius $h$, the maximum $\max_v T(v)$ is closely
related to the \textit{scan statistic} \cite{alm88,kull99}.


\section{Non-Poisson Models}
\label{S:nonPois}

The remainder of the paper deals with the case where the alternative
(and perhaps also the null) is not a Poisson process. Key examples are
stated in Section \ref{S:models}. Non-Poisson models require additional
tools including the Papangelou conditional intensity
(Section \ref{S:papangelou}) and pseudo-likelihood
(Section \ref{S:pseudolik}).

\subsection{Point Process Models with Interaction}
\label{S:models}

We shall frequently consider densities of the form
%
\begin{equation}\label{E:frequent.density}
f(\bx)=c\biggl[\prod_{i}\lambda(x_i)\biggr]\exp(\phi
  V(\bx)),
\end{equation}
where $c$ is a normalizing constant, the first order term $\lambda$ is
a nonnegative function, $\phi$ is a real interaction parameter, and
$V(\bx)$ is a real nonadditive function which specifies the interaction
between the points. We refer to $V$ as the interaction potential. In
general, apart from the Poisson density \eqref{e:poisson} corresponding
to the case $\phi=0$, the normalizing constant is not expressible in
closed form.

Often the definition of $V$ can be extended to all finite point
patterns in $\real^2$ so as to be invariant under rigid motions
(translations and rotations). Then the model for $\bX$ is said to be
homogeneous if $\lambda$ is constant on $W$, and inhomogeneous
otherwise.

Let
\[
d(u,\bx) = \min_j \|u - x_j\|
\]
denote the distance from a location $u$ to its nearest neighbor in the
point configuration $\bx$. For $n(\bx)=n\ge1$ and $i=1,\ldots,n$,
define
\[
\bxminus i=\bx\setminus\{x_i\}.
\]
In many places in this paper we consider the following three
motion-invariant interaction potentials $V(\bx)=V(\bx,r)$ depending on
a parameter $r > 0$\break which specifies the range of interaction. The
\textit{Strauss process} \cite{stra75} has interaction potential
%
\begin{equation}\label{e:straussS}
  V_S(\bx,r) = \sum_{i < j} \indicate{\|x_i - x_j\| \le r},
\end{equation}
the number of $r$-close pairs of points in $\bx$; the \textit{Geyer
saturation model} \cite{geye99} with saturation threshold 1 has
interaction potential
%
\begin{equation}\label{e:geyerS}
  V_G(\bx,r) = \sum_i \indicate{ d(x_i,\bxminus i) \le r},
\end{equation}
the number of points in $\bx$ whose nearest neighbor is closer than
$r$
units; and the Widom--Rowlinson penetrable sphere model
\cite{widorowl70} or \textit{area-interaction process}~\cite{baddlies95a}
has interaction potential
%
\begin{equation}\label{e:areaS}
  V_A(\bx,r) = - \biggl| W \cap \bigcup_i B(x_i, r) \biggr|,
\end{equation}
the negative area of $W$ intersected with the union of balls $B(x_i,r)$
of radius $r$ centered at the points of $\bx$. Each of these densities
favors spatial clustering (positive association)\vadjust{\goodbreak} when $\phi > 0$ and
spatial inhibition (negative association) when $\phi < 0$. The Geyer
and area-interaction models are well-defined point processes for any
value of $\phi$ \cite{baddlies95a,geye99}, but the Strauss density is
integrable only when $\phi \le 0$ \cite{kellripl76}.


\subsection{Conditional Intensity}
\label{S:papangelou}

Consider a parametric model for a point process~$\bX$ in $\real^2$,
with parameter $\theta\in\Theta$. Papangelou \cite{papa74b} defined the
\textit{conditional intensity} of
 $\bX$ as a nonnegative stochastic process $\lambda_\theta(u,\bX)$
indexed by locations $u \in \real^2$ and characterized by the property
that
\begin{eqnarray}\label{e:GNZ}
  &&\expected_\theta\biggl[
    \sum_{x_i \in \bX} h(x_i,\bX\setminus\{x_i\})\biggr]\nonumber\\ [-9pt]\\ [-9pt]
&&\quad=\expected_\theta\biggl[
    \int_{\real^2} h(u,\bX) \lambda_\theta(u,\bX) \dee u\biggr]\nonumber
\end{eqnarray}
for all measurable functions $h$ such that the left or right-hand side
exists. Equation (\ref{e:GNZ}) is known as the
\textit{Georgii}--\textit{Nguyen}--\textit{Zessin (GNZ) formula}
[\cite*{geor76,kall78}, \cite*{kall84,nguyzess79a}]; see also Section 6.4.1 in
\cite{mollwaag04}. Adapting a term from stochastic process theory, we
will call the random integral on the right-hand side of (\ref{e:GNZ}) the
 \textit{(Papangelou) compensator} of the random sum on the left-hand side.

Consider a finite point process $\bX$ in $W$. In the unconditional case
(Section~\ref{S:assump:uncon}) we assume $\bX$ has density
$f_\theta(\bx)$ which is hereditary for all $\theta \in \Theta$. We may
simply define
%
\begin{equation}\label{e:cif:u}
  \lambda_\theta(u,\bx)
  = 
    f_\theta(\bxplus u)
    /f_\theta(\bx)
\end{equation}
for all locations $u \in W$ and point configurations $\bx\subset W$
such that $u \notin \bx$. Here we take $0/0 = 0$. For $x_i \in\bx$ we
set $\lambda_\theta(x_i,\bx) = \lambda_\theta(x_i, \bxminus i)$, and
for $u\notin W$ we set $\lambda_\theta(u,\bx)=0$. Then it may be
verified directly from~(\ref{e:f}) that (\ref{e:GNZ}) holds, so that
(\ref{e:cif:u}) is the Papangelou conditional intensity of $\bX$. Note
that the normalizing constant of $f_\theta$ cancels in \eqref{e:cif:u}.
For a~Poisson process, it follows from \eqref{e:poisson}  and
\eqref{e:cif:u} that the Papangelou conditional intensity is equivalent
to the intensity function of the process.

In the conditional case (Section \ref{S:assump:cond}) we assume that
 the conditional distribution of $\bXfree = \bX \cap \Wfree$
given $\bXfixed = \bX \cap \Wfixed = \bxfixed$ has a hereditary density
$f_\theta(\bxfree \vert \bxfixed)$ with respect to $\PoisProc \Wfree 1$,
for all \mbox{$\theta \in \Theta$}. Then define
%
\begin{equation}\label{e:cif:cond:u}
\lambda_\theta(u,\bxfree \vert \bxfixed)
  = \frac{
    f_\theta(\bxfreeplus u \vert \bxfixed)
  }{f_\theta(\del{\bxfree}{u} \vert \bxfixed)
  }
\end{equation}
if $u \in \Wfree$, and zero otherwise. It can similarly be verified
that this is the Papangelou conditional intensity of the conditional
distribution of $\bXfree$ given $\bXfixed = \bxfixed$.

It is convenient to rewrite (\ref{e:cif:u}) in the form
\[
\lambda_\theta(u,\bx) = \exp(\Delta_u \log f(\bx)),
\]
where $\Delta$ is the one-point difference operator
%
\begin{equation}\label{e:delta}
  \Delta_u h(\bx) = h(\bxplus u) - h(\del{\bx}{u}).
\end{equation}
Note the Poincar\'e inequality for the Poisson process~$\bX$,
%
\begin{equation}\label{e:Poincare}
     \Var[ h(\bX)] \le
    \expected \int_W [ \Delta_u  h(\bX) ]^2 \rho(u) \dee u
\end{equation}
holding for all measurable functionals $h$ such that the right-hand
side is finite; see \cite{lastpenr11,wu00}.


\subsection{Pseudo-Likelihood and Pseudo-Score}
\label{S:pseudolik}

To avoid computational 
problems with point process likelihoods, Besag \cite{besa78} introduced
the \textit{pseudo-likelihood} function
%
\begin{eqnarray}\label{e:PL}
\PL(\theta) &=&
    \biggl[
      \prod_{i} \lambda_\theta(x_i,\bx)
    \biggr]\nonumber\\ [-8pt]\\ [-8pt]
    &&{}\cdot
    \exp\biggl(
        - \int_{W} \lambda_\theta(u,\bx) \dee u\biggr).\nonumber
\end{eqnarray}
This is of the same functional form as the likelihood function of a
Poisson process (\ref{e:poisson}), but has the Papangelou conditional
intensity in place of the Poisson intensity.
The corresponding \textit{pseudo-score}
%
\begin{eqnarray}\label{e:PU}
  \hspace*{22pt}\PU(\theta) &=& \frac{\partial}{\partial\theta} \log
  \PL(\theta)\nonumber\\ [-8pt]\\ [-8pt]
    \hspace*{22pt}&=& \sum_{i} \ddt \log \lambda_\theta(x_i,\bx)
                   - \int_{W} \ddt \lambda_\theta(u,\bx) \dee u\nonumber
\end{eqnarray}
is an unbiased estimating function, $\expected_\theta \PU(\theta) = 0$,
by virtue of (\ref{e:GNZ}). In practice, the pseudo-likelihood is
applicable only if the Papangelou conditional intensity
$\lambda_\theta(u,\bx)$ is tractable.

The pseudo-likelihood function can also be defined in the conditional
case \cite{jensmoll91}. In \eqref{e:PL} the product is instead over
points $x_i\in\bxfree$ and the integral is instead over $\Wfree$; in
\eqref{e:PU} the sum is instead over points $x_i\in\bxfree$ and the
integral is instead over $\Wfree$; and in both places $\bx =
\bxfree\cup \bxfixed$. The Papangelou conditional intensity
$\lambda_\theta(u,\bx)$ must also be replaced by $\lambda_\theta(u,
\bxfree \vert \bxfixed)$.

\subsection{Markov Point Processes}
\label{S:Markov}

For a point process $\bX$ constructed as $\bX = \bY \cap W$ where $\bY$
is a point process in $\real^2$, the density and
Papangelou conditional intensity
of $\bX$ may not be available in simple form. Progress can be made if
$\bY$ is a \textit{Markov point process} of interaction range $R <
\infty$; see \cite{geor76,nguyzess79a,riplkell77,lies00} and
\cite{mollwaag04}, Section 6.4.1. Briefly, this means that\vadjust{\goodbreak} the Papangelou
conditional intensity $\lambda_\theta(u,\bY)$ of $\bY$ satisfies
$\lambda_\theta(u,\bY) = \lambda_\theta(u, \bY \cap\break B(u,R))$, where
$B(u,R)$ is the ball of radius $R$ centered at $u$. Define the erosion
of $W$ by distance $R$,
\[
  W_{\ominus R} = \{ u \in W\dvtx B(u,R) \subset W\},
\]
and assume this has nonzero area. Let $B = W\setminus W_{\ominus R}$ be
the border region. The process satisfies a spatial Markov property: the
processes $\bY \cap W_{\ominus R}$ and $\bY \cap W^c$ are conditionally
independent given  $\bY \cap B$.

In this situation we shall invoke the conditional case with $\Wfree=
W_{\ominus R}$ and $\Wfixed= W\setminus\Wfree$. The conditional
distribution of $\bX \cap \Wfree$ given  $\bX \cap \Wfixed = \bxfixed$
has Papangelou conditional intensity
%
\begin{equation}\label{e:lam=lam}
  \lambda_\theta(u,\bxfree \vert \bxfixed)
  = \cases{
      \lambda_\theta(u,\bxfree \cup \bxfixed)
      & if $u \in \Wfree$, \cr
      0   & otherwise.}\hspace*{-33pt}
\end{equation}
Thus, \textit{the unconditional and conditional versions of a~Markov
point process have the same Papangelou conditional intensity} at
locations in $\Wfree$.

For $\bxfree=\{x_1,\ldots,x_{\free n}\}$, the conditional probability
density given $\bxfixed$ becomes
\begin{eqnarray*}
&&f_\theta(\bxfree \vert \bxfixed)\\
&&\quad = c_\theta(\bxfixed)
           \lambda_\theta(x_1, \bxfree)
            \prod_{i=2}^{\free n} \lambda_\theta(x_i,\{x_1,\ldots,x_{i-1}\}\cup\bxfixed)
\end{eqnarray*}
if $\free n > 0$, and $f_\theta(\emptyset \vert \bxfixed) =
c_\theta(\bxfixed)$, where $\emptyset$ denotes the empty configuration,
and the inverse normalizing constant $c_\theta(\bxfixed)$ depends only
on $\bxfixed$.

For example, instead of \eqref{E:frequent.density} we now consider
\[
 f(\bxfree \vert \bxfixed)=c(\bxfixed)
 \Biggl[\prod_{i=1}^{\free n}\lambda(x_i)\Biggr]\exp\bigl(\phi
  V(\bxfree \cup \bxfixed)\bigr),
\]
assuming $V(\by)$ is defined for all finite $\by\subset\real^2$ such
that for any
 $u\in\real^2\setminus\by$, $\Delta_u V(\by)$ depends only on
 $u$ and $\by\cap B(u,R)$. This condition is satisfied by the
interaction potentials \eqref{e:straussS}--\eqref{e:areaS};
note that the range of interaction is $R=r$ for the Strauss process,
and $R=2r$ for both the Geyer and the area-interaction models.

\section{Score, Pseudo-Score and Residual~Diagnostics}
\label{S:theory}

This section develops the general theory for our diagnostic tools.

By \eqref{e:T=std.S} in Section~\ref{S:scoretest} it is clear that
comparison of a~
summary statistic~$S(\bx)$ to its predicted va\-lue~$\expected S(\bX)$
under a null model is effectively equivalent to the score test under an
exponential family model where $S(\bx)$ is the canonical sufficient
statistic. Similarly, the use of a \textit{functional} summary statistic
$S(\bx,z)$, depending on a function argument $z$, is related to the\vadjust{\goodbreak}
score test under an exponential family model \textit{where $z$ is a
nuisance parameter} and $S(\bx, z)$ is the canonical sufficient
statistic for fixed
$z$.
In this section we construct the corresponding exponential family
models, apply the score test, and propose surrogates for the score test
statistic.

\subsection{Models}
\label{S:theory:models}

Let $f_\theta(\bx)$ be the density of any point process $\bX$ on $W$
governed by a parameter $\theta$. Let $S(\bx, z)$ be a functional
summary statistic of the point pattern data set $\bx$, with function
argument $z$ belonging to any space.

Consider the \textit{extended model} with density
%
\begin{equation}\label{e:extended}
  f_{\theta,\phi,z}(\bx) = c_{\theta,\phi,z} f_\theta(\bx) \exp(\phi S(\bx,z)),
\end{equation}
where $\phi$ is a real parameter, and $c_{\theta,\phi,z}$ is the
normalizing constant. The density is well-defined provided
\[
M(\theta, \phi,z) = \expected[ f_\theta(\bY) \exp(\phi S(\bY,z)) ]
< \infty,
\]
where $\bY \sim \PoisProc W 1$.
 The extended model is constructed by
``exponential tilting'' of the original model by the statistic $S$. By
\eqref{e:T=std.S}, for fixed $\theta$ and $z$, assuming
differentiability of $M$ with respect to $\phi$ in a neighborhood of
$\phi=0$, the signed root of the score test statistic
is approximated by
%
\begin{equation}
  \label{e:scoretestS}
    T = \bigl({S(\bx, z) -
\expected_{\hat\theta}[ S(\bX,z)]}\bigr)/\sqrt{
\Var_{\hat\theta}[S(\bX,z)]},\hspace*{-28pt}
\end{equation}
where $\hat\theta$ is the MLE under the null model, and the expectation
and variance are with respect to the null model with density
$f_{\hat\theta}$.

Insight into the qualitative behavior of the extended model
(\ref{e:extended}) can be obtained by studying  the \textit{perturbing
model}
%
\begin{equation}\label{e:perturbing}
    g_{\phi,z}(\bx) = k_{\phi,z} \exp(\phi S(\bx,z)),
\end{equation}
provided this is a well-defined density with respect to $\PoisProc W
1$, where $k_{\phi,z}$ is the normalizing constant. When the null
hypothesis is a homogeneous Poisson process, the extended model is
identical to the perturbing model, up to a change in the first order
term. In general, the extended model is a qualitative hybrid between
the null and perturbing models.

In this context the score test is equivalent to naive comparison of the
observed and null-expected values of the functional summary statistic
$S$. The test statistic $T$ in \eqref{e:scoretestS} may be difficult to
evaluate; typically, apart from Poisson models,
 the moments (particularly the variance) of $S$ would
not be available in closed form. The null distribution of $T$ would
also typically be unknown. Hence, implementation of the score test
would typically\vadjust{\goodbreak} require moment approximation and simulation from the
null model, which in both cases may be computationally expensive.
Various approximations for the score or the score test statistic can be
constructed, as discussed in the sequel.

\subsection{Pseudo-Score of Extended Model}
\label{S:theory:pseudo-score}


The extended model \eqref{e:extended} is an exponential family with
respect to $\phi$, having Papangelou conditional intensity
\[
  \kappa_{\theta,\phi,z}(u,\bx) = \lambda_\theta(u,\bx) \exp(\phi \Delta_u S(\bx, z)),
\]
where $\lambda_\theta(u,\bx)$ is the Papangelou conditional intensity
of the null model. The pseudo-score function with respect to $\phi$,
evaluated at $\phi = 0$, is
\[
  \PU(\theta,z) = \sum_{i} \Delta_{x_i} S(\bx,z)-\int_{W} \Delta_u S(\bx,z) \lambda_\theta(u,\bx) \dee u,
\]
where the first term
%
\begin{equation}\label{e:pseudo-sum}
  \psum S(\bx,z) = \sum_{i} \Delta_{x_i} S(\bx,z)
\end{equation}
will be called the \textit{pseudo-sum} of $S$. If $\hat\theta$ is the
maximum pseudo-likelihood estimate (MPLE) under $H_0$, the second term
with $\theta$ replaced by $\hat\theta$ becomes
%
\begin{equation}\label{e:pseudocomp}
  \pcom S (\bx, z) = \int_{W} \Delta_u S(\bx,z) \lambda_{\hat\theta}(u,\bx) \dee u
\end{equation}
and will be called the \textit{(estimated) pseudo-compensator} of $S$.
We call 
%
\begin{eqnarray}\label{e:PU:star}
\pres S(\bx,z)&=&\PU(\hat\theta,z)\nonumber\\ [-9pt]\\ [-9pt]
& =&\psum S(\bx,z)-\pcom S (\bx, z)\nonumber
\end{eqnarray}
%
the \textit{pseudo-residual} since it is a weighted residual in the sense
of \cite{baddetal05}.

The pseudo-residual serves as a surrogate for the numerator in the
score test statistic \eqref{e:scoretestS}. For the denominator, we need
the variance of the pseudo-residual. Appendix~\ref{App:var} gives an
exact formula \eqref{e:var.PU} for the variance of the pseudo-score
$\PU(\theta,z)$, which can serve as an approximation to the variance of
the pseudo-residual $\pres S(\bx,z)$. This is likely to be an
overestimate, because the effect of parameter estimation is typically
to deflate the residual variance~\cite{baddetal05}.

The first term in the variance formula \eqref{e:var.PU} is
%
\begin{equation}\label{e:pseudovar}
  \pvar S (\bx, z) = \int_{W} [\Delta_u S(\bx,z)]^2 \lambda_{\hat\theta}(u,\bx) \dee u,\hspace*{-10pt}
\end{equation}
which we shall call the \textit{Poincar\'e pseudo-variance} because of
its similarity to the Poincar\'e upper bound in \eqref{e:Poincare}. It
is easy to compute this quantity alongside the pseudo-residual.\vadjust{\goodbreak} Rough
calculations in Sections \ref{S:varcalc:VS} and~\ref{S:varcalc:VG}
suggest that the Poincar\'e pseudo-variance is likely to be the
dominant term in the variance, except at small $r$ values. The variance
of residuals is also studied in \cite{coeulava10}.

For computational efficiency we propose to use the square root of
\eqref{e:pseudovar} as a surrogate for the denominator in
\eqref{e:scoretestS}. This yields a \textit{``standardized''
pseudo-residual}
%
\begin{equation}\label{e:std.pseudo-res}\qquad
\pstd S(\bx,z) = \pres S(\bx,z)/\sqrt{\pvar
S(\bx, z)}.
\end{equation}
We emphasize that this quantity is not guaranteed to have zero mean and
unit variance (even approximately) under the null hypothesis. It is
merely a~computationally efficient surrogate for the score test
sta\-tistic; its null distribution must be investigated by other means.
Asymptotics of $\pstd S(\bx,z)$ under\break a~large-domain limit
\cite{stei95} could be studied, but limit results are unlikely to hold
uniformly over $r$. In this paper we evaluate null distributions using
Monte Carlo methods.

The pseudo-sum (\ref{e:pseudo-sum}) can be regarded as a~functional
summary statistic for the data in its own right. Its definition depends
only on the choice of the statistic $S$, and it may have a meaningful
interpretation as a nonparametric estimator of a property of the point
process. The pseudo-compensator~(\ref{e:pseudocomp}) might also be
regarded as a functional summary statistic, but its definition involves
the null model. If the null model is true, we may expect the
pseudo-residual to be approximately zero.
Sections~\ref{S:pairwise}--\ref{S:area}
and Appendix~\ref{App:further}
study particular instances of pseudo-residual diagnostics based on
\eqref{e:pseudo-sum}--(\ref{e:PU:star}).

In the conditional case, the Papangelou conditional intensity
$\lambda_{\hat\theta}(u,\bx)$ must be\vspace*{1pt} replaced by
$\lambda_{\hat\theta}(u,\break \bxfree \vert \bxfixed)$ given in
\eqref{e:cif:cond:u} or (\ref{e:lam=lam}). The integral in the
definition of the pseudo-compensator \eqref{e:pseudocomp} must be
restricted to the domain $\Wfree$, and
the summation over data points in \eqref{e:pseudo-sum} must be
restricted to points $x_i \in \Wfree$, that is, to summation over
points of $\bxfree$.

\subsection{Residuals}
\label{S:theory:residual}

A simpler surrogate for the score test is available when the canonical
sufficient statistic $S$ of the perturbing model is naturally
expressible as a sum of local contributions
%
\begin{equation}\label{e:sumoflocal}
  S(\bx,z) = \sum_{i} s(x_i, \bxminus i,z).
\end{equation}
Note that any statistic can be decomposed in this way unless some
restriction is imposed on $s$; such a~decomposition is not necessarily
unique. We call the decomposition ``natural'' if $s(u,\bx,z)$ only
depends on points\vadjust{\goodbreak} of $\bx$ that are close to $u$, as demonstrated in
the examples in Sections~\ref{S:pairwise},
\ref{S:nndist} and \ref{S:area}
and in
Appendix~\ref{App:further}.

Consider a null model with Papangelou conditional intensity
$\lambda_\theta(u,\bx)$. Following \cite{baddetal05}, define the
{($s$-weight\-ed) innovation} by
%
\begin{equation}\label{e:innovlocal}
 \hspace*{25pt} \innov S(\bx,r) =
  S(\bx, z)
  - \int_{W} s(u,\bx,z) \lambda_\theta(u, \bx) \dee u,
\end{equation}
which by the GNZ formula (\ref{e:GNZ}) has mean zero under the null
model. In practice, we replace $\theta$ by an estimate $\hat\theta$
(e.g., the MPLE) and consider the \textit{($s$-weighted) residual}
%
\begin{equation}\label{e:residuallocal}
 \resid S(\bx,z) = S(\bx, z)
  - \int_{W} s(u,\bx,z) \lambda_{\hat\theta}(u, \bx) \dee u.\hspace*{-25pt}
\end{equation}
The residual shares many properties of the score function and can serve
as a computationally efficient surrogate for the score. The
data-dependent integral
%
\begin{equation}\label{e:compensatorlocal}
  \compen S(\bx,z) = \int_{W} s(u,\bx,z) \lambda_{\hat\theta}(u,\bx) \dee u
\end{equation}
is the \textit{(estimated) Papangelou compensator} of $S$. The variance
of $\resid S(\bx, z)$ can be approximated by the innovation variance,
given by the general variance formula \eqref{e:var.innov} of
Appendix~\ref{App:var}. The first term in~\eqref{e:var.innov} is the
\textit{Poincar\'e variance}
%
\begin{equation}\label{e:compvar}
  \compvar S(\bx,z) = \int_{W} s(u,\bx,z)^2 \lambda_{\hat\theta}(u,\bx) \dee u.
\end{equation}
Rough calculations reported in Sections~\ref{S:varcalc:VS} and
\ref{S:varcalc:VG} suggest that the Poincar\'e variance is likely to be
the largest term in the variance for sufficiently large $r$. By analogy
with \eqref{e:pseudovar} we propose to use the Poincar\'e variance as
a~surrogate for the variance of $\resid S(\bx,z)$, and thereby obtain a
``standardized'' residual
%
\begin{equation}\label{e:standcompvar}
\stdres S(\bx,z) = \resid S(\bx,z)/\sqrt{\compvar S(\bx,z)}.
\end{equation}
Once again $\stdres S(\bx,z)$ is not exactly standardized, because
$\compvar S(\bx,z)$ is an approximation to $\Var[\resid S(\bx,z)]$ and
because the numerator and denominator of \eqref{e:standcompvar} are
dependent. The null distribution of $\stdres S(\bx,z)$\break must be
investigated by other means.

In the conditional case, the integral in the definition of the
compensator \eqref{e:compensatorlocal} must be restricted to the domain
$\Wfree$, and the summation over data points in \eqref{e:sumoflocal}
must be restricted to points $x_i \in \Wfree$, that is, to summation
over points of $\bxfree$.

\section{Diagnostics for First Order Trend}
\label{S:firstorder}

Consider any null model with density $f_\theta(\bx)$ and Papangelou
conditional intensity $\lambda_\theta(u,\bx)$. By analogy with\vadjust{\goodbreak}
Section~\ref{S:scorePoisson} we consider alternatives of the form
\eqref{e:extended} where
\[
 S(\bx,z) = \sum_{i} s(x_i,z)
\]
for some function $s$. The perturbing model \eqref{e:perturbing} is
a~Poisson process with intensity $\exp(\phi s(\cdot, z))$,\break where~$z$ is a
nuisance parameter. The score test is a~test for the presence of an
(extra) first order trend. The pseudo-score and residual diagnostics
are both equal to
%
\begin{eqnarray}\label{e:residual:trend}
\resid S(\bx,z) &=&
   \sum_{i} s(x_i,z)\nonumber\\ [-8pt]\\ [-8pt]
  &&{} - \int_W s(u,z) \lambda_{\hat\theta}(u,\bx) \dee u.\nonumber
\end{eqnarray}
This is the $s$-weighted residual described in \cite{baddetal05}. The
variance of \eqref{e:residual:trend} can be estimated by simulation, or
approximated by the Poincar\'e variance \eqref{e:compvar}.

If $Z$ is a real-valued covariate function on $W$, then we may take
$s(u,z) = \indicate{Z(u) \le z}$ for $z \in \real$, corresponding to a
threshold effect (cf. Section~\ref{S:pois:thresh}). A~plot of
(\ref{e:residual:trend}) against $z$ was called a \textit{lurking
variable plot} in \cite{baddetal05}.

If $s(u,z) = k(u - z)$ for $z \in \real^2$, where $k$ is a 
density function on $\real^2$, then 
\[
  \resid S(\bx,z) = \sum_{i} k(x_i - z)
  - \int_W k(u - z) \lambda_{\hat\theta}(u,\bx) \dee u,
\]
which was dubbed the \textit{smoothed residual field} in
\cite{baddetal05}. Examples of application of these techniques have
been discussed extensively in \cite{baddetal05}.

\section{Interpoint Interaction}
\label{S:interaction}

In the remainder of the paper we concentrate on diagnostics for
interpoint interaction.

\subsection{Classical Summary Statistics}

Following Ripley's influential paper \cite{ripl77}, it is standard
practice, when investigating association or dependence between points
in a spatial point pattern, to evaluate functional summary statistics
such as the $K$-function, and to compare graphically the empirical
summaries and theoretical predicted values under a suitable model,
often a stationary Poisson process (``Complete Spatial Randomness,'' CSR)
\cite{ripl77,cres91,digg03}.

The three most popular functional summary sta\-tistics for spatial point
processes are Ripley's $K$-func\-tion, the
nearest neighbor
distance distribution function~$G$ and the empty space function
(spherical contact distance distribution function) $F$. Definitions of\vadjust{\goodbreak}
$K$, $G$ and $F$ and their estimators can be seen in
\cite{badd99b,cres91,digg03,mollwaag04}. Simple empirical estimators of
these functions are of the form
%
\begin{eqnarray}\label{e:Khat}
  \Khat(r) &=& \Khat_\bx(r)\nonumber\hspace*{-25pt}\\ [-8pt]\\ [-8pt]
 &=& \frac{1}{\hat{\rho}{}^2(\bx) |W|}
  \sum_{i\neq j} 
       e_K(x_i,x_j) \indicate{\|x_i - x_j\| \le r}, \nonumber\hspace*{-25pt}\\\label{e:Ghat}
  \Ghat(r) &=& \Ghat_\bx(r)\nonumber\hspace*{-25pt}\\ [-8pt]\\ [-8pt]
 &=& \frac{1}{n(\bx)} \sum_{i}
              e_G(x_i,\bxminus i, r) \indicate{ d(x_i,\bxminus i) \le r},\nonumber\hspace*{-25pt}\\  \label{e:Fhat}
  \Fhat(r) &=& \Fhat_\bx(r)\nonumber\hspace*{-25pt}\\ [-8pt]\\ [-8pt]
& = &\frac{1}{|W|}
                 \int_W e_F(u,r) \indicate{ d(u,\bx) \le r} \dee u,\nonumber\hspace*{-25pt}
\end{eqnarray}
where $e_K(u,v)$, $e_G(u,\bx, r)$ and $e_F(u,r)$ are edge correction
weights, and typically $\hat{\rho}{}^2(\bx) = n(\bx)
(n(\bx)-1)/|W|^2$.

\subsection{Score Test Approach}

The classical approach fits naturally into the\break scheme of
Section~\ref{S:theory}. In order to test for dependence between points,
we choose a perturbing model that exhibits dependence. Three
interesting examples of perturbing models
are the Strauss process, the Geyer saturation model with saturation
threshold 1 and the area-interaction process, with interaction
potentials $V_S(\bx,r)$, $V_G(\bx,r)$ and $V_A(\bx,r)$ given in
\eqref{e:straussS}--\eqref{e:areaS}. The nuisance parameter $r \ge 0$
determines the range of interaction. It is interesting to note that,
although the Strauss density is integrable only when $\phi \le 0$, the
extended model obtained by perturbing $f_\theta$ by the Strauss density
may be well-defined for some $\phi > 0$. This extended model may
support alternatives that are clustered relative to the null, as
originally intended by Strauss \cite{stra75}.

The potentials of these three models are closely related to the summary
statistics $\Khat, \Ghat$ and $\Fhat$
in~\eqref{e:Khat}--\eqref{e:Fhat}. Ignoring the edge correction weights
$e(\cdot)$, we have
\begin{eqnarray}\label{e:K=VS}
  \Khat_\bx(r) &\approx& \frac{2|W|}{n(\bx) (n(\bx)-1)} V_S(\bx,r), \\ \label{e:G=VG}
  \Ghat_\bx(r) &\approx& \frac 1 {n(\bx)} V_G(\bx,r), \\ \label{e:F=VA}
  \Fhat_\bx(r) &\approx& - \frac 1 {|W|} V_A(\bx,r) .
\end{eqnarray}

To draw the closest possible connection with the score test, instead of
choosing  the Strauss, Geyer~or area-interaction process as the
perturbing model,~we shall take the perturbing\vadjust{\goodbreak} model to be defined
through (\ref{e:perturbing}) where $S$ is one of the statistics
$\Khat$, $\Ghat$~or~$\Fhat$. We call these the \textit{(perturbing)
$\Khat$-model}, \textit{$\Ghat$-model} and \textit{$\Fhat$-model},
respectively. The score test is then precisely equivalent to comparing
$\Khat$, $\Ghat$ or $\Fhat$ with its predicted expectation using
\eqref{e:T=std.S}.

Essentially $\Khat$, $\Ghat$, $\Fhat$ are renormalized versions\break
of~$V_S$, $V_G$, $V_A$ as shown in \eqref{e:K=VS}--\eqref{e:F=VA}. In the case
of $\Fhat$ the renormalization is not data-dependent, so the
$\Fhat$-mo\-del is virtually an area-interaction model, ignoring edge
correction. For $\Khat$, the renormalization depends only on $n(\bx)$,
and so, conditionally on $n(\bx) = n$, the $\Khat$-model and the
Strauss process are approximately equivalent. Similarly for $\Ghat$,
the normalization also depends only on $n(\bx)$, so, conditionally on
$n(\bx) = n$, the $\Ghat$-model and Geyer saturation process are
approximately equivalent.
If we follow Ripley's~\cite{ripl77} recommendation to condition on $n$
when testing for interaction, this implies that the use of the~$K$, $G$
or $F$-function is approximately equivalent to the score test of CSR
against a Strauss,
    Geyer  or area-interaction alternative, respectively.

When the null hypothesis is CSR, we saw that the extended model
(\ref{e:extended}) is identical to the perturbing model, up to a change
in intensity, so that the use of the $\Khat$-function is equivalent to
testing the null hypothesis of CSR against the alternative of
a~$\Khat$-model; similarly for $\Ghat$ and $\Fhat$. For a more general
null hypothesis, the use of the $\Khat$-function, for example,
corresponds to adopting an alternative hypothesis that is a hybrid
between the fitted model and a $\Khat$-model.

Note that if the edge correction weight $e_K(u,v)$ is uniformly
bounded, the $\Khat$-model is integrable for all values of $\phi$,
avoiding a difficulty with the Strauss process \cite{kellripl76}.

Computation of the score test statistic \eqref{e:scoretestS} requires
estimation\vspace*{1pt} or approximation of the null variance of $\Khat(r)$,
$\Ghat(r)$ or $\Fhat(r)$. A wide variety of approximations is available
when the null hypothesis is CSR \cite{ripl88,digg03}. For other null
hypotheses, simulation estimates would typically be used. A central
limit theorem is available for  $\Khat(r)$, $\Ghat(r)$ and $\Fhat(r)$
in the large-domain limit, for example,\
\cite{badd80b,hein88,hein88b,joli80,ripl88}. However, convergence is
not uniform in $r$, and the normal approximation will be poor for small
values of $r$. Instead Ripley \cite{ripl76} developed an exact Monte
Carlo test \cite{barn63,hope68} based on simulation envelopes of the
summary statistic under the null hypothesis.

In the following sections we develop the residual and pseudo-residual
diagnostics corresponding to this approach.

\section{Residual Diagnostics for Interaction Using Pairwise
Distances}\vspace*{3pt}
\label{S:pairwise}

This section develops residual \eqref{e:residuallocal} and
pseudo-residual \eqref{e:PU:star} diagnostics derived from a summary
statistic $S$ which is a sum of contributions depending on pairwise
distances.

\subsection{Residual Based on Perturbing Strauss Model}\vspace*{3pt}

\subsubsection{General derivation}

Consider any statistic of the general ``pairwise interaction'' form
%
\begin{equation}\label{E:pairwise}
S(\bx,r) =\sum_{i<j} q(\{x_i,x_j\},r).
\end{equation}
This can be decomposed in the local form \eqref{e:sumoflocal} with
\[
  s(u, \bx,r)=\frac 1 2 \sum_{i}q(\{x_i,u\},r),\quad  u \notin\bx.
\]
%
Hence,
\begin{eqnarray*}
  \Delta_{x_i}S(\bx,r)&=&2s(x_i, \bxminus i,r)\quad
    \mbox{and}\\[2pt]
  \Delta_{u}S(\bx,r)&=&2s(u, \bx,r),\quad  u \notin\bx.
\end{eqnarray*}
Consequently, the pseudo-residual and the pseudo-compensator are just
twice the residual and the Papangelou compensator:
\begin{eqnarray}\label{E:pairwise_relation1}
\hspace*{30pt}\psum S(\bx,r) &=& 2 S(\bx,r) = \sum_{i\neq j} 
q(\{x_i,x_j\},r), \\\label{E:pairwise_relation2}
\hspace*{30pt}\pcom S(\bx,r) &=& 2 \compen
S(\bx,r)\nonumber\\ [-8pt]\\ [-8pt]
\hspace*{30pt}&=&\int_{W} \sum_{i}q(\{x_i,u\},r)
\lambda_{\hat\theta}(u,\bx) \dee u, \nonumber\\\label{E:pairwise_relation3}
 \hspace*{30pt}\pres S(\bx,z) &=& 2\resid S(\bx,r)\nonumber\\ [-8pt]\\ [-8pt]
\hspace*{30pt}&=&2
S(\bx,r)-2 \compen S(\bx,r).\nonumber
\end{eqnarray}

\subsubsection{Residual of Strauss potential}

The Strauss interaction potential $V_S$ of \eqref{e:straussS} is of the
general\break form~\eqref{E:pairwise} with $q(\{x_i,x_j\},r)=\indicate{\|x_i
- x_j\| \le r}$.\break Hence, $V_S$ can be decomposed in the form
\eqref{e:sumoflocal} with $s(u,\bx,r) = \frac 1 2 t(u,\bx,r)$, where
\[
 t(u,\bx,r) = \sum_i \indicate{\| u - x_i \| \le r},\quad  u\notin\bx.
\]
Hence, the Papangelou compensator of $V_S$ is
%
\begin{equation}\label{e:compen.Strauss}
  \compen V_S(\bx, r) = \frac 1 2 \int_W
  t(u,\bx,r) \lambda_{\hat\theta}(u,\bx) \dee u.
\end{equation}

\subsubsection{Case of CSR}
\label{S:Kres:CSR}

If the null model is CSR with intensity $\rho$ estimated by
$\hat\rho=n(\bx)/|W|$ (the MLE, which agrees with the
MPLE in this case),\vadjust{\goodbreak} the Papangelou compensator \eqref{e:compen.Strauss}
becomes
\begin{eqnarray*}
   \compen V_S(\bx,r) &=&
   \frac {\hat\rho} 2 \int_W \sum_i \indicate{\|u - x_i\| \le r} \dee
u\\
   &=& \frac {\hat\rho} 2 \sum_i |W \cap B(x_i,r)|.
\end{eqnarray*}
Ignoring edge effects, we have $|W \cap B(x_i, r)| \approx \pi r^2$
and, applying \eqref{e:K=VS}, the residual is approximately
%
\begin{equation}\label{E:verygood}
   \resid V_S(\bx,r) \approx \frac{n(\bx)^2}{2|W|}
   [ \Khat_{\bx}(r) - \pi r^2 ].
\end{equation}
The term in brackets is a commonly-used measure of departure from CSR,
and is a sensible diagnostic because $K(r) = \pi r^2$ under CSR.




\subsection{\texorpdfstring{Residual Based on Perturbing $\Khat$-Model}{Residual Based on Perturbing K-Model}}

Assuming $\hat{\rho}^2(\bx)= \hat{\rho}^2(n(\bx))$ depends only
on $n(\bx)$, the empirical $K$-function (\ref{e:Khat}) can also be
expressed as a sum of local contributions $
  \Khat_\bx(r) = \sum_i k(x_i,\break \bxminus i, r)
$ with
\[
    k(u,\bx, r) = \frac{t^w(u, \bx,r)}{\hat{\rho}^2(n(\bx)+1)
      |W|},\quad  u\notin \bx,
\]
where
\[
  t^w(u,\bx,r) = \sum_j e_K(u,x_j) \indicate{\|u - x_j\| \le r}
\]
is a weighted count of the points of $\bx$ that are $r$-close to the
location $u$. Hence, the compensator of the $\Khat$-function is
%
\begin{eqnarray}\label{E:com}
\compen \Khat_\bx(r) &=&
\frac{1}{\hat{\rho}^2(n(\bx)+1) |W|}\nonumber\\ [-8pt]\\ [-8pt]
      &&{}\cdot\int_W t^w(u, \bx,r) \lambda_{\hat\theta}(u, \bx) \dee u .\nonumber
\end{eqnarray}

Assume the edge correction weight $e_K(u,v) =\break e_K(v,u)$ is symmetric;
for example, this is satisfied by the Ohser--Stoyan edge correction
weight \cite{ohsestoy81,ohse83} given by $e_K(u,v)=1/|W_u\cap W_v|$
where $W_u=\{u+v\dvtx\break v\in W\}$, but not by Ripley's \cite{ripl76} isotropic
correction weight. Then the increment is, for $u \notin \bx$,
\begin{eqnarray*}
    \Delta_u \Khat_{\bx}(r) 
                &=& \frac{
                  \hat{\rho}^2(\bx)
                  - \hat{\rho}^2(\bxplus u)
                }{
                  \hat{\rho}^2(\bxplus u)
                }
                \hat K_\bx(r)\\
                &&{}+
                \frac{
                  2 t^w(u, \bx, r)}{\hat{\rho}^2(\bxplus u)|W|}
\end{eqnarray*}
and 
when $x_i \in \bx$
\begin{eqnarray*}
    \Delta_{x_i} \Khat_{\bx}(r)
    &=& \frac{
      \hat{\rho}^2(\bxminus i)
      - \hat{\rho}^2(\bx)
    }{
      \hat{\rho}^2(\bxminus i)
    }
    \hat K_\bx(r)\\
    &&{}+
    \frac{
      2 t^w(x_i, \bx_{-i}, r)
    }{
      \hat{\rho}^2(\bxminus i)|W|
    }.
\end{eqnarray*}
Assuming the standard estimator $\hat{\rho}^2(\bx) = n(n-1)/\break|W|^2$
with $n=n(\bx)$, the pseudo-sum is seen to be zero, so the
pseudo-residual is apart from the sign equal to the pseudo-compensator,
which becomes
\[
\pcom \Khat_{\bx}(r) = 2 \compen \Khat_\bx(r) - \biggl[\frac{2}{n-2}
\int_W\lambda_{\hat\theta}(u,\bx)\,\mathrm{d}u\biggr] \Khat_{\bx}(r),
\]
where
$\compen \Khat_\bx(r)$ is given by \eqref{E:com}. So if the null model
is CSR and the intensity is estimated by $n/|W|$, the pseudo-residual
is approximately $2[\hat K_\bx(r)-\compen \Khat_\bx(r)]$, and, hence,
it is equivalent to the residual approximated by \eqref{E:verygood}.
This is also the conclusion in the more general case of a null model
with an activity parameter $\kappa$, that is, where the Papangelou
conditional intensity factorizes as
\[
  \lambda_\theta(u,\bx)=\kappa \xi_\beta(u,\bx),
\]
where $\theta=(\kappa,\beta)$ and $\xi_\beta(\cdot)$ is a Papangelou
conditional intensity, since the pseudo-likelihood equations then imply
that $n=\int_W\lambda_{\hat\theta}(u,\bx)\,\mathrm{d}u$.

In conclusion, the residual diagnostics obtained from the perturbing
Strauss and $\Khat$-models are very similar, the major difference being
the data-depen\-dent normalization of the $\Khat$-function; similarly for
pseudo-residual diagnostics which may be effectively equivalent to the
residual diagnostics. In practice, the popularity of the $K$-function
seems to justify using the residual diagnostics based on the perturbing
$\Khat$-model. Furthermore, due to the familiarity of the $K$-function,
we often choose to plot the compensator(s) of the fitted model(s) in a
plot with the empirical $K$-function rather than the residual(s) for
the fitted model.

\subsection{Edge Correction in Conditional Case}
\label{S:condcase:K}

In the conditional case, the Papangelou conditional intensity
$\lambda_{\hat\theta}(u,\bx)$ is known only at locations $u \in
\Wfree$. The diagnostics must be modified accordingly, by restricting
the domain of summation and integration to $\Wfree$. Appropriate
modifications are discussed in Appendices \ref{App:edge}--\ref{App:G}.

\subsection{Approximate Residual Variance Under CSR}
\label{S:varcalc:VS}

Here we study the residual variance and the accuracy of the Poincar\'e
variance approximation in a~simple case.

We shall approximate the residual variance\break $\Var[\resid V_S(X,r)]$ by
the innovation variance\break $\Var[\innov V_S(X,r)]$, that is, ignoring the
effect of parameter estimation. It is likely that this approximation is
conservative, because the effect of parameter estimation is typically
to deflate the residual variance \cite{baddetal05}. A~more detailed
investigation has been conducted in \cite{coeulava10}.

Assume the null model is CSR with intensity $\rho$ estimated by
$\hat\rho=n(\bx)/|W|$. The exact variance of the innovation for
the Strauss canonical statistic $V_S$ is $
   \Var[ \innov V_S(\bX,r)]
   = I_1 + I_2
$ from equation \eqref{e:var.innov} of Appendix~\ref{App:var}, where
\begin{eqnarray*}
  I_1
  &=&  \frac 1 4 \int_W
  \expected [ t(u, \bX,r)^2 \lambda(u, \bX) ]
\dee u\\
  &=&
  \frac\rho 4 \int_W \expected [ t(u, \bX,r)^2 ] \dee u
\end{eqnarray*}
and
\begin{eqnarray*}
  I_2
  &=&
  \frac 1 4 \int_W \int_W
  \expected[
    \indicate{\| u - v\| \le r}
    \lambda_2(u,v,\bX)
  ] \dee u \dee v\\
  &=&
  \frac{\rho^2}{4} \int_W \int_W \indicate{\| u - v\| \le r} \dee u \dee v
\end{eqnarray*}
as $\lambda(u, \bX)=\rho$ and $\lambda_2(u,v,\bX)=\lambda(u,
\bX)\lambda(v,\break \bX\cup\{u\})=\rho^2$. This is reminiscent of
expressions for the large-domain limiting variance of $\Khat$ under CSR
obtained using the methods of $U$-statistics
\cite{lotwsilv82,chetdigg98,ripl88}, summarized in \cite{digg03}, page 51
ff. Now $Y = t(u,\bX, r)$ is Poisson distributed with mean
$\mu =\break \rho |B(u,r) \cap W|$ so that $\expected(Y^2) = \mu + \mu^2$.
For $u \in W_{\ominus r}$ we have $\mu = \rho \pi r^2$, so ignoring
edge effects
\[
  I_1
  \approx\frac\rho 4 (\nu + \nu^2) |W|
  \quad\mbox{and}\quad
  I_2
  \approx
  \frac \rho 4 \nu |W|,
\]
where $\nu = \rho \pi r^2$. Note that since $\nu$ is the expected
number of points within distance $r$ of a given point, a value of $\nu
= 1$ corresponds to the scale of nearest-neighbor distances in the
pattern, $
     r_{\mathrm{nn}} = 1/\sqrt{\pi \rho}.
$ For the purposes of the $K$ function this is a ``short'' distance.
Hence, it is reasonable to describe $I_1$ as the ``leading term'' in
the variance, since $I_1 \gg I_2$ for $\nu \gg 1$.

Meanwhile, the Poincar\'e variance \eqref{e:compvar} is
\[
  \compvar V_S(\bx,r) = \frac {n(\bx)} {4|W|} \int_W t(u,\bx,r)^2 \dee u,
\]
which is an approximately unbiased estimator of $I_1$
by Fubini's Theorem.
Hence,
\begin{eqnarray*}
  \frac{\expected   \compvar V_S(\bx,r)
  }{
   \Var[ \resid V_S(\bX, r)]
  }
  &\approx&
  \frac{
    \expected   \compvar V_S(\bx,r)
  }{
   \Var[ \innov V_S(\bX, r)]
  }\\
  &\approx&
  \frac{I_1}{I_1 + I_2}
  \approx
  \frac{1 + \nu}{2 + \nu}.
\end{eqnarray*}
Thus, as a rule of thumb,  the Poincar\'e variance underestimates the
true variance; the ratio of means is $(1+\nu)/(2+\nu) \ge 1/2$.
The
ratio falls to $2/3$ when $\nu = 1$,\vadjust{\goodbreak} that is, when $r =
r_{\mathrm{nn}} = 1/\sqrt{\pi \rho}$. We can take this as a
rule-of-thumb indicating the value of $r$ below which the Poincare
variance is a poor approximation to the true variance.

\section{Residual Diagnostics for Interaction Using Nearest Neighbor Distances}
\label{S:nndist}

This section develops residual and pseudo-residual diagnostics derived
from summary statistics based on nearest neighbor distances.

\subsection{Residual Based on Perturbing Geyer Model}

The Geyer interaction potential $V_G(\bx,r)$ given\break by~(\ref{e:geyerS})
is clearly a sum of local statistics \eqref{e:sumoflocal}, and its
compensator is
\[
  \compen V_G(\bx,r) = \int_{W} \indicate{d(u,\bx) \le r}
\lambda_{\hat\theta}(u,\bx) \dee u.
\]
The Poincar\'e variance is equal to the compensator in this case.
Ignoring edge effects, $V_G(\bx,r)$ is approximately $n(\bx)\hat
G_{\bx}(r)$; cf. \eqref{e:Ghat}.

If the null model is CSR with estimated intensity
$\hat\kappa=n(\bx)/|W|$, then
\[
\compen V_G(\bx,r) = \hat\kappa \biggl|W \cap \bigcup_i B(x_i,r)\biggr|;
\]
ignoring edge effects, this is approximately $\hat\kappa |W| \hat
F(r)$; cf. \eqref{e:Fhat}. Thus, the residual diagnostic is
approximately $n(\bx) (\Ghat(r) - \hat F(r))$. This is a reasonable
diagnostic for departure from CSR, since $F \equiv G$ under CSR. This
argument lends support to Diggle's \cite{digg79a}, equation (5.7),
proposal to judge departure from CSR using the quantity $\sup |\hat
G-\hat F|$.

This example illustrates the important point that the compensator of a
functional summary statistic $S$ should not be regarded as an
alternative parametric estimator of the same quantity that $S$ is
intended to estimate. 
In the example just given, under CSR the compensator of $\hat G$ is
approximately $\hat F$, a qualitatively different and in some sense
``opposite'' summary of the point pattern.

We have observed that the interaction potential~$V_G$ of the Geyer
saturation model is closely related to~$\hat G$. However, the
pseudo-residual associated to $V_G$ is a~more complicated statistic,
since a straightforward calculation shows that the pseudo-sum is
\begin{eqnarray*}
   &&\psum V_G(\bx,r)\\
&&\quad  =
   V_G(\bx,r) + \sum_{i} \sum_{j:j\not=i}
\mathbb{I}\{\|x_i - x_j\| \le r \mbox{ and}\\
 &&\hspace*{136pt}d(x_j, \bxminus i) > r\},\vadjust{\goodbreak}
\end{eqnarray*}
and the pseudo-compensator is
\begin{eqnarray*}
\pcom V_G (\bx, r) &=&
\int_W \indicate{d(u,\bx) \le r} \lambda_{\hat\theta}(u,\bx) \dee u\\
&&{}+\sum_{i} \indicate{d(x_i, \bxminus i) > r}\\
  &&\phantom{{}+\sum_{i}}{}\cdot\int_W \indicate{\|u - x_i\| \le r} \lambda_{\hat\theta}(u,\bx) \dee
  u.
\end{eqnarray*}

\subsection{\texorpdfstring{Residual Based on Perturbing $\Ghat$-Model}{Residual Based on Perturbing
G-Model}}

The empirical $G$-function (\ref{e:Ghat}) can be written
%
\begin{equation}\label{e:Ghat=sum}
  \Ghat_\bx(r) = \sum_i g(x_i, \bxminus i, r),
  \end{equation}
where
%
\begin{eqnarray}\label{e:Glocal}
  g(u, \bx, r) &=& \frac{1}{n(\bx)+1} e_G(u,\bx, r)\nonumber\\ [-9pt]\\ [-9pt]
&&{}\cdot \indicate{ d(u, \bx)
    \le r},\quad  u\notin\bx,\nonumber
\end{eqnarray}
so that the Papangelou compensator of the empirical $G$-function is
\begin{eqnarray*}
&&\compen \Ghat_\bx(r)\\
&&\quad=\int_W g(u,\bx,r) \lambda_{\hat\theta}(u,\bx) \dee
u\\
&&\quad= \frac{1}{n(\bx)+1} \int_{W \cap\bigcup_i B(x_i,r)}
                        e_G(u,\bx,r) \lambda_{\hat\theta}(u,\bx) \dee u.
\end{eqnarray*}
The residual diagnostics obtained from the Geyer and $\Ghat$-models are
very similar, and we choose to use the diagnostic based on the popular
$\Ghat$-function. As with the $K$-function, we typically use the
compensator(s) of the fitted model(s) rather than the residual(s), to
visually maintain the close connection to the empirical $G$-function.

The expressions for the pseudo-sum and pseudo-compensator of $\Ghat$
are not of simple form, and we refrain from explicitly writing out
these expressions. For both the $\Ghat$- and Geyer models, the
pseudo-sum and pseudo-compensator are not directly related to a
well-known summary statistic. We prefer to plot the pseudo-residual
rather than the pseudo-sum and pseudo-compensator(s).

\subsection{Residual Variance Under CSR}
\label{S:varcalc:VG}

Again assume a Poisson process of intensity $\rho$ as the null model.
Since $V_G$ is a sum of local statistics,
\[
 V_G(\bx, r) = \sum_i \indicate{d(x_i, \bx\setminus x_i) \le r},
\]
we can again apply the variance formula \eqref{e:var.innov} of
Appendix \ref{App:var}, which gives $
\Var[ \innov V_G(\bX,r)]  = L_1 + L_2,
$ where
\[
L_1 = \rho \int_W \probab \{d(u, \bX) \le r\} \dee u
\]
and
\begin{eqnarray*}
  L_2 &=&  \rho^2 \int_W \int_W
   \probab\{ \| u - v\| \le r,\\
&&\phantom{\rho^2 \int_W \int_W
   \probab\{} d(u,X) > r,\,d(v,X) > r \}
   \dee u
   \dee v .
\end{eqnarray*}
The Poincar\'e variance is equal to the compensator in this case, and
is
\begin{eqnarray*}
\compvar V_G(\bx,r)&=& \int_{W} \indicate{d(u,\bx) \le r}
                          \lambda_{\hat\theta}(u,\bx) \dee u\\
&=& \frac{n(\bx)}{|W|} |W \cap U(\bx, r)|,
\end{eqnarray*}
%
where $U(\bx,r) = \bigcup_i b(x_i, r)$. The Poincar\'e variance is an
approximately unbiased estimator of the term~$L_1$.

For $u \in W_{\ominus r}$ we have $
      \probab \{d(u, \bX) \le r\}
      = 1 -\break \exp(-\rho \pi r^2)$ so that
\[
  L_1 \approx \rho |W| \bigl( 1 - \exp(-\rho \pi r^2) \bigr),
\]
ignoring edge effects. Again, let $\nu = \rho \pi r^2$ so that $L_1
\approx \rho |W| (1 - \exp(-\nu))$. Meanwhile,
\begin{eqnarray*}
    &&\probab\{
    d(u,X) > r,\,d(v, X) > r
    \}\\
    &&\quad= \exp\bigl(- \rho| b(u,r) \cup b(v,r) | \bigr) .
\end{eqnarray*}
This probability lies between $\exp(-\nu)$ and $\exp(-2\nu)$ for all
$u,v$. Thus (ignoring edge effects),
\begin{eqnarray*}
 L_2 &\approx& \rho^2 \pi r^2 |W| \exp\bigl(- (1+\delta)\nu\bigr)\\
&=& \rho \nu |W| \exp\bigl(-(1+\delta)\nu\bigr),
\end{eqnarray*}
where $0 \le \delta \le 1$. Hence,
\[
  \frac{L_2}{L_1}  \le \frac{\nu e^{-\nu}}{1 - e^{-\nu}}.
\]
Let $f(\nu) = \nu e^{-\nu}/(1-e^{-\nu})$. Then $f(\nu)$ is strictly
decreasing and $f(\nu) < 1$ for all $\nu > 0$ so\vspace*{1pt} that $
    L_1/\break(L_1+L_2) \ge \frac 1 2,
$ that is, the variance is underestimated by at most a factor of 2.
Note that $f(1.25) \approx 0.5$, so $
    L_1/(L_1+L_2) \ge \frac 2 3
$ when $r\le r_{\mathrm{crit}}$, where $
        r_{\mathrm{crit}} = \sqrt{1.25/\pi \rho}.
$ The conclusions and rule-of-thumb for $\stdres\Ghat$ are similar to
those obtained for $\stdres\Khat$ in Section~\ref{S:varcalc:VS}.

\section{Diagnostics for Interaction Based on Empty Space Distances}
\label{S:area}

\subsection{Pseudo-Residual Based on Perturbing Area-Interaction Model}

When the perturbing model is the area-interaction process, it is
convenient to reparametrize the density, such that the canonical
sufficient statistic $V_A$ given in (\ref{e:areaS}) is redefined as
\[
  V_A(\bx,r) = \frac{1}{|W|} \biggl| W \cap \bigcup_i B(x_i, r) \biggr|.
\]
This summary statistic is not naturally expressed as a sum of
contributions from each point as in \eqref{e:sumoflocal}, so we shall
only construct the pseudo-residual. Let
\[
  U(\bx, r) = W \cap \bigcup_i B(x_i, r).
\]
The increment
\begin{eqnarray*}
  &&\Delta_u V_A(\bx,r)\\
&&\quad = \frac{1}{|W|} \bigl( |U(\bxplus u,r)| -
    |U(\bx, r)| \bigr),\quad   u \notin \bx,
\end{eqnarray*}
can be thought of as ``unclaimed space''---the proportion of space
around the location $u$ that is not ``claimed'' by the points of $\bx$.
The pseudo-sum
\[
  \psum V_A (\bx,r) = \sum_i \Delta_{x_i} V_A(\bx,r)
\]
is the proportion of the window that has ``single coverage''---the
proportion of locations in $W$ that are covered by exactly one of the
balls $B(x_i,r)$. This can be used in its own right as a functional
summary statistic, and it corresponds to a raw (i.e.,\
not edge corrected)
empirical estimate of a summary function~$F_1(r)$ defined by
\[
  F_1(r) = \mathbb{P}\bigl(\#\{x \in \bX|d(u,x)\le r\} = 1\bigr)
\]
for any stationary point process $\bX$, where $u\in\real^2$ is
arbitrary. Under CSR with intensity $\rho$ we have
\[
  F_1(r) = \rho \pi r^2 \exp(-\rho \pi r^2).
\]
This summary statistic does not appear to be treated in the literature,
and it may be of interest to study it separately, but we refrain from a
more detailed study here.

The pseudo-compensator corresponding to this\break pseudo-sum is
\[
  \pcom V_A(\bx, r) = \int_W \Delta_u V_A(\bx,r)\lambda_{\hat\theta}(u,\bx) \dee u.
\]
This integral does not have a particularly simple interpretation even
when the null model is CSR.

\subsection{\texorpdfstring{Pseudo-Residual Based on
Perturbing~$\Fhat$-Model}{Pseudo-Residual Based on Perturbing F-Model}}

Alternatively, one could use a standard \mbox{empirical} estimator $\Fhat$ of
the empty space function $F$ as the summary statistic in the
pseudo-residual. The pseu\-do-sum associated with the perturbing
$\Fhat$-model is
\[
  \psum \Fhat_\bx(r) = n(\bx)\Fhat_\bx(r) - \sum_i \Fhat_{\bxminus i}(r),
\]
with pseudo-compensator
\[
  \pcom \Fhat_\bx(r) =
   \int_W \bigl(\Fhat_{\bxplus u}(r) - \Fhat_\bx(r)\bigr)\lambda_{\hat\theta}(u,\bx) \dee u.
\]
Ignoring edge correction weights, $\Fhat_{\bxplus u}(r) - \Fhat_\bx(r)$
is approximately equal to $\Delta_u V_A(\bx,r)$, so the pseudo-sum and
pseudo-compensator
 associated with the perturbing $\Fhat$-model are
approximately equal to the pseudo-sum and pseudo-compensator associated
with the perturbing area-interaction model. Here, we usually prefer
graphics using the pseudo-compensator(s) and the pseudo-sum since this
has an intuitive interpretation as explained above.


\section{Test Case: Trend with Inhibition}
\label{S:TrendInhib}

In Sections \ref{S:TrendInhib}--\ref{S:jap} we demonstrate the
diagnostics on the point pattern data sets shown in
Figure \ref{fig:data}. This section concerns the synthetic point
pattern in Figure \ref{fig:data}(b).

\subsection{Data and Models}
\label{S:TrendInhib:data}

Figure \ref{fig:data}(b) shows a simulated realization of the
inhomogeneous Strauss process with first order term $\lambda(x,y) = 200
\exp(2 x + 2 y + 3x^2)$, interaction range $R=0.05$, interaction
parameter $\gamma = \exp(\phi)=0.1$ and $W$ equal to the unit square;
see \eqref{E:frequent.density} and \eqref{e:straussS}. This is an
example of extremely strong inhibition (negative association) between
neighboring points, combined with a~spatial trend. Since it is easy to
recognize spatial trend in the data (either visually or using existing
tools such as kernel smoothing \cite{digg85}), the main challenge here
is to detect the inhibition after accounting for the trend.

\begin{figure*}[t]
  \centering
\begin{tabular}{@{}cc@{}}

\includegraphics{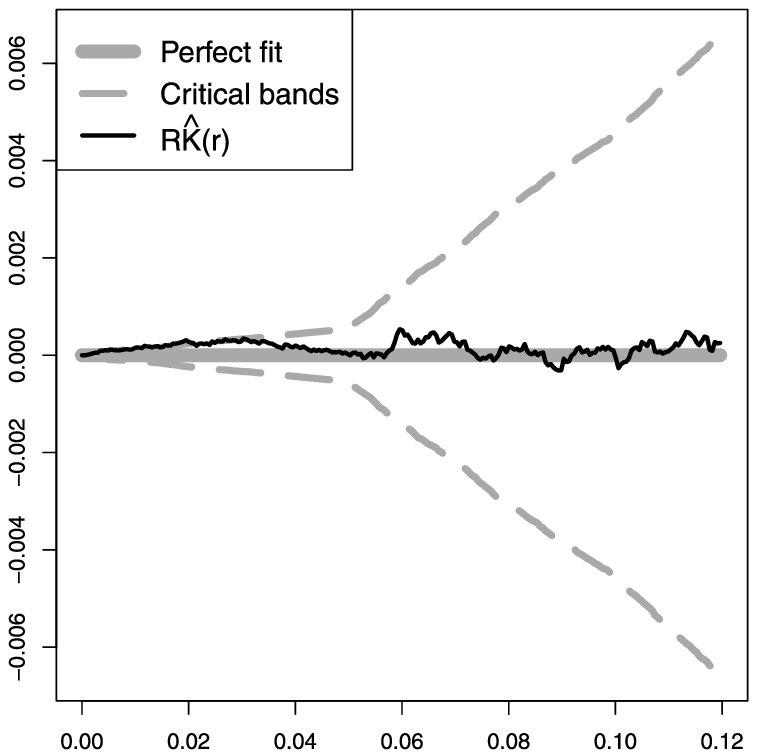}&
\includegraphics{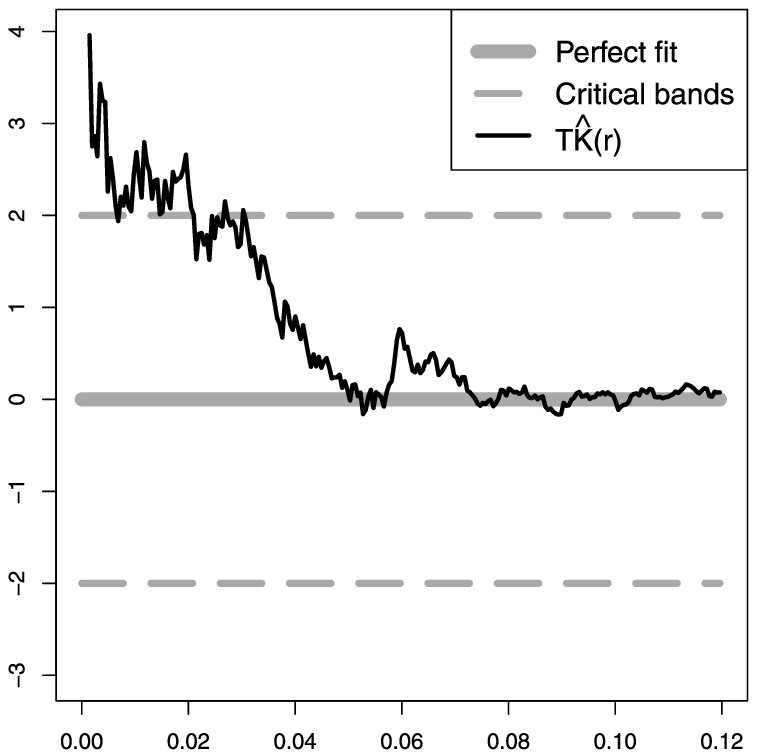} \\
\scriptsize{(a)} & \scriptsize{(b)}\\[-6pt]
\end{tabular}
\caption{Residual diagnostics based on pairwise distances,
    for a model of the correct form fitted to the data in
    Figure \protect\ref{fig:data}\textup{(b)}.
    \textup{(a)}~Residual $\Khat$-function
    and two-standard-deviation limits under the fitted model of the
    correct form.
   \textup{(b)} Standardized residual $\Khat$-function
    under the fitted model of the correct form.}\label{fig:Kright}
    \vspace*{-3pt}
\end{figure*}

We fitted four point process models to the data in
Figure \ref{fig:data}(b). They were \modelOneCSR\ a homogeneous
Poisson process (CSR); \modelOneInh\ an inhomogeneous Poisson process
with the correct form of the first order term, that is, with intensity
%
\begin{equation} \label{e:trend4b}
  \rho(x,y) = \exp(\beta_0 + \beta_1 x + \beta_2 y + \beta_3 x^2),
\end{equation}
where $\beta_0,\ldots,\beta_3$ are real parameters; \modelOneStr\ a
homogeneous Strauss process with the correct interaction range\vadjust{\goodbreak} $R =
0.05$; and \modelOneInS\ a process of the correct form, that is,
inhomogeneous Strauss with the correct interaction range $R = 0.05$ and
the correct form of the first order potential (\ref{e:trend4b}).

\subsection{Software Implementation}

The diagnostics defined in Sections~\ref{S:pairwise}--\ref{S:area} were
implemented in the \textsf{R} language, and has been publicly released in
the \texttt{spatstat} library \cite{baddturn05}. Unless otherwise
stated, models were fitted by approximate maximum pseudo-likelihood
using the algorithm of \cite{baddturn00} with the default quadrature
scheme in \texttt{spatstat}, having an $m \times m$ grid of dummy
points where $m = \max(25, 10[1+2\sqrt{n(\bx)}/10])$ was equal to 40
for most of our examples. Integrals over the domain $W$ were
approximated by finite sums over the quadrature points. Some models
were refitted using a finer grid of dummy points, usually $80 \times
80$. In addition to maximum pseudo-likelihood estimation, the software
also supports the Huang--Ogata \cite{huanogat99} approximate maximum
likelihood.

\subsection{\texorpdfstring{Application of $\Khat$ Diagnostics}{Application of K Diagnostics}}
\label{S:TrendInhib:K}

\subsubsection{Diagnostics for correct model}
\label{S:TrendInhib:K:correct}

First we fitted a point process model of the correct form \modelOneInS.
The fitted parameter values were $\hat\gamma = 0.217$ and
$\hat\beta = (5.6, -0.46, 3.35, 2.05)$ using the
coarse grid of dum\-my points, and $\hat\gamma = 0.170$ and
$\hat\beta =
  (5.6, -0.64, 4.06,\break 2.44)$ using the finer grid of dummy points, as
against the true values $\gamma = 0.1$ and $\beta = (5.29,
2, 2, 3)$.\vadjust{\goodbreak}

Figure \ref{F:KrightKcom} in Section \ref{S:intro} shows $\Khat$ along
with its compensator for the fitted model, together with the
theoretical $K$-function under CSR.  The empirical $K$-function and its
compensator coincide very closely, suggesting correctly that the model
is a good fit.
Figure \ref{fig:Kright}(a) shows the residual $\Khat$-function and
the two-standard-deviation limits, where the surrogate standard
deviation is the square root of \eqref{e:compvar}.
Figure~\ref{fig:Kright}(b) shows the corresponding standardized
residual $\Khat$-function obtained by dividing by the surrogate
standard deviation.

\begin{figure*}
  \centering
 \begin{tabular}{@{}cc@{}}

\includegraphics{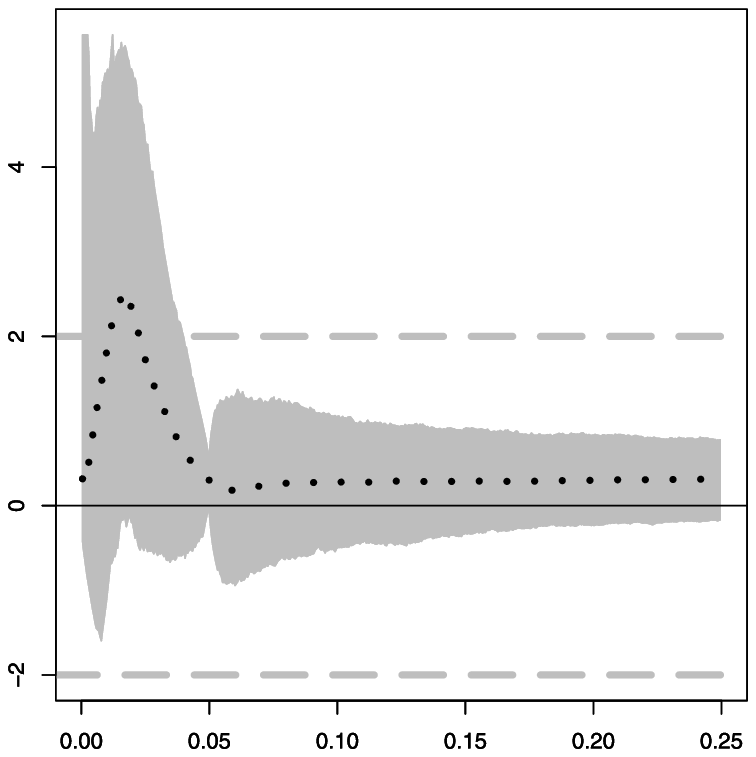}&
\includegraphics{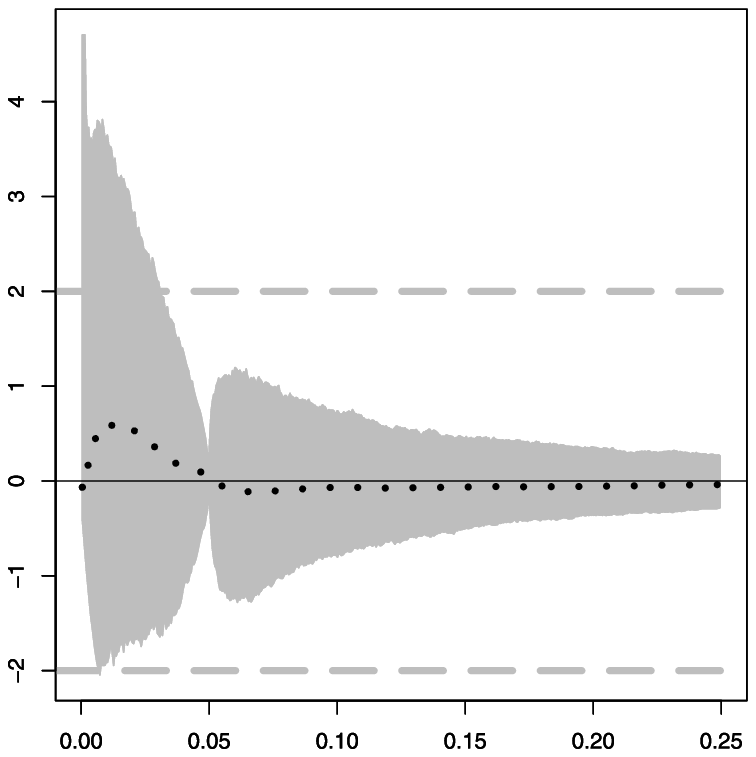}\\
\scriptsize{(a)}&\scriptsize{(b)}\\[-6pt]
\end{tabular}
  \caption{Null distribution of standardized residual of $\hat K$.
    Pointwise 2.5\% and 97.5\% quantiles (grey shading)
    and sample mean (dotted lines) of $\stdres \Khat$
    from 1000 simulated realizations
    of model \modelOneInS\ with estimated parameter values
    \textup{(a)} $\hat\gamma = 0.217$
    and $\hat\beta =\break (5.6, -0.46, 3.35, 2.05)$ using a
    $40\times 40$ grid of dummy points;
    \textup{(b)} $\hat\gamma = 0.170$ and
   $\hat\beta = (5.6, -0.64, 4.06, 2.44)$ using a $80 \times
    80$~grid.
  }
  \label{fig:Kstdres:env}
\end{figure*}

Although this model is of the correct form, the standardized residual
exceeds 2 for small values of $r$. This is consistent with the
prediction in Section~\ref{S:varcalc:VS} that the variance
approximation would be inaccurate for small $r$. The null model is a
nonstationary Poisson process; the minimum value of the intensity is
$200$. Taking $\rho = 200$ and applying the rule of thumb in
Section \ref{S:varcalc:VS} gives
\[
r_{\mathrm{nn}} = \frac 1 {\sqrt{200\pi}} = 0.04,
\]
suggesting that the Poincar\'e variance estimate becomes unreliable for
$r \le 0.04$ approximately.

Formal significance interpretation of the critical bands in
Figure \ref{fig:Kright}(b) is limited, because the null distribution
of the standardized residual is not known exactly, and the values $\pm
2$ are approximate \textit{pointwise}  critical values, that is,
critical values for the score test based on fixed $r$. The usual
problems of multiple testing arise when the test statistic is
considered as a function of $r$; see \cite{digg03}, page 14. For very
small $r$ there are small-sample effects so that a normal approximation
to the null distribution of the standardized residual is inappropriate.

To confirm this, Figure \ref{fig:Kstdres:env} shows the pointwise 2.5\%
and 97.5\% quantiles of
the null distribution of $\stdres \Khat$,
obtained by extensive simulation. The sample mean of the simulated
$\stdres \Khat$ is also shown, and indicates that the expected
standardized residual is nonzero for small values of $r$. Repeating the
computation with a finer grid of quadrature points (for approximating
integrals over $W$ involved in the pseudo-likelihood and the residuals)
reduces the bias,
suggesting
that this is a discretization artefact.

\subsubsection{Comparison of competing models}
\begin{figure*}
  \centering
\begin{tabular}{@{}cc@{}}

\includegraphics{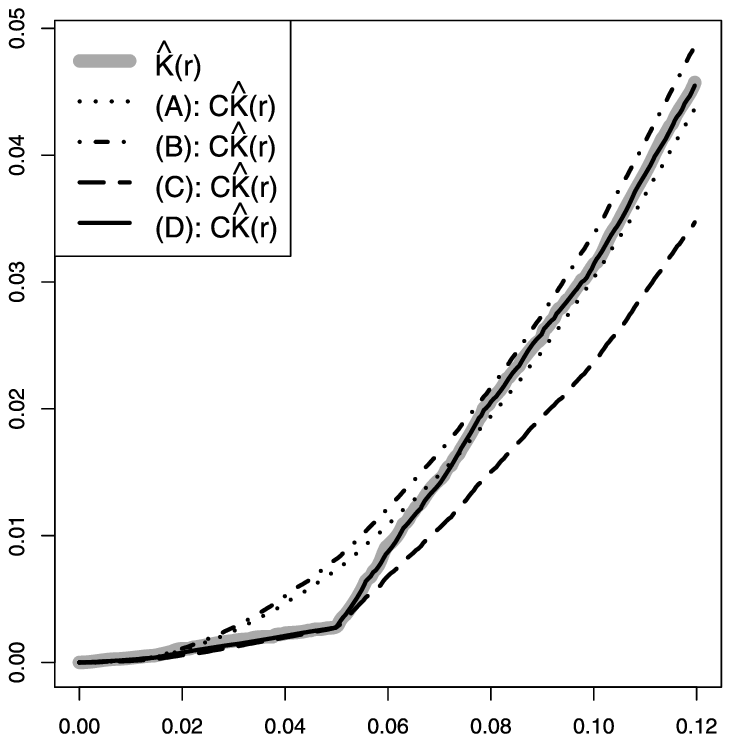}&
\includegraphics{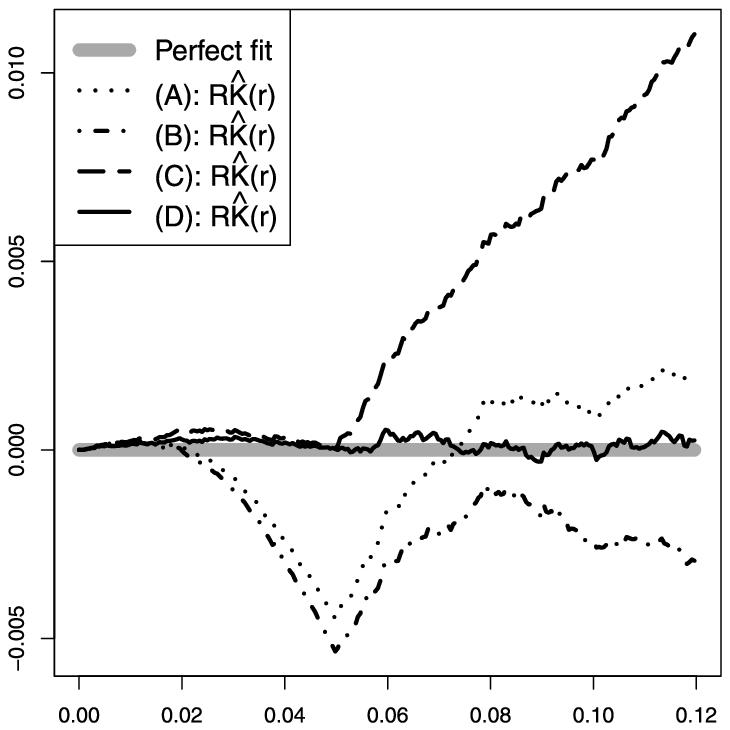}\\
\scriptsize{(a)}& \scriptsize{(b)}\\ [6pt]
\multicolumn{2}{@{}c@{}}{
\includegraphics{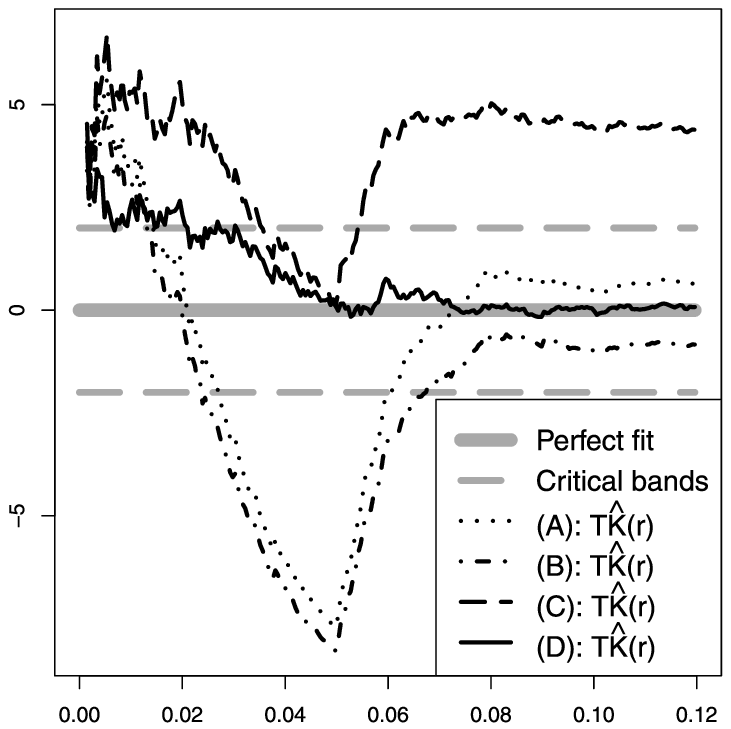}
}\\
\multicolumn{2}{@{}c@{}}{\scriptsize{(c)}}
\end{tabular}
  \caption{
    Model
    diagnostics based on pairwise distances,
    for each of the models \modelOneCSR--\modelOneInS\
    fitted to the data in Figure \protect\ref{fig:data}\textup{(b)}.
    \textup{(a)} $\Khat$
    and its compensator under each model.
   \textup{(b)} Residual $\Khat$-function (empirical
    minus compensator) under each model.
   \textup{(c)}~Standardized residual $\Khat$-function
    under each model.}\label{fig:Synth1:K}
\end{figure*}

Figu-\break re~\ref{fig:Synth1:K}(a) shows the empirical $K$-func\-tion and
its compensator for each of the models \modelOneCSR--\modelOneInS\ in
Section~\ref{S:TrendInhib:data}. Figure \ref{fig:Synth1:K}(b) shows
the corresponding residual plots, and Figure \ref{fig:Synth1:K}(c)
the standardized residuals. A~positive or negative value of the
residual suggests that the data are more clustered or more inhibited,
respectively, than the model. The clear inference is that the Poisson
models \modelOneCSR\ and \modelOneInh\ fail to capture interpoint
inhibition at range $r \approx 0.05$, while the homogeneous Strauss
model~\modelOneStr\ is less clustered than the data at very\vadjust{\goodbreak} large
scales, suggesting that it fails to capture spatial trend. The correct
model \modelOneInS\ is judged to be a good fit.

The interpretation of this example requires some caution, because the
residual $\Khat$-function of the fitted Strauss models \modelOneStr\
and \modelOneInS\ is constrained to be approximately zero at $r = R =
0.05$. The maximum pseudo-likelihood fitting algorithm solves an
estimating equation that is approximately equivalent to this
constraint, because of \eqref{e:K=VS}.

\begin{figure*}
\centering
\begin{tabular}{@{}cc@{}}

\includegraphics{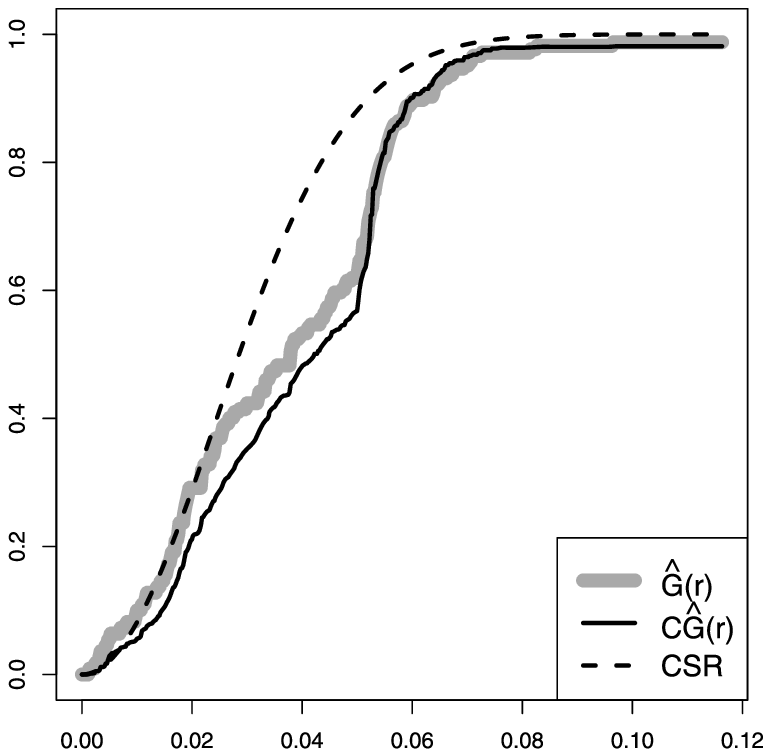}&
\includegraphics{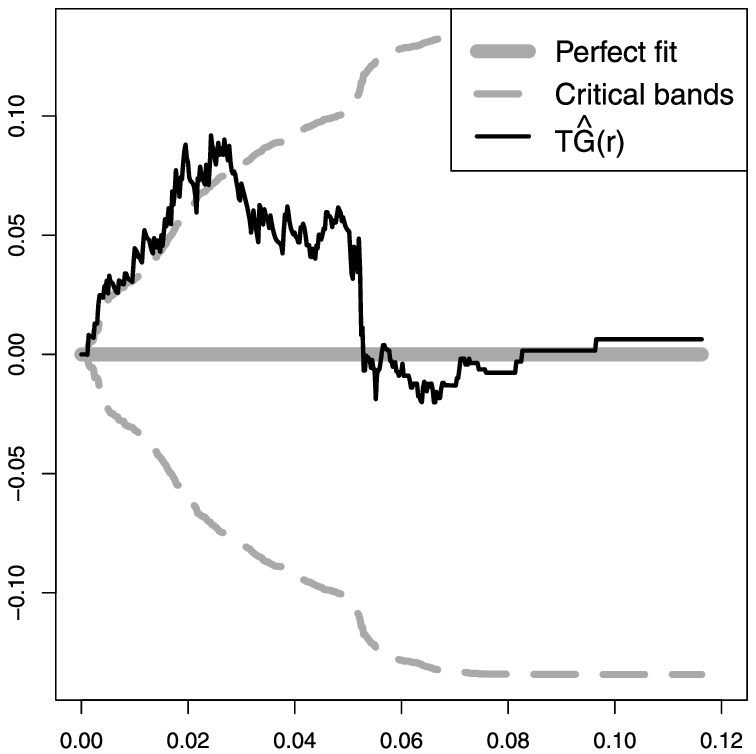}\\
\scriptsize{(a)}& \scriptsize{(b)}\vspace*{-1pt}
\end{tabular}
  \caption{Residual diagnostics obtained from the perturbing $\Ghat$-model
    when the data pattern is a realization of an inhomogeneous Strauss process.
    \textup{(a)} $\Ghat$ 
    and its
    compensator 
    under a fitted model of the
    correct form, and theoretical $G$-function for a Poisson process.
    \textup{(b)} Residual $\Ghat$-function
    and two-standard-deviation limits under the fitted model of the
    correct form.
  }
  \label{fig:Gright}
\end{figure*}

\begin{figure*}[b]
\centering
\begin{tabular}{@{}cc@{}}

\includegraphics{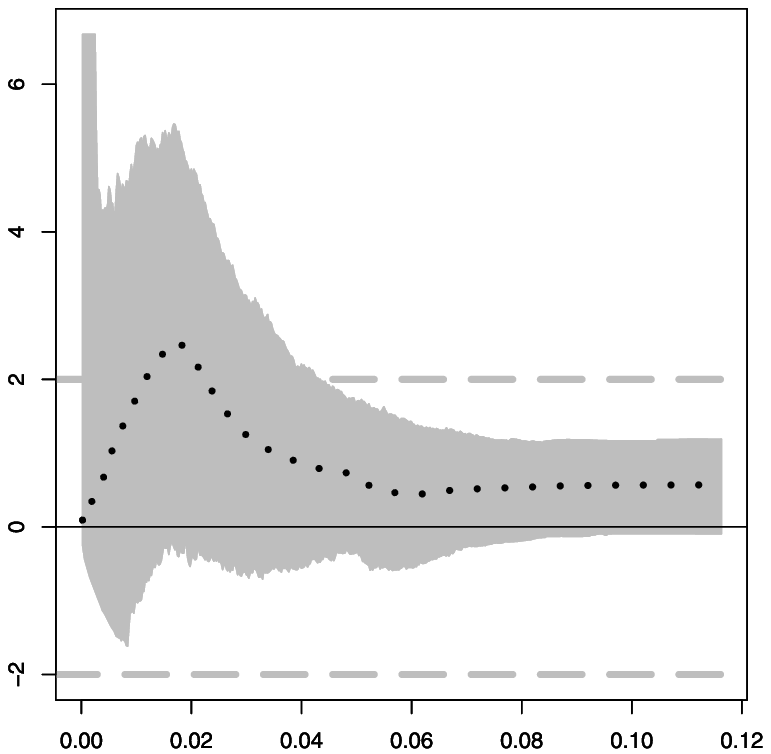}&
\includegraphics{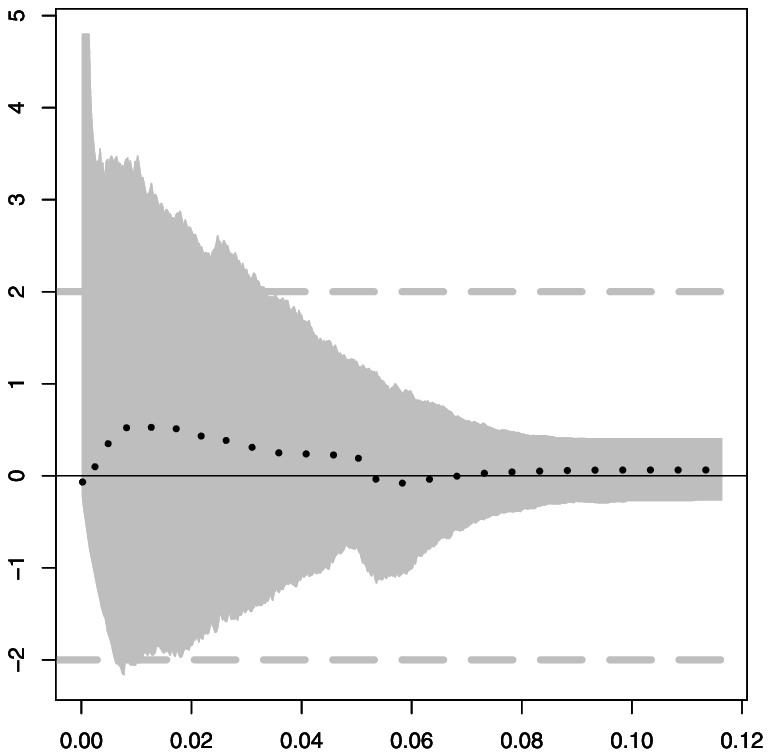}\\
\scriptsize{(a)} & \scriptsize{(b)}\vspace*{-1pt}
\end{tabular}
\caption{Null distribution of standardized residual of $\hat G$.
    Pointwise 2.5\% and 97.5\% quantiles (grey shading)
    and sample mean (dotted lines)
    from 1000 simulated realizations
    of model \modelOneInS\ with estimated parameter values
   \textup{(a)} $\hat\gamma = 0.217$ and
   $\hat\beta = (5.6, -0.46, 3.35, 2.05)$ using a $40\times
    40$ grid of dummy points;
    \textup{(b)} $\hat\gamma = 0.170$ and
    $\hat\beta = (5.6, -0.64, 4.06, 2.44)$ using a~$80 \times
    80$~grid.}\label{fig:Gstdres:env}
\end{figure*}

It is debatable which of the presentations in Figure~\ref{fig:Synth1:K}
is more effective at revealing lack of fit. A~compensator plot such as
Figure \ref{fig:Synth1:K}(a) seems best at capturing the main
differences between competing models. It is particularly useful for
recognizing a~gross lack of fit. A~residual plot such as
Figure \ref{fig:Synth1:K}(b) seems better for making finer
comparisons of
model fit, for example, assessing models with slightly different ranges
of interaction. A standardized residual plot such as
Figure~\ref{fig:Synth1:K}(c) tends to be highly irregular for small
values of $r$, due to discretization effects in the computation and the
inherent nondifferentiability of the empirical statistic. In difficult
cases we may apply smoothing to the standardized residual.

\subsection{\texorpdfstring{Application of $\Ghat$ Diagnostics}{Application of G Diagnostics}}
\label{S:TrendInhib:G}

\subsubsection{Diagnostics for correct model}

Consider\break again the model of the correct form \modelOneInS. The residual
and compensator of the empirical nearest neighbor function $\Ghat$ for
the fitted model are shown in Figu\-re~\ref{fig:Gright}. The residual
plot suggests a marginal lack of fit for $r < 0.025$. This may be
correct, since the fitted model parameters
(Section \ref{S:TrendInhib:K:correct}) are marginally poor\ estimates of
the true values, in particular, of the interaction parameter. This was
not reflected so strongly in the $\Khat$ diagnostics. This suggests
that the residual of $\Ghat$ may be particularly sensitive to lack of
fit of interaction.

Applying the rule of thumb in Section \ref{S:varcalc:VG}, we have
$r_{\mathrm{crit}} = 0.044$, agreeing with the interpretation
that the $\pm 2$ limits are not trustworthy for $r < 0.05$
approximately.

Figure \ref{fig:Gstdres:env} shows the pointwise 2.5\% and 97.5\%
quantiles of
the null distribution of $\stdres \Ghat$.
Again, there is a~suggestion of bias for small values of $r$ which
appears to be a discretization artefact.\vspace*{3pt}

\subsubsection{Comparison of competing models}

\begin{figure*}
  \centering
\begin{tabular}{@{}cc@{}}

\includegraphics{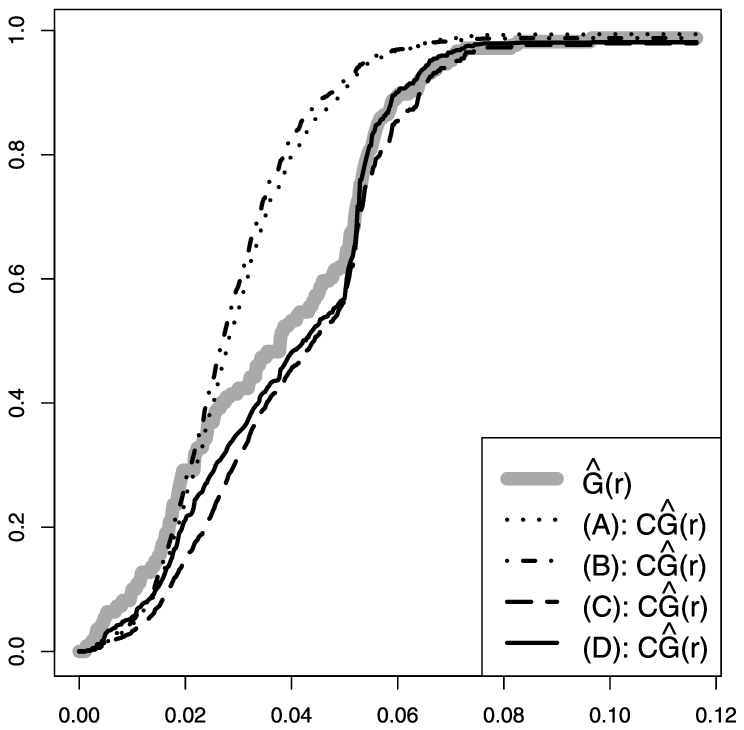}&
\includegraphics{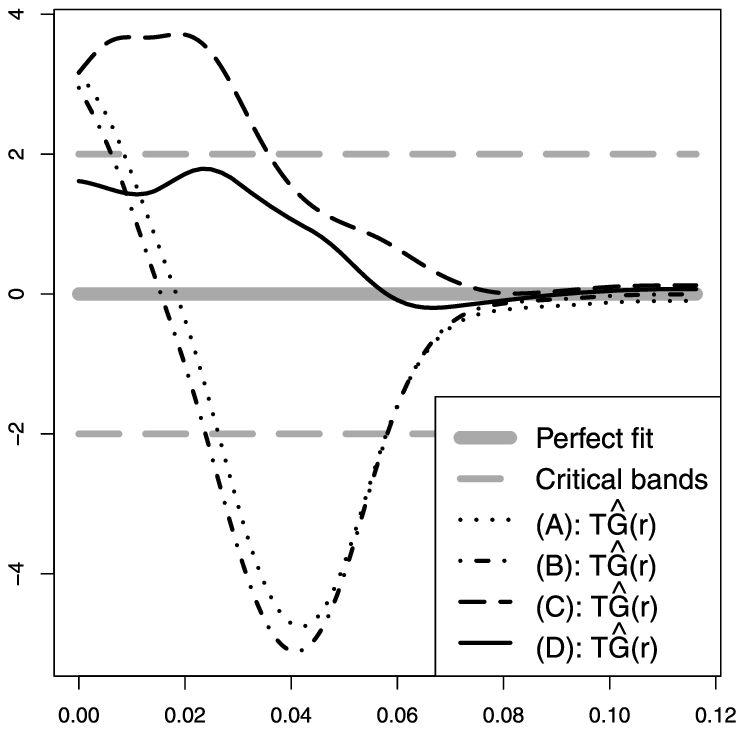}\\
\scriptsize{(a)}& \scriptsize{(b)}\\ [6pt]
\multicolumn{2}{@{}c@{}}{
\includegraphics{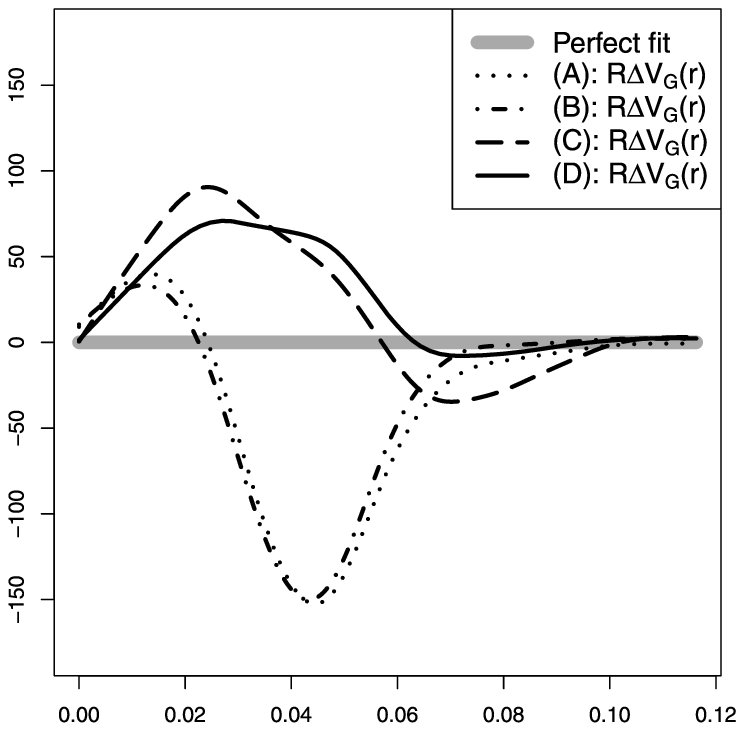}
}\\
\multicolumn{2}{@{}c@{}}{\scriptsize{(c)}}
\end{tabular}
  \caption{Diagnostics
    based on nearest neighbor distances,
    for the models \modelOneCSR--\modelOneInS\
    fitted to the data in Figure \protect\ref{fig:data}\textup{(b)}.
   \textup{(a)}
     Compen\-sator for $\Ghat$.
    \textup{(b)}
     Smoothed standardized residual of $\Ghat$.
    \textup{(c)}
     Smoothed pseudo-residual derived from a
perturbing Geyer model.}\label{fig:Synth1:G}
\end{figure*}

For each of the four models, Figure \ref{fig:Synth1:G}(a) shows
$\Ghat$ and its Papangelou compensator. This clearly shows that the
Poisson models \modelOneCSR\ and~\modelOneInh\ fail to capture interpoint
inhibition in the data. The Strauss models~\modelOneStr\ and~\modelOneInS\
appear virtually equivalent in
Figure~\ref{fig:Synth1:G}(a).\looseness=1

Figure \ref{fig:Synth1:G}(b) shows the standardized residual of~$\Ghat$,
and Figure \ref{fig:Synth1:G}(c) the pseudo-residual of
$V_G$ (i.e., the pseudo-residual based on the pertubing Geyer mo\-del),
with spline smoothing applied to both plots. The Strauss models
\modelOneStr\ and \modelOneInS\ appear virtually equivalent in
Figure \ref{fig:Synth1:G}(c). The standardized residual plot
Figure~\ref{fig:Synth1:G}(b) correctly suggests a~slight lack of fit
for model~\modelOneStr\ while model \modelOneInS\ is judged to be
a~reasonable fit.

\subsection{\texorpdfstring{Application of $\Fhat$ Diagnostics}{Application of F Diagnostics}}

\begin{figure*}
  \centering
  \begin{tabular}{@{}cc@{}}

\includegraphics{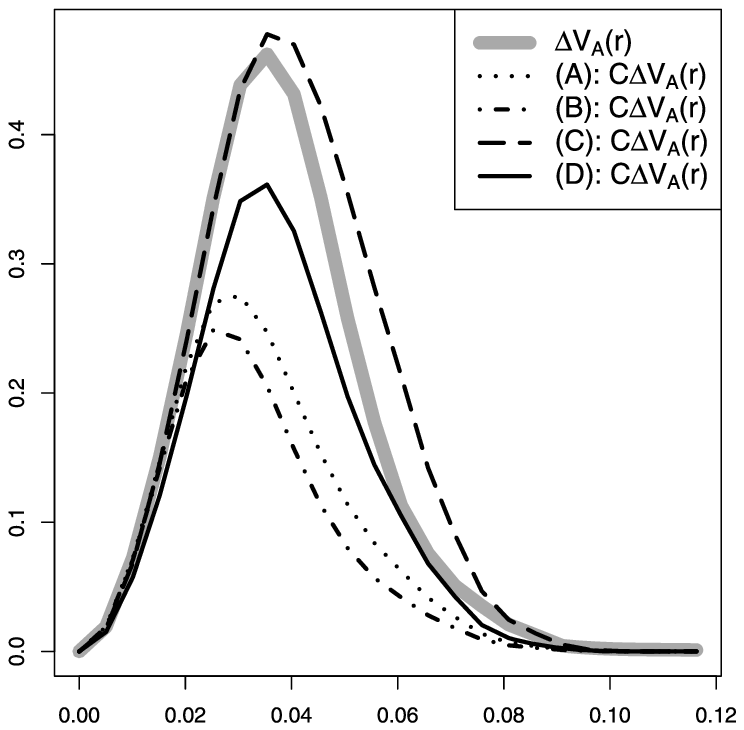}&
\includegraphics{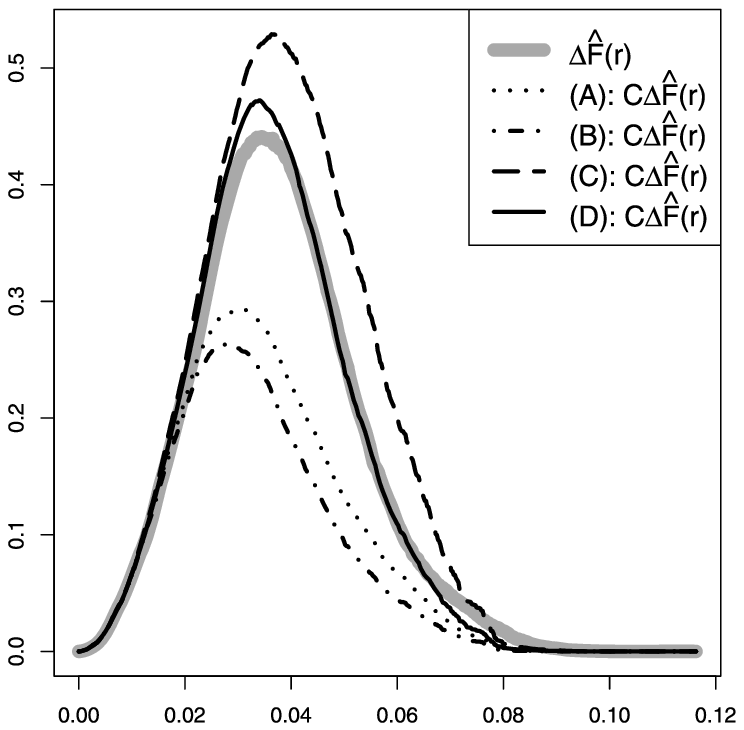}\\
\scriptsize{(a)}&\scriptsize{(b)}
  \end{tabular}
  \caption{Pseudo-sum and pseudo-compensators
    for the models \modelOneCSR--\modelOneInS\ fitted to the data in
    Figure \protect\ref{fig:data}\textup{(b)} when the perturbing model is
   \textup{(a)} the area-interaction
    process (null fitted on a fine grid)
    and \textup{(b)}
    the $\Fhat$-model (null fitted on a coarse grid).}
  \label{fig:Synth1:F}
\end{figure*}

Figure~\ref{fig:Synth1:F} shows the pseudo-residual diagnostics\break based
on empty space distances. Both diagnostics clearly show models
\modelOneCSR--\modelOneInh\ are poor fits to data. However, in
Figure \ref{fig:Synth1:F}(a) it is hard to decide which\break of the
models \modelOneStr--\modelOneInS\ provide a better fit. Despite the
close connection between the area-interaction process and the
$\Fhat$-model, the diagnostic in Figu-\break re~\ref{fig:Synth1:F}(b) based
on the $\Fhat$-model performs better in this particular example and
correctly shows \modelOneInS\ is the best fit to data. In both cases it
is noticed that the pseudo-sum has a much higher peak than the
pseudo-com\-pensators for the Poisson models \modelOneCSR--\modelOneInh,
correctly suggesting that these models do not capture the strength of
inhibition present in the data.\looseness=1

\section{Test Case: Clustering Without Trend}\vspace*{3pt}
\label{S:cluster}

\subsection{Data and Models}
\label{S:cluster:data}

Figure \ref{fig:data}(c) is a realization of a homogeneous Geyer
saturation process \cite{geye99} on the unit square, with first order
term $\lambda=\exp(4)$, saturation threshold $s=4.5$ and interaction
parameters $r = 0.05$ and $\gamma =\break \exp(0.4) \approx 1.5$, that is,
the density is
%
\begin{equation}\label{E:general-geyer}\qquad
f(\bx) \propto \exp\bigl( n(\bx) \log\lambda + V_{G,s}(\bx,r)
\log\gamma\bigr),
\end{equation}
where
\[
  V_{G,s}(\bx,r) =
  \sum_{i} \min\biggl\{
               s,
               \sum_{j:j\neq i} \indicate{\| x_i - x_j \| \le r}
               \biggr\}.
\]
This is an example of moderately strong clustering (with interaction
range $R = 2r=0.1$) without trend. The main challenge
here is to correctly identify the range and type of interaction.

We fitted three point process models to the data: \modelTwoCSR\ a~homogeneous
Poisson process (CSR); \modelTwoAre\ a~homogeneous
area-interaction process with disc radius $r = 0.05$; \modelTwoGey\ a
homogeneous Geyer saturation process of the correct form, with
interaction parameter $r = 0.05$ and saturation threshold $s = 4.5$
while the parameters $\lambda$ and $\gamma$ in \eqref{E:general-geyer}
are unknown. The parameter estimates for \modelTwoGey\ were
$\log\hat\lambda = 4.12$ and $\log\hat\gamma = 0.38$.

\subsection{\texorpdfstring{Application of $\Khat$ Diagnostics}{Application of K Diagnostics}}
\label{S:cluster:K}

A plot (not shown) of the $\Khat$-function and its compensator, under
each of the three models \modelTwoCSR--\modelTwoGey, demonstrates
clearly that the homogeneous Poisson model~\modelTwoCSR\ is a poor fit,
but does not discriminate between the other models.

\begin{figure*}
\centering
\begin{tabular}{@{}cc@{}}

\includegraphics{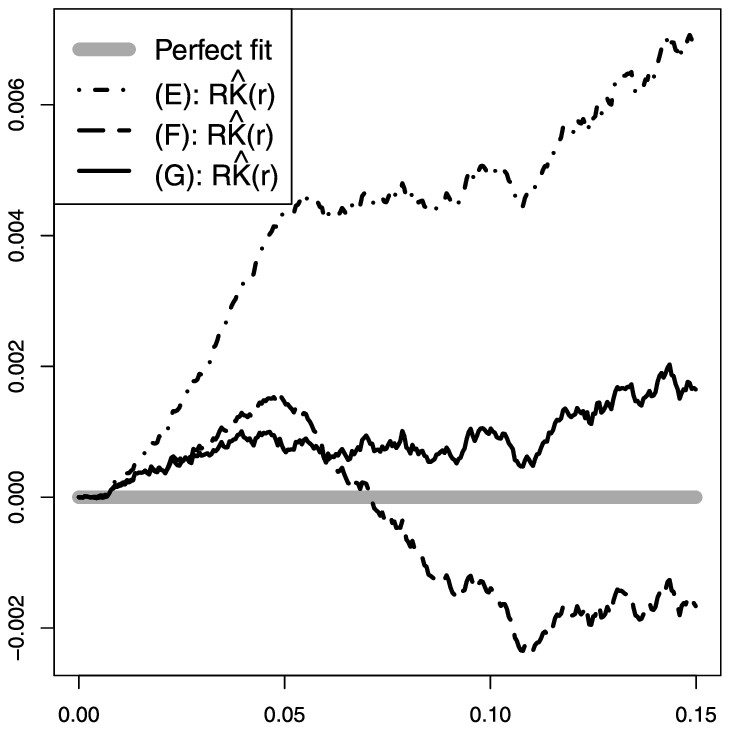}&
\includegraphics{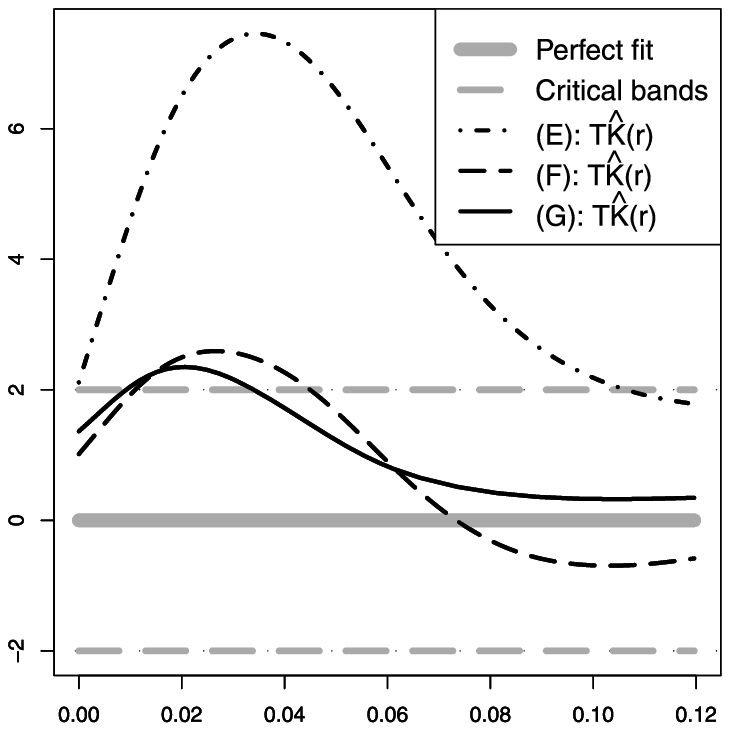}\\
\scriptsize{(a)}&\scriptsize{(b)}
\end{tabular}
  \caption{
    Model
    diagnostics based on pairwise distances
    for each of the models \modelTwoCSR--\modelTwoGey\ fitted to the data in
    Figure \protect\ref{fig:data}\textup{(c)}.
    \textup{(a)}~Residual $\Khat$;
    \textup{(b)}~smoothed standardized residual $\Khat$.}
  \label{fig:Synth2:K}
  \vspace*{3pt}
\end{figure*}

Figure \ref{fig:Synth2:K} shows the residual $\Khat$ and the smoothed
standardized residual $\Khat$ for the three models. These diagnostics
show that the homogeneous Poisson mo\-del~\modelTwoCSR\ is a poor fit,
with a positive residual suggesting correctly that the data are more
clustered than the Poisson process. The plots suggests that both models
\modelTwoAre\ and \modelTwoGey\ are considerably better fits to the data
than a Poisson model. They show that \modelTwoGey\ is a better fit than
\modelTwoAre\ over a range of $r$ values, and suggest that \modelTwoGey\
captures the correct form of the interaction.

\begin{figure*}[b]
\vspace*{3pt}
    \centering
\begin{tabular}{@{}cc@{}}

\includegraphics{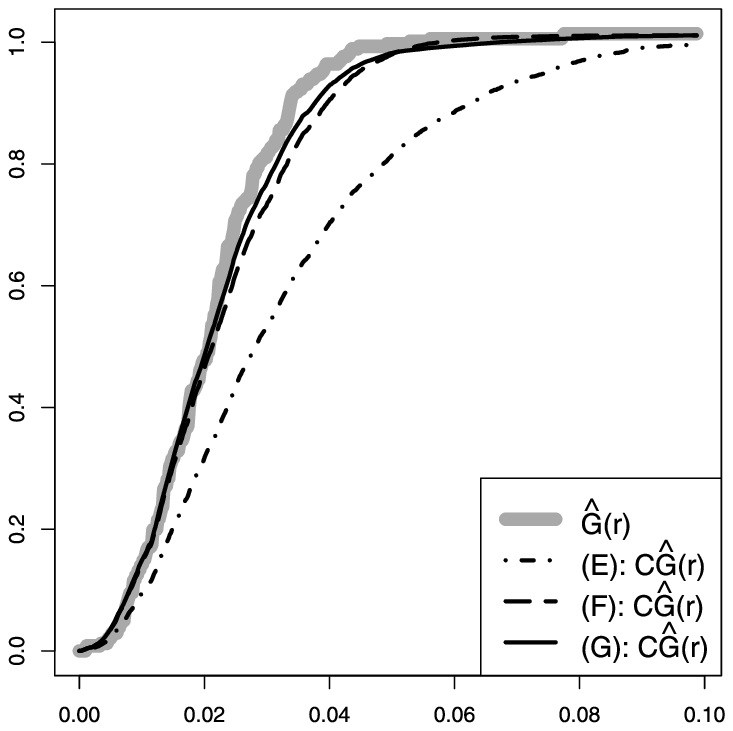}&
\includegraphics{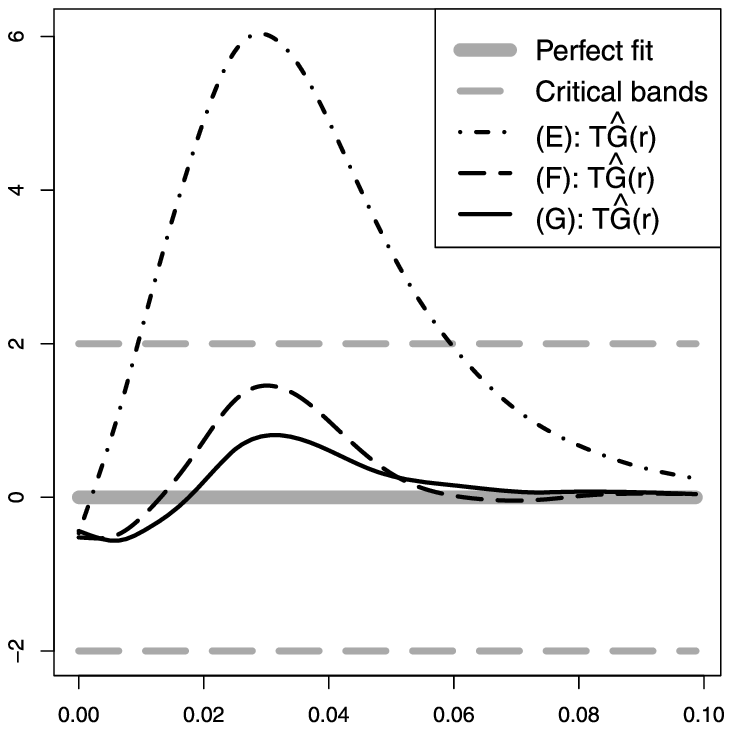}\\
\scriptsize{(a)}&\scriptsize{(b)}
\end{tabular}
  \caption{
    Model
    diagnostics based on nearest neighbor distances
    for each of the models \modelTwoCSR--\modelTwoGey\ fitted to the data in
    Figure~\protect\ref{fig:data}\textup{(c)}.
    \textup{(a)} $\Ghat$
    and its compensator under each model;
    \textup{(b)} smoothed standardized residual $\Ghat$.}
  \label{fig:Synth2:G}
\end{figure*}

\subsection{\texorpdfstring{Application of $\Ghat$ Diagnostics}{Application of G Diagnostics}}
\label{S:cluster:G}

\begin{figure*}
    \centering
\begin{tabular}{@{}cc@{}}

\includegraphics{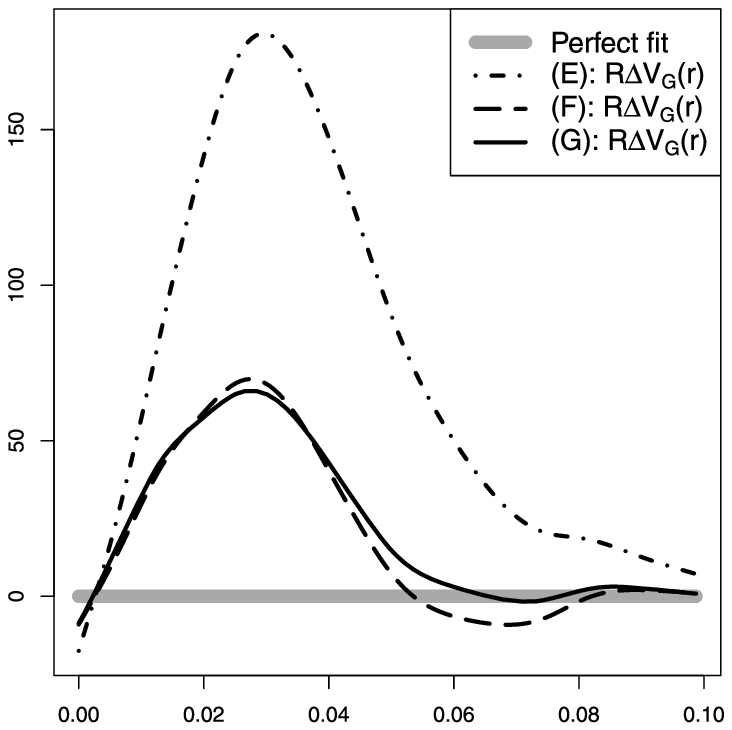}&
\includegraphics{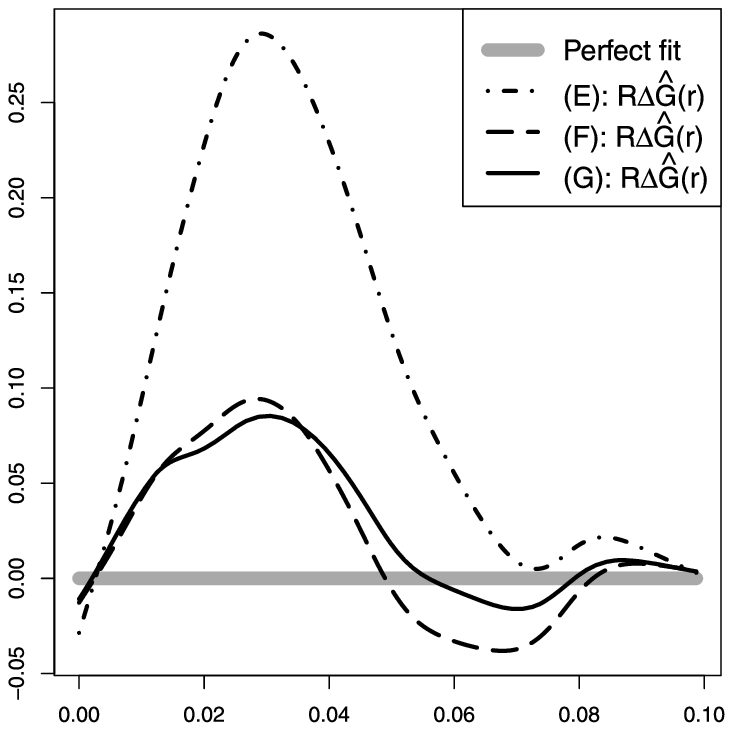}\\
\scriptsize{(a)}&\scriptsize{(b)}\vspace*{-3pt}
\end{tabular}
  \caption{Smoothed pseudo-residuals for each of the models
    \modelTwoCSR--\modelTwoGey\ fitted to the
    clustered point pattern in Figure \protect\ref{fig:data}\textup{(c)} when the
    perturbing model is \textup{(a)} the Geyer saturation
    model with saturation 1 and \textup{(b)} the
    $\Ghat$-model.
  }
  \label{fig:Synth2:Gps}\vspace*{-3pt}
\end{figure*}

Figure~\ref{fig:Synth2:G} shows $\hat G$ and its compensator, and the
corresponding residuals and standardized residuals, for each of the
models \modelTwoCSR--\modelTwoGey\ fitted to the clustered point pattern
in
Figure~\ref{fig:data}(c). 
The conclusions obtained from Figure \ref{fig:Synth2:G}(a) are the same
as those in Section \ref{S:cluster:K} based on $\hat K$ and its
compensator. Figure \ref{fig:Synth2:Gps} shows the smoothed
pseudo-residual diagnostics based on the
nearest neighbor
distances. The message from these diagnostics is very similar to that
from Figure~\ref{fig:Synth2:G}.\vadjust{\goodbreak}

Models \modelTwoAre\ and \modelTwoGey\ have the same range of interaction
$R = 0.1$. Comparing Figures~\ref{fig:Synth2:K} and \ref{fig:Synth2:G},
we might conclude that the $\Ghat$-compensator appears less sensitive
to the \textit{form} of interaction than the $\Khat$-compensator. Other
experiments suggest that $\Ghat$ is more sensitive than $\Khat$ to
discrepancies in the \textit{range} of interaction.\vadjust{\goodbreak}



\subsection{\texorpdfstring{Application of $\Fhat$ Diagnostics}{Application of F Diagnostics}}\vspace*{4pt}

\begin{figure*}
    \centering
\begin{tabular}{@{}cc@{}}

\includegraphics{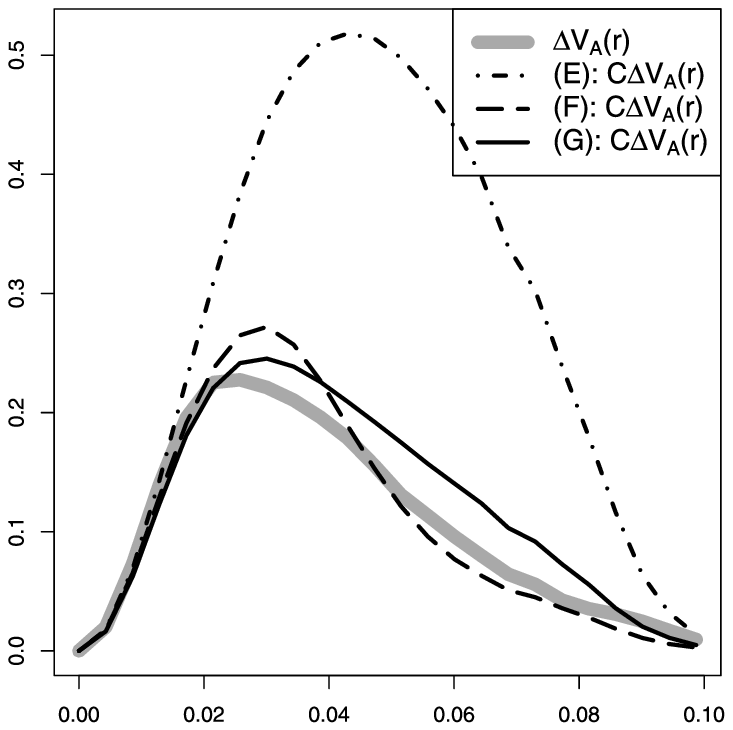}&
\includegraphics{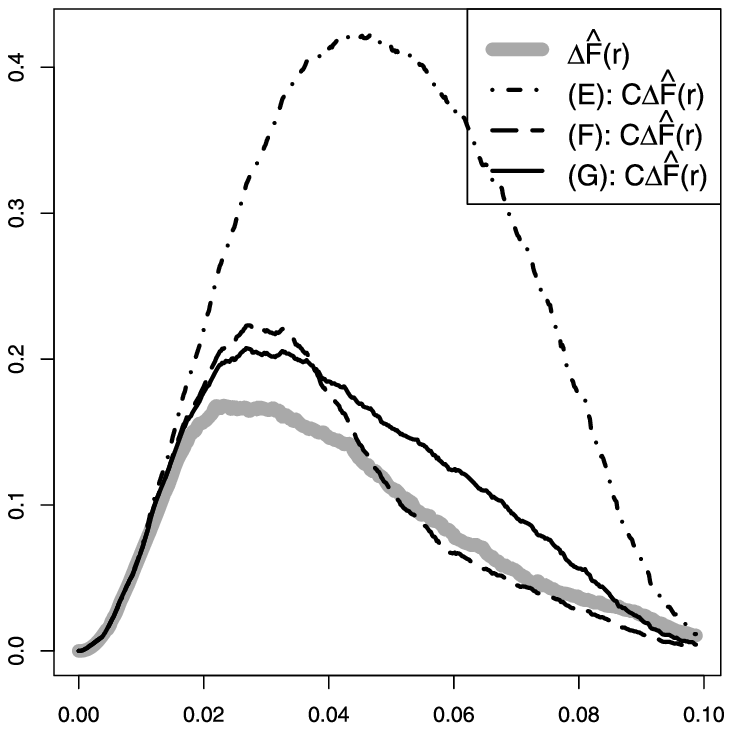}\\
\scriptsize{(a)}&\scriptsize{(b)}\vspace*{-3pt}
\end{tabular}
  \caption{Pseudo-sum and pseudo-compensators
    for the models \modelTwoCSR--\modelTwoGey\
    fitted to the clustered point pattern in
    Figure \protect\ref{fig:data}\textup{(c)}
    when the perturbing model is
    \textup{(a)} the area-interaction process and
    \textup{(b)} the $\Fhat$-model.}
  \label{fig:Synth2:A}
\end{figure*}

Figure~\ref{fig:Synth2:A} shows the pseudo-residual diagnostics based
on the empty space distances, for the three models fitted to the
clustered point pattern in Figure \ref{fig:data}(c). In this
case diagnostics based on the area-interaction process and the
$\Fhat$-model are very similar, as expected due to the close connection
between the two diagnostics. Here it is very noticeable that the
pseudo-compensator for the Poisson model has a higher peak than the
pseudo-sum, which correctly indicates that the data is more clustered
than a Poisson process.

\section{Test Case: Japanese Pines}
\label{S:jap}

\subsection{Data and Models}
\label{S:jap:data}

Figure \ref{fig:data}(a) shows the locations of seedlings and saplings
of Japanese black pine, studied by Numata \cite{numa61,numa64} and
analyzed extensively by Ogata and Tanemura
\cite{ogattane81,ogattane86}. In their definitive analysis
\cite{ogattane86} the fitted model was an inhomogeneous ``soft core''
pairwise interaction process with log-cubic first order term
$\lambda_\beta(x,y) = \exp(P_\beta(x,y))$, where $P_\beta$ is a cubic
polynomial in $x$ and $y$ with coefficient vector $\beta$, and density
%
\begin{eqnarray}\label{e:softcore}
\hspace*{15pt}f_{(\beta,\sigma^2)}(\bx)&=&c_{(\beta,\sigma^2)}\exp\biggl(\sum_i
P_\beta(x_i)\nonumber\\[-8pt]\\ [-8pt]
 \hspace*{15pt}&&\phantom{c_{(\beta,\sigma^2)}\exp\biggl(}{} -\sum_{i<j}(\sigma^4/\|x_i-x_j\|^4)\biggr),\nonumber
\end{eqnarray}
where 
$\sigma^2$ is a positive parameter. 

\begin{figure*}[b]
    \centering
\begin{tabular}{@{}cc@{}}

\includegraphics{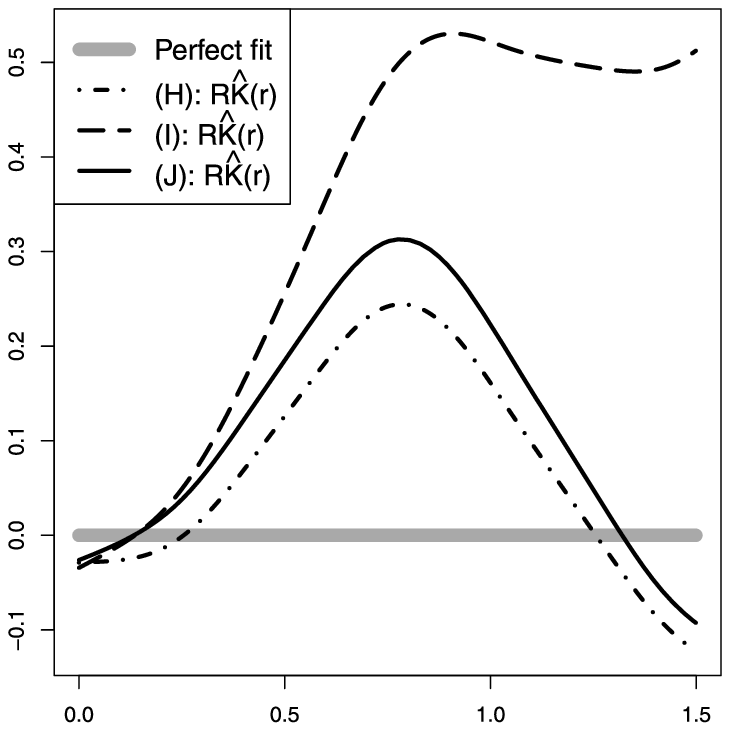}&
\includegraphics{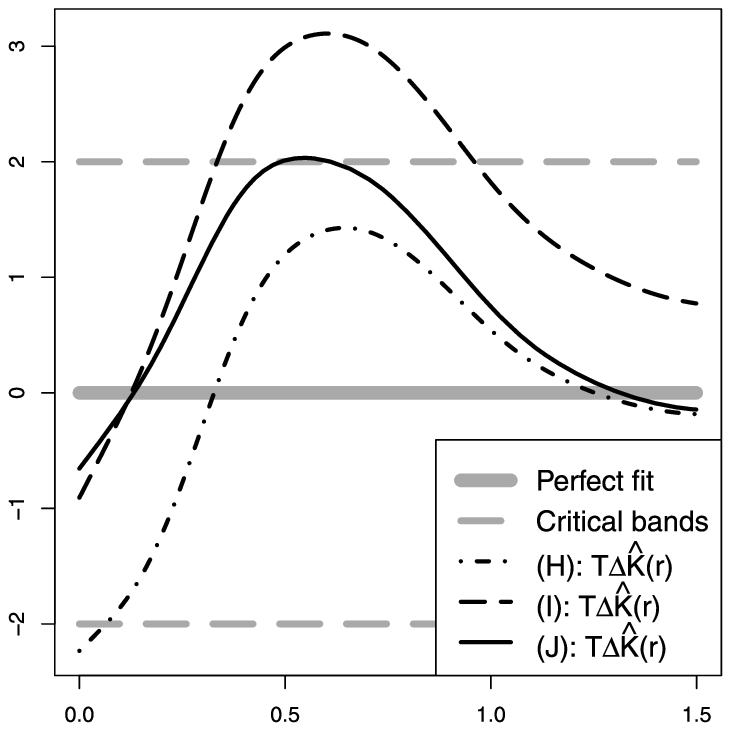}\\
\scriptsize{(a)}&\scriptsize{(b)}\vspace*{-3pt}
\end{tabular}
  \caption{
    Model
    diagnostics based on pairwise distances
    for each of the models \modelJapPoi--\modelJapOgT\
     fitted to the Japanese pines data in Figure~\protect\ref{fig:data}\textup{(a)}.
    \textup{(a)}~Smoothed residual $\Khat$;
    \textup{(b)} smoothed standardized residual $\Khat$.}
\label{fig:Jap:K}
\end{figure*}

Here we evaluate 
three models: \modelJapPoi\ an inhomogeneous Poisson process with
log-cubic intensity;\break \modelJapSof~a~homogeneous soft core pairwise
interaction process,
 that is, when $P_\beta(x,y)$ in \eqref{e:softcore} is
 replaced by a~real parameter;\vadjust{\goodbreak}
\modelJapOgT\ the Ogata--Tanemura mo-\break del~\eqref{e:softcore}. For more
detail on the data set and the fitted inhomogeneous soft core model,
see \cite{ogattane86,baddetal05}.

A complication in this case is that the soft core process
\eqref{e:softcore} is not Markov, since the pair potential $c(u,v) =
\exp(- \sigma^4/\|u-v\|^4)$ is always positive. Nevertheless, since
this function decays rapidly, it seems reasonable to apply the residual
and pseudo-residual diagnostics, using a cutoff distance $R$
such that $|\log c(u,v)| \le \epsilon$ when $\| u - v\| \le R$, for a
specified
tolerance $\epsilon$. The cutoff depends on the fitted parameter value
$\sigma^2$. We chose $\epsilon = 0.0002$, yielding $R = 1$. Estimated
interaction parameters were $\hat\sigma^2 = 0.11$ for model
\modelJapSof\ and $\hat\sigma^2 = 0.12$ for model \modelJapOgT.\vspace*{-1pt}

\subsection{\texorpdfstring{Application of $\Khat$ Diagnostics}{Application of K Diagnostics}}\vspace*{-1pt}
\label{S:jap:K}

A plot (not shown) of  $\Khat$ and its compensator for each of the
models \modelJapPoi--\modelJapOgT\ suggests that the homogeneous soft
core model \modelJapSof\ is inadequate, while the inhomogeneous models
\modelJapPoi\ and \modelJapOgT\ are reasonably good fits to the data.
However, it does not discriminate between the  models \modelJapPoi\ and
\modelJapOgT.

Figure~\ref{fig:Jap:K} shows smoothed versions of the residual and
standardized residual of $\Khat$ for each model. The Ogata--Tanemura
model \modelJapOgT\ is judged to be the best fit.\vspace*{-1pt}

\subsection{\texorpdfstring{Application of $\Ghat$ diagnostics}{Application of G diagnostics}}\vspace*{-1pt}
\label{S:jap:G}

\begin{figure*}
\centering
\begin{tabular}{@{}cc@{}}

\includegraphics{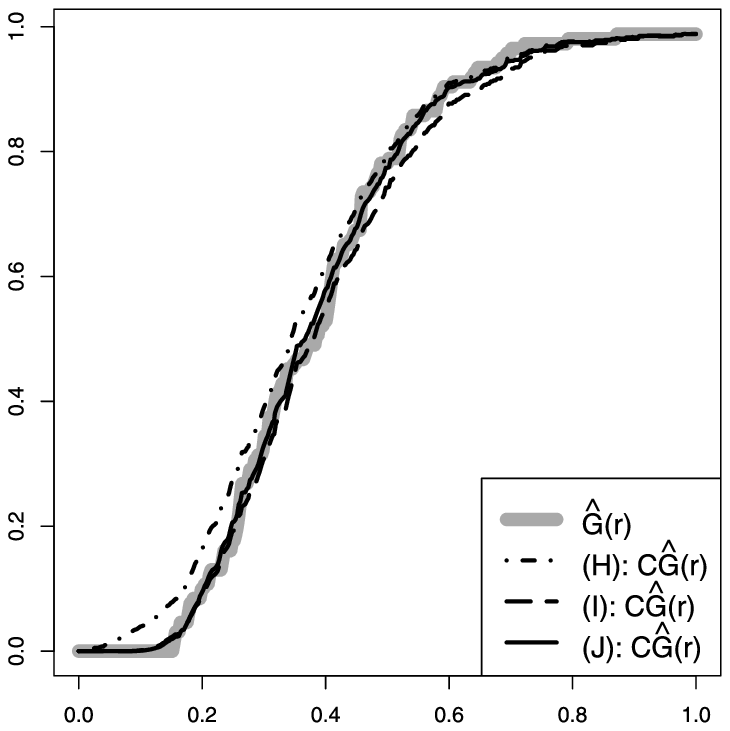}&
\includegraphics{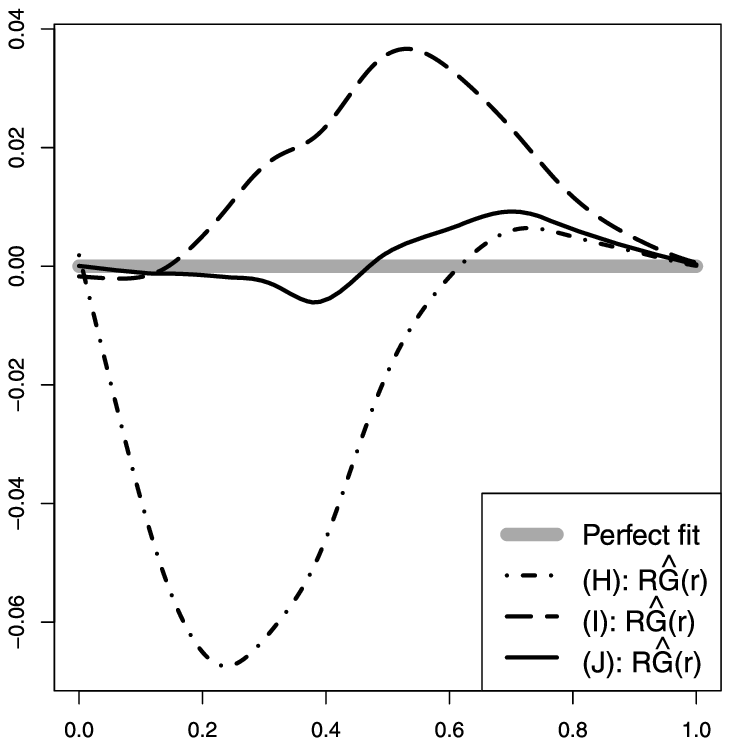}\\
\scriptsize{(a)}&\scriptsize{(b)}\\ [6pt]
\multicolumn{2}{@{}c@{}}{
\includegraphics{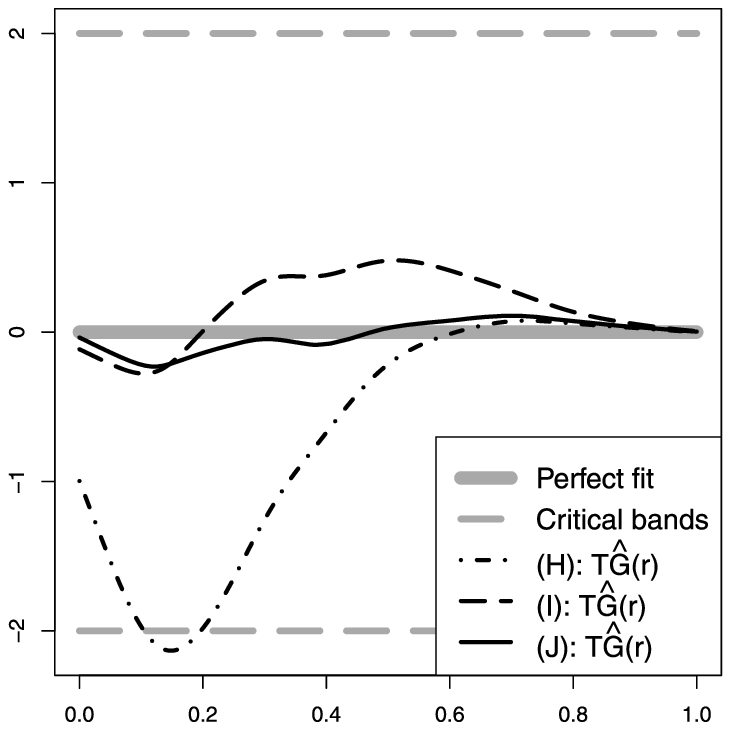}
}\\
\multicolumn{2}{@{}c@{}}{\scriptsize{(c)}}\vspace*{-3pt}
\end{tabular}
  \caption{
    Model
    diagnostics based on nearest neighbour distances
    for each of the models \modelJapPoi--\modelJapOgT\
     fitted to the Japanese pines data in Figure~\protect\ref{fig:data}\textup{(a)}.
    \textup{(a)} $\Ghat$
    and its compensator;
    \textup{(b)} smoothed residual $\Ghat$;
    \textup{(c)} smoothed standardised residual $\Ghat$.}
\label{fig:Jap:G}
\end{figure*}

\begin{figure*}
 \centering
\begin{tabular}{@{}cc@{}}

\includegraphics{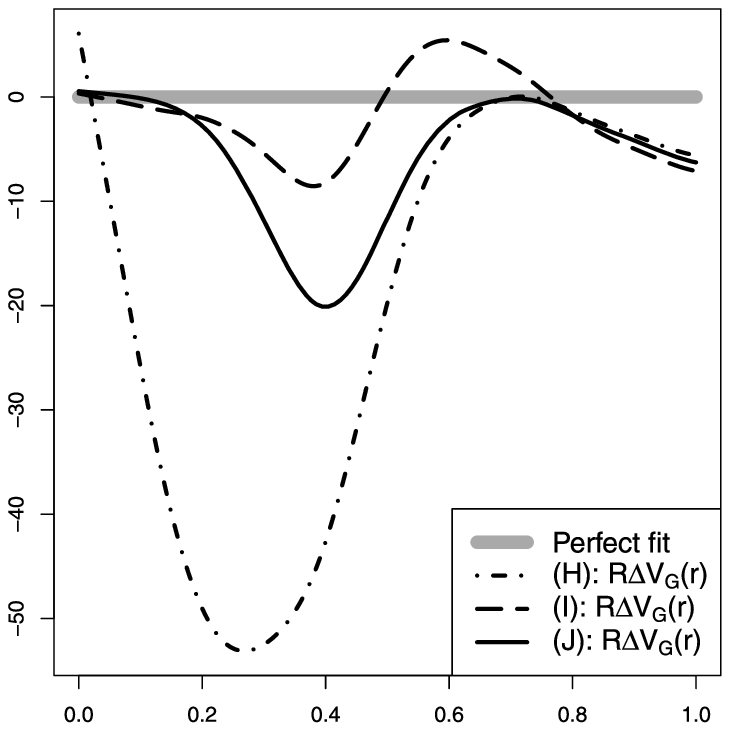}&
\includegraphics{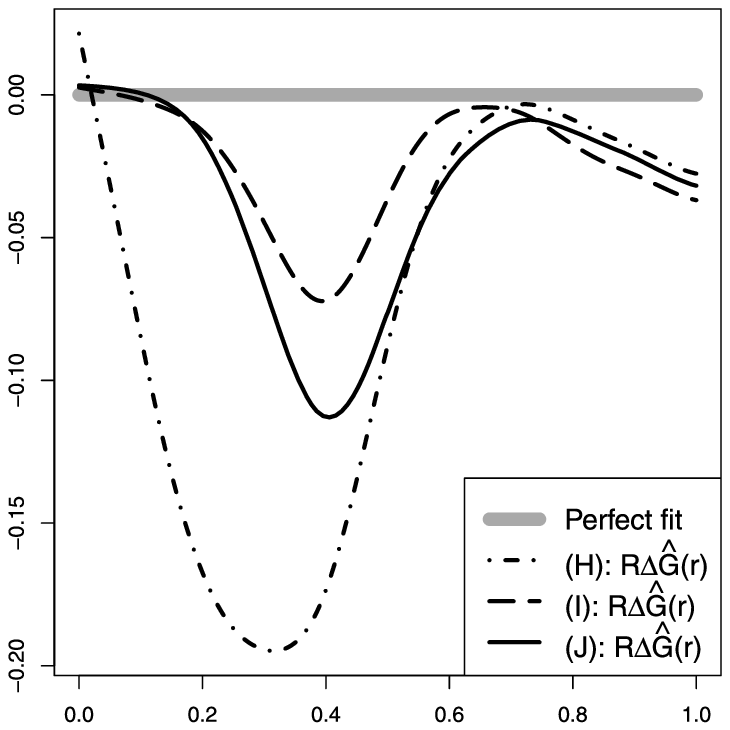}\\
\scriptsize{(a)}&\scriptsize{(b)}\vspace*{-3pt}
\end{tabular}
  \caption{Smoothed pseudo-residuals
    for each of the models \modelJapPoi--\modelJapOgT\
    fitted to the Japanese pines data in
Figure \protect\ref{fig:data}\textup{(a)}
    when the perturbing model is
    \textup{(a)} the Geyer saturation
    model with saturation 1 (null fitted on a fine grid) and
    \textup{(b)} the $\Ghat$-model.
  }
  \label{fig:Jap:Gps}\vspace*{-1pt}
\end{figure*}

Finally, for each of the models \modelJapPoi--\modelJapOgT\ fitted to
the Japanese pines data in Figure~\ref{fig:data}(a),
Figure \ref{fig:Jap:G}(a) shows\vadjust{\goodbreak} $\hat G$ and its compensator. The
conclusions are the same as those based on $\hat K$ shown in
Figure~\ref{fig:Jap:K}. Figure \ref{fig:Jap:Gps} shows the
pseudo-residuals when using either a perturbing Geyer model
[Figure \ref{fig:Jap:Gps}(a)] or a~perturbing $\hat G$-model
[Figure~\ref{fig:Jap:Gps}(b)].
Figures~\ref{fig:Jap:Gps}(a)--(b) tell almost the
same story: the inhomogeneous Poisson model \modelJapPoi\ provides the
worst fit, while it is difficult to discriminate between the fit for
the soft core models \modelJapSof\ and\vadjust{\goodbreak} \modelJapOgT. In conclusion,
considering Figures~\ref{fig:Jap:K}, \ref{fig:Jap:G} and
\ref{fig:Jap:Gps}, the Ogata--Tanemura model~\modelJapOgT\ provides the
best fit.

\subsection{\texorpdfstring{Application of $\Fhat$ diagnostics}{Application of F diagnostics}}

\begin{figure*}[b]
 \vspace*{-1pt}\centering
\begin{tabular}{@{}cc@{}}

\includegraphics{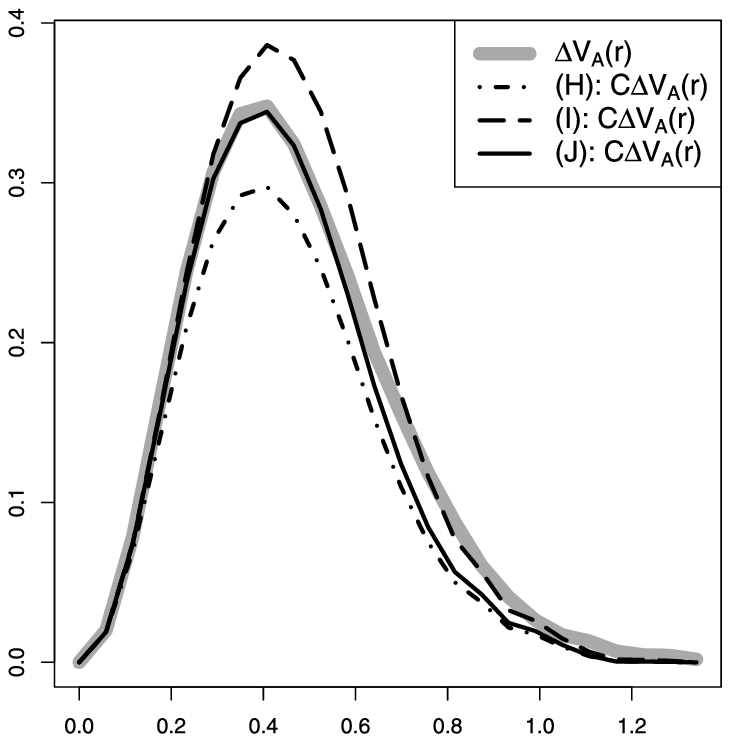}&
\includegraphics{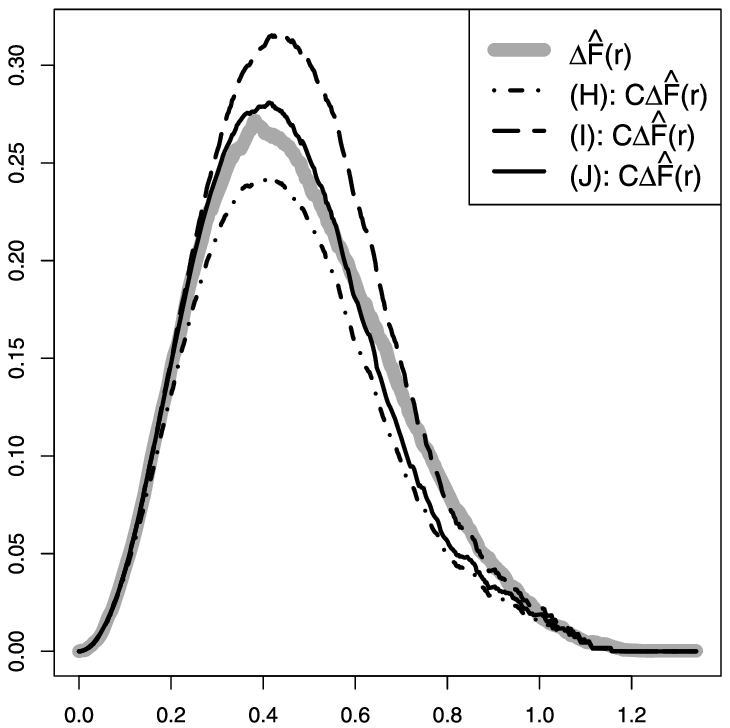}\\
\scriptsize{(a)}&\scriptsize{(b)}\vspace*{-3pt}
\end{tabular}
  \caption{
    Pseudo-sum and pseudo-compensators for the models
    \modelJapPoi--\modelJapOgT\ fitted to the real
    data pattern in Figure~\protect\ref{fig:data}\textup{(a)} when the
    perturbing model is \textup{(a)} the area-interaction
    process and \textup{(b)} the $\Fhat$-model.
  }
  \label{fig:Jap:A}
\end{figure*}

Finally, the empty space pseudo-residual diagnostics are shown in
Figure~\ref{fig:Jap:A} for the Japanese Pines data in
Figure~\ref{fig:data}(a). This gives a clear indication that the
Ogata--Tanemura model \modelJapOgT\ is the best fit to the data, and the
data pattern appears to be too regular compared to the Poisson model
\modelJapPoi\ and not regular enough for the homogeneous softcore
model~\modelJapSof.\looseness=1

\section{Summary of Test Cases}
\label{S:summary}

In this section we discuss which of the diagnostics we prefer to use
based on their behavior for the three test cases in
Sections~\ref{S:TrendInhib}--\ref{S:jap}.

Typically, the various diagnostics supplement each other well, and it
is recommended to use more than one diagnostic when
validating a model. It is well known that $\Khat$ is sensitive to
features at a larger scale than $\Ghat$ and $\Fhat$. Compensator and
pseudo-compensa\-tor plots are informative for gaining an overall picture
of
model validity, and tend to make it easy to recognize a poor fit when
comparing competing models. To compare models which fit closely, it may
be more informative to use (standardized) residuals or
pseudo-residuals. We prefer to use the standardized residuals, but it
is important not to over-interpret the significance of departure from
zero.

Based on the test cases here, it is not clear whether diagnostics based
on pairwise distances, nearest\break neighbor distances, or empty space
distances are preferable. However, for each of these we prefer to work
with compensators and residuals rather than pseudo-compensators and
pseudo-residuals when possible (i.e., it is only necessary to use
pseudo-versions for\vadjust{\goodbreak} diagnostics based on empty space distances). For
instance, for the first test case (Section~\ref{S:TrendInhib}) the best\vspace*{-0.5pt}
compensator plot is that in Figure~\ref{fig:Synth1:K}(a)\vspace*{-1pt} based on
pairwise distances ($\Khat$ and $\compen \Khat$) which makes it easy to
identify the correct model. On the other hand, in this test case the
best residual type plot is that in Figure~\ref{fig:Synth1:G}(b)\vspace*{-1pt}
based on nearest neighbor distances ($\stdres \Ghat$) where the correct
model is the only one within the critical\vadjust{\goodbreak} bands. However, in the third
test case (Section~\ref{S:jap}) the best compensator plot is one of the
plots in Figure~\ref{fig:Jap:A} with pseudo-compensators based on empty
space distances ($\psum V_A$ and $\pcom V_A$ or $\psum \Fhat$ and
$\pcom \Fhat$, respectively) which clearly indicates which model is
correct.

In the first and third test cases (Sections~\ref{S:TrendInhib}\break and~\ref{S:jap}),
which both involve inhomogeneous models, it is clear that
$\Khat$ and its compensator are more sensitive to lack of fit in the
first order term than~$\Ghat$ and its compensator [compare, e.g., the
results for the homogeneous model \modelOneStr\ in
Figures~\ref{fig:Synth1:K}(a) and~\ref{fig:Synth1:G}(b)]. It is
our general experience that diagnostics based on~$\Khat$ are
particularly well suited to assess the presence of interaction and to
identify the general form of interaction. Diagnostics based on $\Khat$
and, in particular, on $\Ghat$ seem to be good for assessing the range
of interaction.

Finally, it is worth mentioning the computational difference between
the various diagnostics (timed on a 2.5~GHz laptop). The calculations
for $\Khat$ and $\compen \Khat$ used
in Figure~\ref{F:KrightKcom}
are carried out in approximately five seconds, whereas the
corresponding calculations for $\Ghat$ and $\compen \Ghat$ only take a
fraction of a second. For  $\psum \Fhat$ and $\pcom \Fhat$, for
example, the calculations take about 45~seconds.

\section{Possible Extensions}

The definition of residuals and pseudo-residuals should extend
immediately to marked point processes. For space--time point processes,
residual diagnostics can be defined using the spatiotemporal
conditional intensity (i.e., given the past history). Pseudo-residuals
are unnecessary because the likelihood of a general space--time point
process is a~product integral (Mazziotto--Szpirglas identity). In the
space--time case there is a martingale structure in time, which gives
more hope of rigorous asymptotic results in the temporal (long-run)
limit regime.

Residuals can be derived from many other summary statistics. Examples
include third-order and higher-order moments
(Appendix~\ref{App:kthmoment}),
tessellation statistics (Appendix~\ref{App:tess}),
and various combinations of $F$, $G$ and~$K$.

In the definition of the extended model \eqref{e:extended} the
canonical statistic $S$ could have been allowed to depend on the
nuisance parameter $\theta$, but this would have complicated our
notation and some analysis.







\appendix
\renewcommand{\theequation}{\arabic{equation}}

\section{Further Diagnostics}
\label{App:further}
\setcounter{equation}{57}
In this appendix we present other diagnostics\break which we have not
implemented in software, and which therefore are not accompanied by
experimental results.

\subsection{Third and Higher Order Functional Summary~Statistics}
\label{App:kthmoment}

While the intensity and $K$-function are frequently-used summaries for
the first and second order moment properties of a spatial point
process, third and higher order summaries have been less used
\cite{schlbadd00,mollsyvewaag98,stiletal00,stoystoy95}.

\subsubsection*{Statistic of order $k$}

For a functional summary sta\-tistic of $k$th order, say,
%
\begin{equation}\label{E:kth}\qquad
S(\bx,r) =
\sum_{\{x_{i_1},\ldots,x_{i_k}\}\subseteq\bx}
q(\{x_{i_1},\ldots,x_{i_k}\},r),
\end{equation}
we obtain
\begin{eqnarray}\label{E:pairwise_relation4}
&&\psum S(\bx,r)\nonumber\hspace*{-28pt}\\
 &&\quad= k! S(\bx,r)\hspace*{-28pt}\\
 &&\quad=k!
 \sum_{\{x_{i_1},\ldots,x_{i_k}\}\subseteq\bx}
q(\{x_{i_1},\ldots,x_{i_k}\},r),\nonumber\hspace*{-28pt}\\
\label{E:pairwise_relation5}
&&\pcom S(\bx,r)\nonumber\hspace*{-28pt}\\
&&\quad = k! \compen
S(\bx,r)\nonumber\hspace*{-28pt}\\
&&\quad=(k-1)!\hspace*{-28pt}\\
&&\qquad{}\cdot\int_{W}
\sum_{\{x_{i_1},\ldots,x_{i_{k-1}}\}\subseteq\bx}
q(\{x_{i_1},\ldots,x_{i_{k-1}},u\},r)\nonumber\hspace*{-28pt}\\
&&\phantom{\qquad{}\cdot\int_{W}}{}\cdot \lambda_{\hat\theta}(u,\bx) \dee
u,\nonumber\hspace*{-28pt}\\\label{E:pairwise_relation6}
&&\PU(\hat\theta,r) \nonumber\hspace*{-28pt}\\ [-8pt]\\ [-8pt]
&&\quad= k!\resid S(\bx,r)=k!
S(\bx,r)-k! \compen S(\bx,r),\nonumber\hspace*{-28pt}
\end{eqnarray}
where  $i_1,i_2,\ldots$ are pairwise distinct in the sums
in~\eqref{E:pairwise_relation4}--\eqref{E:pairwise_relation5}.
In this case again, pseudo-residual diagnostics are equivalent to those
based on residuals.

\subsubsection*{Third order example}

For a stationary and isotropic point process (i.e., when the
distribution of $\bX$ is invariant under translations and rotations),
the intensity and $K$-function of the process completely determine its
first and second order moment properties. However, even in this case,
the simplest description of third order moments depends on a
three-dimensional vector specified from triplets $(x_i,x_j,x_k)$ of
points from $\bX$ such as the lengths and angle between the vectors
$x_i-x_j$ and $x_j-x_k$. This is often considered too complex, and
instead one considers a certain one-dimensional\vadjust{\goodbreak} property of the
triangle $T(x_i,x_j,x_k)$ as exemplified below, where $L(x_i,x_j,x_k)$
denotes the largest side in $T(x_i,x_j,x_k)$.

Let the canonical sufficient statistic of the perturbing density
\eqref{e:perturbing} be
%
\begin{eqnarray}\label{e:wwww}
S(\bx,r)&=&V_T(\bx,r)\nonumber\\ [-8pt]\\ [-8pt]
&=&\sum_{i<j<k}\indicate{L(x_i,x_j,x_k)\le r}.\nonumber
\end{eqnarray}
The perturbing model is a special case of the \textit{triplet interaction
point process} studied in \cite{geye99}. It is also a~special case of
\eqref{E:kth} with
\[
q(\{x_i,x_j,x_k\},r)=\indicate{L(x_i,x_j,x_k)\le r} ;
\]
residual and pseudo-residual diagnostics are equivalent and given by
 \eqref{E:pairwise_relation4}--\eqref{E:pairwise_relation6}.

\subsection{Tessellation Functional Summary Statistics}
\label{App:tess}

Some authors have suggested the use of tessellation methods for
characterizing spatial point processes \cite{illietal08}.
A planar tessellation is a subdivision of planar region such as $W$ or
the entire plane $\mathbb R^2$.
For example, consider the Dirichlet tessellation of~$W$ generated by
$\bx$, that is, the tessellation with cells
\begin{eqnarray*}
  C(x_i|\bx)&=&\{u\in W|\|u-x_i\|\le \|u-x_j\|\\
&&\hspace*{70pt}\mbox{for all $x_j$
    in $\bx$}\},\\
&&\hspace*{135pt} i=1,\ldots,n.
\end{eqnarray*}
Suppose the canonical sufficient statistic of the perturbing density
\eqref{e:perturbing} is
%
\begin{equation}\label{E:soft.ord}
\hspace*{10pt}S(\bx,r) =  V_O(\bx,r) =
\sum_{i}\indicate{|C(x_i|\bx)|\le r}.
\end{equation}
This is a sum of local contributions as in \eqref{e:sumoflocal},
although not of local statistics in the sense mentioned in
Section~\ref{S:theory:residual}, since $\indicate{|C(x_i|\bx)|\le r}$
depends on those points in $\bx_{-i}$ which are Dirichlet neighbors to
$x_i$ and such points may of course not be $r$-close to $x_i$ (unless
$r$ is larger than the diameter of $W$).
We call this perturbing model 
a \textit{soft Ord process}; Ord's process as defined in
\cite{baddmoll89} is the limiting case $\phi\rightarrow-\infty$ in
\eqref{e:perturbing}, that is, when $r$ is the lower bound on the size
of cells. Since $V_O(\bx)\le n(\bx)$, the perturbing model is
well-defined for all $\phi\in\mathbb
R$. 

Let $\sim_{\bx}$ denote the Dirichlet neighbor relation for the points
in $\bx$, that is, $x_i\sim_{\bx}x_j$ if $C(x_i|\bx)\cap\break
C(x_j|\bx)\not=\emptyset$. Note that $x_i\sim_{\bx}x_i$. Now,
\begin{eqnarray}\label{e:fffffffff}
  \Delta_u S(\bx,r)&=&\mathbb{I}\bigl\{|C(u|\bxplus u)|\le
r\bigr\}\nonumber\\
  &&{}+\sum_{v\not=u:v\sim_{\bxplus u}u}\bigl[
    \mathbb{I}\bigl\{|C(v|\bxplus u)| \le r\bigr\}\\
   &&\hspace*{64pt}{} - \mathbb{I}\bigl\{|C(v|\bx\setminus\{u\})|\le r\bigr\}\bigr]\nonumber
\end{eqnarray}
depends not only on the points in $\bx$ which are Dirichlet neighbors
to $u$ (with respect to $\sim_{\bxplus u}$) but also on the Dirichlet
neighbors to those points (with respect to $\sim_{\bxplus u}$ or with
respect to $\sim_{\bx\setminus\{u\}}$). In other words, if we define
the iterated Dirichlet neighbor relation by that $x_i\sim_{\bx}^2x_j$
if there exists some $x_k$ such that $x_i\sim_{\bx}x_k$ and
$x_j\sim_{\bx}x_k$, then $t(u,\bx)$ depends on those points in $\bx$
which are iterated Dirichlet neighbors to $u$ with respect to
$\sim_{\bxplus u}$ or with respect to $\sim_{\bx\setminus\{u\}}$. The
pseudo-sum associated to the soft Ord process is
\begin{eqnarray*}
\psum V_O(\bx,r) &=& V_O(\bx,r)\\
   &&{}+ \sum_{i}\sum_{j\not=i:x_j\sim_\bx x_i}
    [\indicate{|C(x_j|\bx)|\le r}\\
   &&\hspace*{72pt}{}-\indicate{|C(x_j|\bxminus i)|\le r}
   ]
\end{eqnarray*}
and from \eqref{e:pseudocomp} and \eqref{e:fffffffff} we obtain the
pseudo-com\-pensator. From \eqref{e:compensatorlocal}
   and \eqref{E:soft.ord}, we obtain
the Papangelou compensator
\[
\compen V_O(\bx,r)=\int_W \mathbb{I}\bigl\{|C(u|\bx\cup\{u\})|\le r\bigr\}
\lambda_{\hat\theta}(u,\bx)\,\mathrm{d}u.
\]
%

Many other examples of tessellation characteristics may be of interest.
For example, often the Delaunay tessellation is used instead of the
Dirichlet tessellation. This is the dual tessellation to the Dirichlet
tessellation, where the Delaunay cells generated by $\bx$ are given by
those triangles $T(x_i,x_j,x_k)$ such that the disc containing
$x_i,x_j,x_k$ in its bounda\-ry does not contain any further points from
$\bx$ (strict\-ly speaking we need to assume a regularity condition,
namely, that $\bx$ has to be in general quadratic position; for such
details, see \cite{baddmoll89}). For instance, the summary statistic
$t(\bx,r)$ given by the number of Delaunay cells
$T(x_i,x_j,x_k)$ with $L(x_i,x_j,x_k)\le r$,
where $L(u,v,w)$ is the length of the triangle with vertices
$u,v,w$,
is a kind of third order statistic%
is related to \eqref{e:wwww} but concerns only the maximal cliques
of Dirichlet neighbors (assuming again the general quadratic position
condition). The corresponding\break perturbing model has not been studied in
the literature, to the best of our knowledge.

\section{Variance Formulae}
\label{App:var}

\setcounter{equation}{64}
This appendix concerns the variance of diagnostic quantities of the
form
\begin{eqnarray*}
I &=& \sum_i h(x_i, \bXminus i)
      - \int_W h(u,\bx) \lambda_\theta(u,\bX) \dee u, \\
R &=& \sum_i h(x_i, \bXminus i)
      - \int_W h(u,\bx) \lambda_{\hat\theta}(u,\bX) \dee u,
\end{eqnarray*}
where $h(\cdot)$ is a functional for which these quantities are almost
surely finite, $\bX$ is a point process on~$W$ with Papangelou
conditional intensity $\lambda_\theta(u,\bX)$ and~$\hat\theta$ is an
estimate of $\theta$ (e.g., the MPLE).

\subsection{General Identity}

Exact formulae for the variance of the innovation~$I$ and residual $R$
are given in \cite{baddmollpake08}. Expressions for $\Var R$ are
unwieldy \cite{baddmollpake08}, Section 6, but to a first approximation
we may ignore the effect of estimating~$\theta$ and consider the
variance of $I$. Suppressing the dependence on $\theta$, this is
(\cite{baddmollpake08}, Proposition 4),
%
\begin{eqnarray}\label{e:var.innov}
   \Var I
   &=& \int_W \expected[ h(u,\bX)^2 \lambda(u,\bX)] \dee u \nonumber\hspace*{-28pt}\\ [-8pt]\\ [-8pt]
     \hspace*{15pt}&&{} + \int_{W^2} \expected[
        A(u,v,\bX) + B(u,v,\bX)
        ] \dee u \dee v,\nonumber\hspace*{-28pt}
 \end{eqnarray}
where
\begin{eqnarray*}
  A(u,v,\bX) &=& \Delta_u h(v,\bX)  \Delta_v h(u,\bX) \lambda_2(u,v,\bX), \\
  B(u,v,\bX) &=& h(u,\bX) h(v,\bX)\\
&&{}\cdot \{
                  \lambda(u,\bX) \lambda(v,\bX) - \lambda_2(u,v,\bX)
                  \},
\end{eqnarray*}
where $\lambda_2(u,v,\bx) = \lambda(u,\bx) \lambda(v, \augm{\bx}{u})$
is the second order
Papangelou conditional intensity.
%
Note that for a Poisson process $B(u,v,\bX)$ is identically zero.

\subsection{Pseudo-Score}

Let $S(\bx,z)$ be a functional summary statistic with function argument
$z$. Take $h(u,\bX) = \Delta_u S(\bx, z)$.\break 
Then the innovation $I$ is the pseudo-score (\ref{e:PU}), and the
variance formula \eqref{e:var.innov} becomes
\begin{eqnarray}\label{e:var.PU}
 &&\Var[{\PU(\theta)}]\nonumber\hspace*{-10pt}\\[2pt]
 &&\quad=
  \int_W \expected [
    (\Delta_u S(\bX,z) )^2 \lambda(u,\bX)
  ] \dee u \nonumber\hspace*{-10pt}\\[2pt]
   &&\qquad{}+  \int_{W^2} \expected[
   (\Delta_u \Delta_v S(\bX,z) )^2 \lambda_2(u,v,\bX)
  ] \dee u \dee v \hspace*{-10pt}\\[2pt]
   &&\qquad{}+  \int_{W^2} \expected[
          \Delta_u S(\bx,z)  \Delta_v S(\bx,z)\nonumber\hspace*{-10pt}\\[2pt]
&&\qquad{}\cdot
          \{
            \lambda(u,\bX) \lambda(v,\bX) - \lambda_2(u,v,\bX)
          \}
        ] \dee u \dee v,\nonumber\hspace*{-10pt}
\end{eqnarray}
where for $u\not=v$ and $\{u,v\}\cap\bx=\emptyset$,
\begin{eqnarray*}
  \Delta_u \Delta_v S(\bx,z)
&=&   S(\bxplus {u,v}, z) - S(\bxplus u,z)\\[2pt]
&&{} - S(\bxplus v, z) + S(\bx,z)
\end{eqnarray*}
satisfies $\Delta_u \Delta_v S(\bx,z) = \Delta_v \Delta_u S(\bx,z)$.

\section{Modified Edge Corrections}
\label{App:edge}
\setcounter{equation}{66}

Appendices~\ref{App:edge}--\ref{App:G} describe modifications to the
standard edge corrected estimators of $K(r)$ and $G(r)$
required in the conditional case (Section~\ref{S:assump:cond}) because
the Papangelou conditional intensity $\lambda(u,\bx)$ can or should
only be evaluated at locations $u \in \Wfree$ where $\Wfree \subset W$.
Corresponding compensators are also given.

Assume the point process is Markov and we are in the conditional case
as described in Section~\ref{S:Markov}. Consider an empirical
functional statistic of the form
%
\begin{equation}\label{e:sumS}
  S_W(\bx,r) = \sum_{x_i \in \bx} s_W(x_i, \bx\setminus\{x_i\}, r)
\end{equation}
with compensator (in the unconditional case)
\[
  \compen S_W(\bx,r) = \int_W s_W(u,\bx,r) \lambda_{\hat\theta}(u,\bx) \dee u.
\]
We explore two different strategies for modifying the edge correction.

In the \textit{restriction approach}, we replace $W$ by $\Wfree$ and
$\bx$ by $\bxfree = \bx \cap \Wfree$, yielding
%
\begin{eqnarray}\label{e:sumSrestrict}
 \hspace*{20pt} S_{\Wfree}(\bx,r) &=& \sum_{x_i \in \bxfree} s_{\Wfree}(x_i,
    \bxfree\setminus\{x_i\}, r),\nonumber \\ [-8pt]\\ [-8pt]
   \hspace*{20pt} \compen S_{\Wfree}(\bx,r) &=& \int_{\Wfree} s_{\Wfree}(u,\bxfree,r)
\lambda_{\hat\theta}(u,\bxfree \vert \bxfixed) \dee u .\nonumber
\end{eqnarray}
%
Data points in the boundary region $\Wfixed$ are ignored in the
calculation of the empirical statistic $S_{\Wfree}$. The boundary
configuration $\bxfixed = \bx \cap \Wfixed$ contributes only to the
estimation of $\theta$ and the calculation of the Papangelou
conditional intensity $\lambda_{\hat\theta}(\cdot,\cdot \vert
\bxfixed)$. This has the advantage that the modified empirical
statistic~(\ref{e:sumSrestrict}) is identical to the standard statistic
$S$ computed on the subdomain $\Wfree$; it can be computed using
existing software, and requires no new theoretical justification.
The disadvantage 
is that we lose information by discarding some of the data.

In the \textit{reweighting approach} we retain the boundary points and
compute
\begin{eqnarray*}
  S_{\Wfree,W}(\bx,r) &=& \sum_{x_i \in \bxfree}
    s_{\Wfree,W}(x_i, \bx\setminus\{x_i\}, r), \\
  \compen S_{\Wfree,W}(\bx,r) &=& \int_{\Wfree} s_{\Wfree,W}(u,\bx,r)
                     \lambda_{\hat\theta}(u,\bxfree \vert \bxfixed) \dee u,
\end{eqnarray*}
where $s_{\Wfree,W}(\cdot)$ is a modified version of
$s_W(\cdot)$.
Boundary points contribute to the computation of the modified summary
statistic $S_{\Wfree,W}$ and its compensator. The modification is
designed so that $S_{\Wfree,W}$ has properties
analogous to $S_W$.

The $K$-function and $G$-function of a point process~$\bY$ in $\real^2$
are defined \cite{ripl76,ripl77} under the assumption that $\bY$ is
second order stationary and strictly stationary, respectively. The
standard estima-\break tors~$\Khat_W(r)$ and $\Ghat_{\bx}(r)$ of the
$K$-function and $G$-function,  respectively, are designed to be
approximately pointwise unbiased estimators when applied to $\bX = \bY
\cap W$.

We do not necessarily assume stationarity, but when constructing
modified summary statistics\break $\Khat_{\Wfree,W}(r)$ and
$\Ghat_{\Wfree,W}(r)$, we shall require that they are also
approximately pointwise unbiased estimators of $K(r)$ and $G(r)$,
respectively, when $\bY$ is stationary. This greatly simplifies the
interpretation of plots of $\Khat_{\Wfree,W}(r)$ and
$\Ghat_{\Wfree,W}(r)$ and their compensators.

\section{Modified Edge Corrections for the $K$-Function}
\label{App:K}
\setcounter{equation}{68}

\subsection{Horvitz--Thompson Estimators}

The most common nonparametric estimators of the $K$-function
\cite{ripl76,ohse83,badd99b} are continuous Horvitz--Thompson type
estimators \cite{badd93a,cord93} of the form
%
\begin{eqnarray}\label{e:Khat:app}
  \Khat(r) &=& \Khat_W(r)\nonumber\hspace*{-20pt}\\ [-8pt]\\ [-8pt]
&=& \frac{1}{\hat{\rho}^2(\bx) |W|}
  \sum_{i\neq j} 
  e_W(x_i,x_j) \indicate{\|x_i - x_j\| \le r} .\nonumber\hspace*{-20pt}
\end{eqnarray}
Here $\hat{\rho}^2 = \hat{\rho}^2(\bx)$ should be an
approximately unbiased estimator of the squared intensity
$\rho^2$ under stationarity. 
Usually $\hat{\rho}^2(\bx) = n(n-1)/|W|^2$ where \mbox{$n =
n(\bx)$.}

The term $e_W(u,v)$ is an edge correction weight, depending on the
geometry of $W$, designed so that the double sum in (\ref{e:Khat:app}),
say, $\hat Y(r) = \hat{\rho}^2(\bx) |W| \Khat(r)$, is an unbiased
estimator of $Y(r) = \rho^2 |W| K(r)$. Popular examples are the
Ohser--Stoyan translation edge correction with
%
\begin{eqnarray} \label{e:edgetrans}
        e_W(u,v) &=& e_W^{\mathrm{trans}}(u,v)\nonumber\\ [-8pt]\\ [-8pt]
& =& \frac{|W|}{|W \cap (W+(u-v))|}\nonumber
\end{eqnarray}
and Ripley's isotropic correction with
\begin{eqnarray}\label{e:edgeRipley}
  e_W(u,v)
  &=& e_W^{\mathrm{iso}}(u,v)\nonumber\\ [-8pt]\\ [-8pt]
  &=& \frac{
      2 \pi \Vert u-v\Vert
    }{\mathsf{length}(\partial B(u,\Vert u-v\Vert) \cap W)}.\nonumber
\end{eqnarray}
Estimators of the form (\ref{e:Khat:app}) satisfy the local
decomposition (\ref{e:sumS}) where
\begin{eqnarray*}
     s_W(u,\bx,r) &=&
          \frac 1 {\hat{\rho}^2(\bx \cup \{u\}) |W|}\\
          &&{}\cdot\sum_j  e_W(u,x_j) \indicate{ \Vert u - x_j\Vert \le r },\quad  u \notin \bx.
\end{eqnarray*}
Now we wish to modify (\ref{e:Khat:app}) so that the outer summation is
restricted to data points $x_i$ in $\Wfree \subset W$, while retaining
the property of unbiasedness for stationary and isotropic point
processes.
The \textit{restriction estimator} is
%
\begin{eqnarray}\label{e:Khat:restrict}
 &&\hspace*{20pt}\Khat_{\Wfree}(r)\nonumber\\
 &&\hspace*{20pt}\quad=
  \frac{1}{\hat{\rho}^2(\bxfree) |\Wfree|}\\
&&\hspace*{20pt}\qquad\cdot{} \sum_{x_i \in\bxfree}
  \sum_{x_j \in \bxfreeminus i} e_{\Wfree}(x_i,x_j) \indicate{\|x_i - x_j\| \le r},\nonumber
\end{eqnarray}
where the edge correction weight is given by (\ref{e:edgetrans}) or
(\ref{e:edgeRipley}) with $W$ replaced by $\Wfree$.
%
A more efficient alternative is to replace (\ref{e:Khat:app}) by the
\textit{reweighting estimator}
%
\begin{eqnarray}\label{e:Khat:smart}
     &&\hspace*{18pt}\Khat_{\Wfree,W}(r)\nonumber\\
&&\hspace*{18pt}\quad=
     \frac{1}{\hat{\rho}^2(\bx) |\Wfree|}\\
&&\hspace*{18pt}\qquad{}\cdot \sum_{x_i \in\bxfree}
    \sum_{x_j \in \bxminus i} e_{\Wfree,W}(x_i,x_j) \indicate{ \Vert x_i - x_j\Vert \le r },\nonumber
\end{eqnarray}
where $e_{\Wfree,W}(u,v)$ is a modified version of $e_W(\cdot)$
constructed so that the double sum in (\ref{e:Khat:smart}) is unbiased
for $Y(r)$. Compared to the restriction estimator~\eqref{e:Khat:restrict},
the reweighting estimator \eqref{e:Khat:smart}
contains additional contributions from point pairs $(x_i,x_j)$ where
$x_i \in \bxfree$ and $x_j \in \bxfixed$.

The modified edge correction factor  $e_{\Wfree,W}(\cdot)$\break for~\eqref{e:Khat:smart}
is the Horvitz--Thompson weight \cite{badd99b} in
an appropriate sampling context. Ripley's \cite{ripl76,ripl77}
iso\-tropic correction \eqref{e:edgeRipley} is derived assuming isotropy,
by Palm conditioning on the location of the first point $x_i$, and
determining the probability that $x_j$ would be observed inside $W$
after a random rotation about $x_i$. Since the constraint on $x_j$ is
unchanged, no modification of the edge correction weight is required,
and we take $e_{\Wfree,W}(\cdot) = e_W(\cdot)$ as in
\eqref{e:edgeRipley}. Note, however, that the denominator in
(\ref{e:Khat:smart}) is\break changed from $|W|$ to $|\Wfree|$.

The Ohser--Stoyan \cite{ohsestoy81} translation correction
(\ref{e:edgetrans}) is derived by considering two-point sets\vadjust{\goodbreak}
$(x_i,x_j)$ sampled under the constraint that both $x_i$ and $x_j$ are
inside $W$. Under the modified constraint that $x_i \in \Wfree$ and
${x_j \in W}$, the appropriate edge correction weight is
\begin{eqnarray*}
   e_{\Wfree,W}(u,v) &=& e_{\Wfree,W}(u-v)\\
&=& \frac{|W \cap (\Wfree + (u-v))|}{|\Wfree|}
\end{eqnarray*}
so that $1/e_{\Wfree,W}(z)$ is the fraction of locations $u$ in~$\Wfree$ such that $u + z \in W$.

\subsection{Border Correction}

A slightly different creature is the border corrected estimator [using
usual intensity estimator $\hat\rho=n(\bx)/\break|W|$]
\begin{eqnarray*}
   \Khat_W(r) &=& \frac{|W|
                     }{
                     n(\bx) n(\bx \cap W_{\ominus r})}\\
&&{}\cdot
                  \sum_{x_i\in\bx} \sum_{x_j\in\bxminus i}
                  \indicate{x_i \in W_{\ominus r}}
                      \indicate{\Vert x_i - x_j\Vert \le r}
\end{eqnarray*}
with compensator (in the unconditional case)
\begin{eqnarray*}
     \compen \Khat_W(r) &=&
     \int_{W_{\ominus r}}
     \frac{
       |W|
       \sum_{x_j\in\bx}
       \indicate{\Vert u - x_j\Vert \le r }
     }{
       (n(\bx)+1)(n(\bx \cap W_{\ominus r})+1)
     }\\
      &&\hphantom{\int_{W_{\ominus r}}}{}\cdot \lambda_{\hat\theta}(u,\bxfree \vert \bxfixed)
       \dee u
.
\end{eqnarray*}
%

\noindent The
\textit{restriction estimator} is
\begin{eqnarray*}
    \Khat_{\Wfree}(r) &=&
    \frac{|\Wfree|}{n(\bxfree)n(\bx \cap \Wfree_{\ominus r})
}\\
 &&{}\cdot   \sum_{x_i\in\bxfree} \sum_{x_j\in\bxfreeminus i}\!
      \indicate{x_i \in \Wfree_{\ominus r}}
      \indicate{ \Vert x_i - x_j\Vert \le r }
\end{eqnarray*}
and the compensator is
\begin{eqnarray*}
     \compen \Khat_{\Wfree}(r) &=&
       \int_{\Wfree_{\ominus r}}
     \frac{
       |\Wfree|
       \sum_{x_j\in\bxfree}
       \indicate{\Vert u - x_j\Vert \le r }
     }{
      (n(\bxfree)+1)(n(\bx \cap \Wfree_{\ominus r})+1)
     }\\
&&\hphantom{\int_{\Wfree_{\ominus r}}}{}\cdot
       \lambda_{\hat\theta}(u,\bxfree \vert \bxfixed) \dee u.
\end{eqnarray*}
Typically, $\Wfree = W_{\ominus R}$, so $\Wfree_{\ominus r}$ is equal
to $W_{\ominus(R+r)}$.
The \textit{reweighting estimator} is
\begin{eqnarray*}
    \Khat_{\Wfree,W}(r) &\!=\!&
    \frac{|W|}{n(\bx)n(\bxfree \cap W_{\ominus r})
}\\
  &&\!{}\cdot  \sum_{x_i\in\bxfree} \sum_{x_j\in\bxminus i}\!
      \indicate{x_i\!\in\!W_{\ominus r}}
      \indicate{ \Vert x_i\!-\!x_j\Vert\!\le\!r }
\end{eqnarray*}
and the compensator is
\begin{eqnarray*}
     \compen \Khat_{\Wfree,W}(r) &\!=\!&
       \int_{\Wfree \cap W_{\ominus r}}\!
     \frac{
       |W|
       \sum_{x_j\in\bx}\!
       \indicate{\Vert u\!-\!x_j\Vert\!\le\!r }
     }{
      (n(\bx)\!+\!1)(n(\bxfree\!\cap\!W_{\ominus r})\!+\!1)
     }\\
&&\hphantom{\int_{\Wfree \cap W_{\ominus r}}}{}\cdot
       \lambda_{\hat\theta}(u,\bxfree \vert \bxfixed) \dee u.
\end{eqnarray*}
Usually, $\Wfree = W_{\ominus R}$, so $\Wfree \cap W_{\ominus r}$ is
equal to\break $W_{\ominus \max(R,r)}$. From this we conclude that when using
border correction we should always use the reweighting estimator since
the restriction estimator discards additional information and neither
the implementation nor the interpretation is easier.

\section{Modified Edge Corrections for Nearest Neighbor Function $G$}
\vspace*{3pt}
\label{App:G}

\setcounter{equation}{73}
\subsection{Hanisch Estimators}


Hanisch \cite{hani84a} considered estimators for $G(r)$ of the form
$\Ghat_W(r) = \hat D_{\bx}(r)/\hat \rho$, where $\hat\rho$ is some
estimator of the intensity $\rho$, and
%
\begin{equation}\label{e:Dhat}
  \hat D_{\bx}(r) = \sum_{x_i \in \bx}
  \frac{
    \indicate{x_i \in W_{\ominus d_i}} \indicate{d_i \le r}
  }{
    |W_{\ominus d_i}|
  },
\end{equation}
where $d_i = d(x_i, \bx\setminus \{x_i\})$ is the nearest neighbor
distance for $x_i$. If $\hat\rho$ were replaced by $\rho$, then
$\Ghat_W(r)$ would be an unbiased, Horvitz--Thompson estimator of
$G(r)$. See \cite{stoykendmeck87}, pages 128--129, \cite{badd99b}.
Hanisch's recommended estimator $D_4$ is the one in which $\hat\rho$ is
taken to be
\[
  \hat D_{\bx}(\infty)
  = \sum_{x_i \in \bx}
  \frac{
    \indicate{x_i \in W_{\ominus d_i}}
  }{
    |W_{\ominus d_i}|
  }.
\]
This is sensible because $\hat D_{\bx}(\infty)$ is an unbiased
estimator of $\rho$ and is positively correlated with $\hat
D_{\bx}(r)$. The resulting estimator $\Ghat_W(r)$ can be decomposed in
the form \eqref{e:sumS} where
\[
      s_W(u,\bx,r) =
      \frac{
        \indicate{u \in W_{\ominus d(u, \bx)}} \indicate{d(u,\bx) \le r}
      }{
        \hat D_{\bx\cup\{u\}}(\infty) |W_{\ominus d(u,\bx)}|
      }
\]
for $u \notin\bx$, where $d(u, \bx)$ is the (``empty space'') distance
from location $u$ to the nearest point of $\bx$. Hence, the
corresponding compensator is
\begin{eqnarray*}
  \compen \Ghat_W(r)
  &=&
\int_W
  \frac{
    \indicate{u \in W_{\ominus d(u,\bx)}} \indicate{d(u,\bx) \le r}
  }{
    \hat D_{\bx \cup\{u\}}(\infty) |W_{\ominus d(u,\bx)}|
  }\\
  &&\hphantom{\int_W}{}\cdot\lambda_{\hat\theta}(u,\bx)
  \dee u.
\end{eqnarray*}
This is difficult to evaluate, since the denominator of the integrand
involves a summation over all data points: $D_{\bx\cup\{u\}}(\infty)$
is not related in a simple way to $D_{\bx}(\infty)$.
Instead, we choose $\hat\rho$ to be the conventional estimator
$\hat\rho = n(\bx)/|W|$. Then
\[
  \Ghat_W(r) = \frac{|W|}{n(\bx)} \hat D_{\bx}(r),
\]
which can be decomposed in the form (\ref{e:sumS}) with
\[
  s_W(u,\bx,r)\!=\!\frac{|W|}{n(\bx)\!+\!1}
  \frac{
    \indicate{u\!\in\!W_{\ominus d(u,\bx)}} \indicate{d(u,\bx)\!\le\!r}
  }{
    |W_{\ominus d(u,\bx)}|
  }
\]
for $u \notin\bx$, so that the compensator is
\begin{eqnarray}
\label{e:Gcom:modif}
  &&\compen \Ghat_W(r)\nonumber\\
  &&\quad{}= \frac{|W|}{n(\bx)+1}
  \int_W
  \frac{
    \indicate{u\in W_{\ominus d(u,\bx)}} \indicate{d(u,\bx)\le r}
  }{
    |W_{\ominus d(u,\bx)}|
  }\\
&&\quad\hphantom{\frac{|W|}{n(\bx)+1}
  \int_W}{}\cdot
  \lambda_{\hat\theta}(u,\bx) \dee u.\nonumber
\end{eqnarray}




In the \textit{restriction estimator} we exclude the boundary points and
take $\free{d}_i = d(x_i,\bxfreeminus i)$, effectively
replacing the data set $\bx$ by its restriction $\bxfree = \bx \cap
\Wfree$:
\[
  \Ghat_{\Wfree}(r) = \frac{|\Wfree|}{n(\bxfree)}
  \sum_{x_i \in \bxfree}
  \frac{
    \indicate{x_i \in \Wfree_{\ominus \free{d}_i}} \indicate{\free{d}_i \le r}
  }{
    |\Wfree_{\ominus \free{d}_i}|
  }.
\]
The compensator is (\ref{e:Gcom:modif}) but computed for the point
pattern $\bxfree$ in the window $\Wfree$:
\begin{eqnarray*}
 && \compen \Ghat_{\Wfree}(r)\\
 &&\quad= \frac{|\Wfree|}{n(\bxfree)+1}
  \int_{\Wfree}
  \frac{
    \indicate{u \in \Wfree_{\ominus d(u,\bxfree)}} \indicate{d(u,\bxfree) \le r}
  }{
    |\Wfree_{\ominus d(u,\bxfree)}|
  }\\
  &&\qquad\hphantom{\frac{|\Wfree|}{n(\bxfree)+1}
  \int_{\Wfree}}{}\cdot\lambda_{\hat\theta}(u,\bxfree \vert \bxfixed) \dee u.
\end{eqnarray*}
In the usual case $\Wfree=W_{\ominus R}$, we have $\Wfree_{\ominus
d}=\break W_{\ominus (R+d)}$.



In the \textit{reweighting estimator} we take $d_i = d(x_i, \bx\setminus
\{x_i\})$. To retain the Horvitz--Thompson property, we must replace the
weights $1/|W_{\ominus d_i}|$ in (\ref{e:Dhat}) by
$1/|\Wfree \cap W_{\ominus d_i}|$. Thus, the modified
statistics are
%
\begin{equation}\label{e:GhatR}
  \Ghat_{\Wfree,W}(r) = \frac{|W|}{n(\bx)}
  \sum_{x_i \in \bxfree}
  \frac{
    \indicate{x_i \in W_{\ominus d_i}} \indicate{d_i \le r}
  }{
    |\Wfree \cap W_{\ominus d_i}|
  }\hspace*{-25pt}
\end{equation}
and
\begin{eqnarray}\label{e:GcomR}
  &&\hspace*{10pt}\compen \Ghat_{\Wfree,W}(r)\nonumber\\
 &&\hspace*{10pt}\quad= \frac{|W|}{n(\bx)+1}
  \int_{\Wfree}
  \frac{
    \indicate{u \in W_{\ominus d(u,\bx)}} \indicate{d(u,\bx) \le r}
  }{
    |\Wfree \cap W_{\ominus d(u,\bx)}|
  }\\
  &&\hspace*{10pt}\qquad\hphantom{\frac{|W|}{n(\bx)+1}
  \int_{\Wfree}}{}\cdot\lambda_{\hat\theta}(u,\bxfree \vert \bxfixed) \dee u.\nonumber
\end{eqnarray}
In the usual case where $\Wfree = W_{\ominus R}$, we have $\Wfree \cap
W_{\ominus d_i} = W_{\ominus \max(R,d_i)}$.

Optionally, we may also replace $|W|/n(\bx)$ in (\ref{e:GhatR}) by
$|\Wfree|/n(\bx \cap \Wfree)$, and, correspondingly, replace
$\frac{|W|}{n(\bx)+1}$ in (\ref{e:GcomR}) by $|\Wfree|/(n(\bx \cap
\Wfree) + 1)$.

\subsection{Border Correction Estimator}

The classical border correction estimate of $G$ is
%
\begin{eqnarray}\label{e:Ghat:bord}
    \hspace*{15pt}\Ghat_W(r) &=&
    \frac{1
    }{
      n(\bx\cap W_{\ominus r})
    }\nonumber\\ [-8pt]\\ [-8pt]
    \hspace*{15pt}&&{}\cdot  \sum_{x_i\in\bx} \indicate{x_i \in W_{\ominus r}}
      \indicate{d(x_i, \bxminus i) \le r}\nonumber
\end{eqnarray}
with compensator (in the unconditional case)
%
\begin{eqnarray}
  \label{e:Gcom:bord}
     \hspace*{15pt}\compen \Ghat_W(r) &=&
      \frac{1
      }{
        1 + n(\bx\cap W_{\ominus r})
      }\nonumber\\ [-8pt]\\ [-8pt]
       \hspace*{15pt}&&{}\cdot \int_{W_{\ominus r}}
        \indicate{d(u, \bx) \le r}
        \lambda_{\hat\theta}(u,\bx)
        \dee u.\nonumber
\end{eqnarray}
In the conditional case, the Papangelou conditional intensity
$\lambda_{\hat\theta}(u,\bx)$ must be replaced by
$\lambda_{\hat\theta}(u, \bxfree \vert \bxfixed)$ given in
(\ref{e:lam=lam}). The \textit{restriction estimator} is obtained by
replacing $W$ by $\Wfree$ and $\bx$ by $\bxfree$ in
(\ref{e:Ghat:bord})--(\ref{e:Gcom:bord}), yielding
\begin{eqnarray*}
    \Ghat_{\Wfree}(r) &=&
    \frac{1
    }{
      n(\bx\cap \Wfree_{\ominus r})}\\
     &&{}\cdot \sum_{x_i\in\bxfree} \indicate{x_i \in \Wfree_{\ominus
r}}
      \indicate{d(x_i, \bxfreeminus i) \le r},
      \\
     \compen \Ghat_{\Wfree}(r) &=&
      \frac{1
      }{
        1 + n(\bx\cap \Wfree_{\ominus r})}\\
      &&{}\cdot  \int_{\Wfree_{\ominus r}}
        \indicate{d(u, \bxfree) \le r}
        \lambda_{\hat\theta}(u,\bxfree \vert \bxfixed)
        \dee u.
\end{eqnarray*}
Typically, $\Wfree = W_{\ominus R}$ so that $\Wfree_{\ominus r} =
W_{\ominus (R+r)}$. The \textit{reweighting estimator} is obtained by
restricting $x_i$ and $u$ in (\ref{e:Ghat:bord})--(\ref{e:Gcom:bord})
to lie in $\Wfree$, yielding
\begin{eqnarray*}
    \Ghat_{\Wfree,W}(r) &=&
    \frac{1
    }{
      n(\bxfree\cap W_{\ominus r})
    }\\
    &&{}\cdot  \sum_{x_i\in\bxfree} \indicate{x_i \in W_{\ominus r}}
      \indicate{d(x_i, \bxminus i) \le r},\\
     \compen \Ghat_{\Wfree,W}(r) &=&
      \frac{1
      }{
        1 + n(\bxfree\cap W_{\ominus r})
      }\\
       &&{}\cdot \int_{\Wfree\cap W_{\ominus r}}
       \mathbb{I}\{d(u, \bx) \le r\}\\
       &&{}\hspace*{48pt}\cdot \lambda_{\hat\theta}(u,\bxfree \vert \bxfixed)
        \dee u.
\end{eqnarray*}
In the usual case where $\Wfree = W_{\ominus R}$, we have $\Wfree \cap
W_{\ominus r} = W_{\ominus \max(R,r)}$.
%
Again, the reweighting approach is preferable to the restriction
approach.

The border corrected estimator $\hat G(r)$
has relatively poor performance and sample properties \cite{illietal08},\vadjust{\goodbreak} pa\-ge~209.
Its main advantage is its computational efficiency in large data sets.
Similar considerations should apply to its compensator.

\section*{Acknowledgments}
This paper has benefited from very fruitful discussions with Professor
Rasmus P. Waagepetersen. We also thank the referees for insightful
comments. The research was supported by the University of Western
Australia, the Danish Natural Science Research Council (Grants
272-06-0442 and 09-072331, \textit{Point process
modeling and statistical inference}), the Danish Agency for Science,
Technology and Innovation (Grant 645-06-0528,
\textit{International Ph.D. student}) and by the Centre for Stochastic
Geometry and Advanced Bioimaging, funded by a grant from the Villum
Foundation.



\begin{thebibliography}{29}

\bibitem{alm88}
\begin{barticle}[mr]
\bauthor{\bsnm{Alm},~\bfnm{Sven~Erick}\binits{S.~E.}}
(\byear{1998}).
\btitle{Approximation and simulation of the distributions of scan statistics
  for {P}oisson processes in higher dimensions}.
\bjournal{Extremes}
\bvolume{1}
\bpages{111--126}.
\bid{doi={10.1023/A:1009965918058}, issn={1386-1999}, mr={1652932}}
\bptnote{check year}%
\bptok{imsref}%
\end{barticle}
\endbibitem

\bibitem{atki82}
\begin{barticle}[mr]
\bauthor{\bsnm{Atkinson},~\bfnm{A.~C.}\binits{A.~C.}}
(\byear{1982}).
\btitle{Regression diagnostics, transformations and constructed variables (with
  discussion)}.
\bjournal{J. Roy. Statist. Soc. Ser. B}
\bvolume{44}
\bpages{1--36}.
\bid{issn={0035-9246}, mr={0655369}}
\bptnote{check related}%
\bptok{imsref}%
\end{barticle}
\endbibitem

\bibitem{badd80b}
\begin{barticle}[mr]
\bauthor{\bsnm{Baddeley},~\bfnm{Adrian}\binits{A.}}
(\byear{1980}).
\btitle{A limit theorem for statistics of spatial data}.
\bjournal{Adv. in Appl. Probab.}
\bvolume{12}
\bpages{447--461}.
\bid{doi={10.2307/1426605}, issn={0001-8678}, mr={0569436}}
\bptok{imsref}%
\end{barticle}
\endbibitem

\bibitem{baddmollpake08}
\begin{barticle}[mr]
\bauthor{\bsnm{Baddeley},~\bfnm{A.}\binits{A.}},
  \bauthor{\bsnm{M{\o}ller},~\bfnm{J.}\binits{J.}} \AND
  \bauthor{\bsnm{Pakes},~\bfnm{A.~G.}\binits{A.~G.}}
(\byear{2008}).
\btitle{Properties of residuals for spatial point processes}.
\bjournal{Ann. Inst. Statist. Math.}
\bvolume{60}
\bpages{627--649}.
\bid{doi={10.1007/s10463-007-0116-6}, issn={0020-3157}, mr={2434415}}
\bptok{imsref}%
\end{barticle}
\endbibitem

\bibitem{baddturn00}
\begin{barticle}[mr]
\bauthor{\bsnm{Baddeley},~\bfnm{Adrian}\binits{A.}} \AND
  \bauthor{\bsnm{Turner},~\bfnm{Rolf}\binits{R.}}
(\byear{2000}).
\btitle{Practical maximum pseudolikelihood for spatial point patterns (with
  discussion)}.
\bjournal{Aust. N. Z. J. Stat.}
\bvolume{42}
\bpages{283--322}.
\bid{doi={10.1111/1467-842X.00128}, issn={1369-1473}, mr={1794056}}
\bptok{imsref}%
\end{barticle}
\endbibitem

\bibitem{baddturn05}
\begin{barticle}[author]
\bauthor{\bsnm{Baddeley},~\bfnm{A.}\binits{A.}} \AND
  \bauthor{\bsnm{Turner},~\bfnm{R.}\binits{R.}}
(\byear{2005}).
\btitle{Spatstat: An \textsf{R} package for analyzing spatial point patterns}.
\bjournal{J.~Statist. Software}
\bvolume{12}
\bpages{1--42}.
\bptok{imsref}%
\end{barticle}
\endbibitem

\bibitem{baddetal05}
\begin{barticle}[mr]
\bauthor{\bsnm{Baddeley},~\bfnm{A.}\binits{A.}},
  \bauthor{\bsnm{Turner},~\bfnm{R.}\binits{R.}},
  \bauthor{\bsnm{M{\o}ller},~\bfnm{J.}\binits{J.}} \AND
  \bauthor{\bsnm{Hazelton},~\bfnm{M.}\binits{M.}}
(\byear{2005}).
\btitle{Residual analysis for spatial point processes (with discussion)}.
\bjournal{J. R. Stat. Soc. Ser. B Stat. Methodol.}
\bvolume{67}
\bpages{617--666}.
\bid{doi={10.1111/j.1467-9868.2005.00519.x}, issn={1369-7412}, mr={2210685}}
\bptnote{check related}%
\bptok{imsref}%
\end{barticle}
\endbibitem

\bibitem{badd93a}
\begin{barticle}[author]
\bauthor{\bsnm{Baddeley},~\bfnm{A.~J.}\binits{A.~J.}}
(\byear{1993}).
\btitle{Stereology and survey sampling theory}.
\bjournal{Bull. Int. Statist. Inst.}
\bvolume{50}
\bpages{435--449}.
\bptok{imsref}%
\end{barticle}
\endbibitem

\bibitem{badd99b}
\begin{bincollection}[mr]
\bauthor{\bsnm{Baddeley},~\bfnm{Adrian~J.}\binits{A.~J.}}
(\byear{1999}).
\btitle{Spatial sampling and censoring}.
In \bbooktitle{Stochastic Geometry ({T}oulouse, 1996)}.
\bseries{Monogr. Statist. Appl. Probab.}
\bvolume{80}
\bpages{37--78}.
\bpublisher{Chapman \& Hall/CRC, Boca Raton, FL}.
\bid{mr={1673114}}
\bptok{imsref}%
\end{bincollection}
\endbibitem

\bibitem{baddmoll89}
\begin{barticle}[author]
\bauthor{\bsnm{Baddeley},~\bfnm{A.~J.}\binits{A.~J.}} \AND
  \bauthor{\bsnm{M{\o}ller},~\bfnm{J.}\binits{J.}}
(\byear{1989}).
\btitle{Nearest-neighbour {M}arkov point processes and random sets}.
\bjournal{Int. Stat. Rev.}
\bvolume{57}
\bpages{89--121}.
\bptok{imsref}%
\end{barticle}
\endbibitem

\bibitem{baddlies95a}
\begin{barticle}[mr]
\bauthor{\bsnm{Baddeley},~\bfnm{A.~J.}\binits{A.~J.}} \AND
  \bauthor{\bparticle{van} \bsnm{Lieshout},~\bfnm{M.~N.~M.}\binits{M.~N.~M.}}
(\byear{1995}).
\btitle{Area-interaction point processes}.
\bjournal{Ann. Inst. Statist. Math.}
\bvolume{47}
\bpages{601--619}.
\bid{doi={10.1007/BF01856536}, issn={0020-3157}, mr={1370279}}
\bptok{imsref}%
\end{barticle}
\endbibitem

\bibitem{barn63}
\begin{barticle}[author]
\bauthor{\bsnm{Barnard},~\bfnm{G.}\binits{G.}}
(\byear{1963}).
\btitle{Discussion of ``The spectral analysis of point
  processes'' by M. S. Bartlett}.
\bjournal{J. R. Stat. Soc. Ser. B Stat. Methodol.}
\bvolume{25}
\bpages{294}.
\bptnote{check related}%
\bptok{imsref}%
\end{barticle}
\endbibitem

\bibitem{berm86}
\begin{barticle}[author]
\bauthor{\bsnm{Berman},~\bfnm{M.}\binits{M.}}
(\byear{1986}).
\btitle{Testing for spatial association between a point process and another
  stochastic process}.
\bjournal{J. Roy. Statist. Soc. Ser. C}
\bvolume{35}
\bpages{54--62}.
\bptok{imsref}%
\end{barticle}
\endbibitem

\bibitem{besa78}
\begin{barticle}[author]
\bauthor{\bsnm{Besag},~\bfnm{J.}\binits{J.}}
(\byear{1978}).
\btitle{Some methods of statistical analysis for spatial data}.
\bjournal{Bull. Int. Statist. Inst.}
\bvolume{44}
\bpages{77--92}.
\bptok{imsref}%
\end{barticle}
\endbibitem

\bibitem{chen83}
\begin{barticle}[author]
\bauthor{\bsnm{Chen},~\bfnm{C.}\binits{C.}}
(\byear{1983}).
\btitle{Score tests for regression models}.
\bjournal{J.~Amer. Statist. Assoc.}
\bvolume{78}
\bpages{158--161}.
\bptok{imsref}%
\end{barticle}
\endbibitem

\bibitem{chetdigg98}
\begin{barticle}[mr]
\bauthor{\bsnm{Chetwynd},~\bfnm{A.~G.}\binits{A.~G.}} \AND
  \bauthor{\bsnm{Diggle},~\bfnm{P.~J.}\binits{P.~J.}}
(\byear{1998}).
\btitle{On estimating the reduced second moment measure of a~stationary spatial
  point process}.
\bjournal{Aust. N. Z. J. Stat.}
\bvolume{40}
\bpages{11--15}.
\bid{doi={10.1111/1467-842X.00002}, issn={1369-1473}, mr={1628212}}
\bptok{imsref}%
\end{barticle}
\endbibitem

\bibitem{coeulava10}
\begin{bunpublished}[author]
\bauthor{\bsnm{Coeurjolly},~\bfnm{J.~F.}\binits{J.~F.}} \AND
  \bauthor{\bsnm{Lavancier},~\bfnm{F.}\binits{F.}}
(\byear{2010}).
\btitle{Residuals for stationary marked {G}ibbs point processes}.
\bnote{Preprint}.
Available at
\texttt{\href{http://arxiv.org/abs/1002.0857}{http://arxiv.org/abs/}
\href{http://arxiv.org/abs/1002.0857}{1002.0857}}.
\bptok{imsref}%
\end{bunpublished}
\endbibitem

\bibitem{conn01}
\begin{barticle}[mr]
\bauthor{\bsnm{Conniffe},~\bfnm{Denis}\binits{D.}}
(\byear{2001}).
\btitle{Score tests when a nuisance parameter is unidentified under the null
  hypothesis}.
\bjournal{J.~Statist. Plann. Inference}
\bvolume{97}
\bpages{67--83}.
\bid{doi={10.1016/S0378-3758(00)00346-3}, issn={0378-3758}, mr={1851375}}
\bptok{imsref}%
\end{barticle}
\endbibitem

\bibitem{cookweis83}
\begin{barticle}[mr]
\bauthor{\bsnm{Cook},~\bfnm{R.~Dennis}\binits{R.~D.}} \AND
  \bauthor{\bsnm{Weisberg},~\bfnm{Sanford}\binits{S.}}
(\byear{1983}).
\btitle{Diagnostics for heteroscedasticity in regression}.
\bjournal{Biometrika}
\bvolume{70}
\bpages{1--10}.
\bid{doi={10.1093/biomet/70.1.1}, issn={0006-3444}, mr={0742970}}
\bptok{imsref}%
\end{barticle}
\endbibitem

\bibitem{cord93}
\begin{barticle}[mr]
\bauthor{\bsnm{Cordy},~\bfnm{Clifford~B.}\binits{C.~B.}}
(\byear{1993}).
\btitle{An extension of the {H}orvitz--{T}hompson theorem to point sampling
  from a continuous universe}.
\bjournal{Statist. Probab. Lett.}
\bvolume{18}
\bpages{353--362}.
\bid{doi={10.1016/0167-7152(93)90028-H}, issn={0167-7152}, mr={1247446}}
\bptok{imsref}%
\end{barticle}
\endbibitem

\bibitem{cox72pp}
\begin{bincollection}[mr]
\bauthor{\bsnm{Cox},~\bfnm{D.~R.}\binits{D.~R.}}
(\byear{1972}).
\btitle{The statistical analysis of dependencies in point processes}.
In \bbooktitle{Stochastic Point Processes: Statistical Analysis, Theory, and
  Applications ({C}onf., {IBM} {R}es. {C}enter, {Y}orktown {H}eights, {N}.{Y}.,
  1971)}
\bpages{55--66}.
\bpublisher{Wiley-Interscience, New York}.
\bid{mr={0375705}}
\bptok{imsref}%
\end{bincollection}
\endbibitem

\bibitem{coxhink74}
\begin{bbook}[mr]
\bauthor{\bsnm{Cox},~\bfnm{D.~R.}\binits{D.~R.}} \AND
  \bauthor{\bsnm{Hinkley},~\bfnm{D.~V.}\binits{D.~V.}}
(\byear{1974}).
\btitle{Theoretical Statistics}.
\bpublisher{Chapman \& Hall}, \baddress{London}.
\bid{mr={0370837}}
\bptok{imsref}%
\end{bbook}
\endbibitem

\bibitem{cres91}
\begin{bbook}[mr]
\bauthor{\bsnm{Cressie},~\bfnm{Noel A.~C.}\binits{N.~A.~C.}}
(\byear{1991}).
\btitle{Statistics for Spatial Data}.
\bpublisher{Wiley}, \baddress{New York}.
\bid{mr={1127423}}
\bptok{imsref}%
\end{bbook}
\endbibitem

\bibitem{dalevere03}
\begin{bbook}[mr]
\bauthor{\bsnm{Daley},~\bfnm{D.~J.}\binits{D.~J.}} \AND
  \bauthor{\bsnm{Vere-Jones},~\bfnm{D.}\binits{D.}}
(\byear{2003}).
\btitle{An Introduction to the Theory of Point Processes. Volume I: Elementary
  Theory and Methods},
\bedition{2nd ed.}
\bpublisher{Springer}, \baddress{New York}.
\bid{mr={1950431}}
\bptok{imsref}%
\end{bbook}
\endbibitem

\bibitem{davi77}
\begin{barticle}[mr]
\bauthor{\bsnm{Davies},~\bfnm{R.~B.}\binits{R.~B.}}
(\byear{1977}).
\btitle{Hypothesis testing when a nuisance parameter is present only under the
  alternative}.
\bjournal{Biometrika}
\bvolume{64}
\bpages{247--254}.
\bid{issn={0006-3444}, mr={0501523}}
\bptok{imsref}%
\end{barticle}
\endbibitem

\bibitem{davi87}
\begin{barticle}[mr]
\bauthor{\bsnm{Davies},~\bfnm{Robert~B.}\binits{R.~B.}}
(\byear{1987}).
\btitle{Hypothesis testing when a nuisance parameter is present only under the
  alternative}.
\bjournal{Biometrika}
\bvolume{74}
\bpages{33--43}.
\bid{issn={0006-3444}, mr={0885917}}
\bptok{imsref}%
\end{barticle}
\endbibitem

\bibitem{digg79a}
\begin{barticle}[author]
\bauthor{\bsnm{Diggle},~\bfnm{P.~J.}\binits{P.~J.}}
(\byear{1979}).
\btitle{On parameter estimation and goodness-of-fit testing for spatial point
  patterns}.
\bjournal{Biometrika}
\bvolume{35}
\bpages{87--101}.
\bptok{imsref}%
\end{barticle}
\endbibitem

\bibitem{digg85}
\begin{barticle}[author]
\bauthor{\bsnm{Diggle},~\bfnm{P.~J.}\binits{P.~J.}}
(\byear{1985}).
\btitle{A kernel method for smoothing point process data}.
\bjournal{J. Roy. Statist. Soc. Ser. C}
\bvolume{34}
\bpages{138--147}.
\bptok{imsref}%
\end{barticle}
\endbibitem

\bibitem{digg03}
\begin{bbook}[author]
\bauthor{\bsnm{Diggle},~\bfnm{P.~J.}\binits{P.~J.}}
(\byear{2003}).
\btitle{Statistical Analysis of Spatial Point Patterns}, \bedition{2nd} ed.
\bpublisher{Hodder Arnold}, \baddress{London}.
\bptok{imsref}%
\end{bbook}
\endbibitem

\bibitem{geor76}
\begin{barticle}[mr]
\bauthor{\bsnm{Georgii},~\bfnm{Hans-Otto}\binits{H.-O.}}
(\byear{1976}).
\btitle{Canonical and grand canonical {G}ibbs states for continuum systems}.
\bjournal{Comm. Math. Phys.}
\bvolume{48}
\bpages{31--51}.
\bid{issn={0010-3616}, mr={0411497}}
\bptok{imsref}%
\end{barticle}
\endbibitem

\bibitem{geye99}
\begin{bincollection}[mr]
\bauthor{\bsnm{Geyer},~\bfnm{C.}\binits{C.}}
(\byear{1999}).
\btitle{Likelihood inference for spatial point processes}.
In \bbooktitle{Stochastic Geometry ({T}oulouse, 1996)}.
\bseries{Monogr. Statist. Appl. Probab.}
\bvolume{80}
\bpages{79--140}.
\bpublisher{Chapman \& Hall/CRC, Boca Raton, FL}.
\bid{mr={1673118}}
\bptok{imsref}%
\end{bincollection}
\endbibitem

\bibitem{hani84a}
\begin{barticle}[mr]
\bauthor{\bsnm{Hanisch},~\bfnm{K.~H.}\binits{K.~H.}}
(\byear{1984}).
\btitle{Some remarks on estimators of the distribution function of nearest
  neighbour distance in stationary spatial point processes}.
\bjournal{Math. Operationsforsch. Statist. Ser. Statist.}
\bvolume{15}
\bpages{409--412}.
\bid{issn={0323-3944}, mr={0756346}}
\bptok{imsref}%
\end{barticle}
\endbibitem

\bibitem{hans96}
\begin{barticle}[mr]
\bauthor{\bsnm{Hansen},~\bfnm{Bruce~E.}\binits{B.~E.}}
(\byear{1996}).
\btitle{Inference when a nuisance parameter is not identified under the null
  hypothesis}.
\bjournal{Econometrica}
\bvolume{64}
\bpages{413--430}.
\bid{doi={10.2307/2171789}, issn={0012-9682}, mr={1375740}}
\bptok{imsref}%
\end{barticle}
\endbibitem

\bibitem{hein88b}
\begin{barticle}[mr]
\bauthor{\bsnm{Heinrich},~\bfnm{Lothar}\binits{L.}}
(\byear{1988}).
\btitle{Asymptotic behaviour of an empirical nearest-neighbour distance
  function for stationary {P}oisson cluster processes}.
\bjournal{Math. Nachr.}
\bvolume{136}
\bpages{131--148}.
\bid{doi={10.1002/mana.19881360109}, issn={0025-584X}, mr={0952468}}
\bptok{imsref}%
\end{barticle}
\endbibitem

\bibitem{hein88}
\begin{barticle}[mr]
\bauthor{\bsnm{Heinrich},~\bfnm{L.}\binits{L.}}
(\byear{1988}).
\btitle{Asymptotic {G}aussianity of some estimators for reduced factorial
  moment measures and product densities of stationary {P}oisson cluster
  processes}.
\bjournal{Statistics}
\bvolume{19}
\bpages{87--106}.
\bid{doi={10.1080/02331888808802075}, issn={0233-1888}, mr={0921628}}
\bptok{imsref}%
\end{barticle}
\endbibitem

\bibitem{hope68}
\begin{barticle}[author]
\bauthor{\bsnm{Hope},~\bfnm{A.~C.~A.}\binits{A.~C.~A.}}
(\byear{1968}).
\btitle{A simplified {M}onte {C}arlo significance test procedure}.
\bjournal{J. R. Stat. Soc. Ser. B Stat. Methodol.}
\bvolume{30}
\bpages{582--598}.
\bptok{imsref}%
\end{barticle}
\endbibitem

\bibitem{huanogat99}
\begin{barticle}[author]
\bauthor{\bsnm{Huang},~\bfnm{F.}\binits{F.}} \AND
  \bauthor{\bsnm{Ogata},~\bfnm{Y.}\binits{Y.}}
(\byear{1999}).
\btitle{Improvements of the maximum pseudo-likelihood estimators in various
  spatial statistical models}.
\bjournal{J. Comput. Graph. Statist.}
\bvolume{8}
\bpages{510--530}.
\bptok{imsref}%
\end{barticle}
\endbibitem

\bibitem{illietal08}
\begin{bbook}[mr]
\bauthor{\bsnm{Illian},~\bfnm{Janine}\binits{J.}},
  \bauthor{\bsnm{Penttinen},~\bfnm{Antti}\binits{A.}},
  \bauthor{\bsnm{Stoyan},~\bfnm{Helga}\binits{H.}} \AND
  \bauthor{\bsnm{Stoyan},~\bfnm{Dietrich}\binits{D.}}
(\byear{2008}).
\btitle{Statistical Analysis and Modelling of Spatial Point Patterns}.
\bpublisher{Wiley}, \baddress{Chichester}.
\bid{mr={2384630}}
\bptok{imsref}%
\end{bbook}
\endbibitem

\bibitem{jensmoll91}
\begin{barticle}[mr]
\bauthor{\bsnm{Jensen},~\bfnm{Jens~Ledet}\binits{J.~L.}} \AND
  \bauthor{\bsnm{M{\o}ller},~\bfnm{Jesper}\binits{J.}}
(\byear{1991}).
\btitle{Pseudolikelihood for exponential family models of spatial point
  processes}.
\bjournal{Ann. Appl. Probab.}
\bvolume{1}
\bpages{445--461}.
\bid{issn={1050-5164}, mr={1111528}}
\bptok{imsref}%
\end{barticle}
\endbibitem

\bibitem{joli80}
\begin{bincollection}[mr]
\bauthor{\bsnm{Jolivet},~\bfnm{E.}\binits{E.}}
(\byear{1981}).
\btitle{Central limit theorem and convergence of empirical processes for
  stationary point processes}.
In \bbooktitle{Point Processes and Queuing Problems ({C}olloq., {K}eszthely,
  1978)}.
\bseries{Colloq. Math. Soc. J\'anos Bolyai}
\bvolume{24}
\bpages{117--161}.
\bpublisher{North-Holland}, \baddress{Amsterdam}.
\bid{mr={0617406}}
\bptnote{check year}%
\bptok{imsref}%
\end{bincollection}
\endbibitem

\bibitem{kall78}
\begin{barticle}[mr]
\bauthor{\bsnm{Kallenberg},~\bfnm{Olav}\binits{O.}}
(\byear{1978}).
\btitle{On conditional intensities of point processes}.
\bjournal{Z. Wahrsch. Verw. Gebiete}
\bvolume{41}
\bpages{205--220}.
\bid{mr={0461654}}
\bptnote{check year}%
\bptok{imsref}%
\end{barticle}
\endbibitem

\bibitem{kall84}
\begin{barticle}[mr]
\bauthor{\bsnm{Kallenberg},~\bfnm{Olav}\binits{O.}}
(\byear{1984}).
\btitle{An informal guide to the theory of conditioning in point processes}.
\bjournal{Internat. Statist. Rev.}
\bvolume{52}
\bpages{151--164}.
\bid{doi={10.2307/1403098}, issn={0306-7734}, mr={0967208}}
\bptok{imsref}%
\end{barticle}
\endbibitem

\bibitem{kellripl76}
\begin{barticle}[mr]
\bauthor{\bsnm{Kelly},~\bfnm{F.~P.}\binits{F.~P.}} \AND
  \bauthor{\bsnm{Ripley},~\bfnm{B.~D.}\binits{B.~D.}}
(\byear{1976}).
\btitle{A note on {S}trauss's model for clustering}.
\bjournal{Biometrika}
\bvolume{63}
\bpages{357--360}.
\bid{issn={0006-3444}, mr={0431375}}
\bptok{imsref}%
\end{barticle}
\endbibitem

\bibitem{kull99}
\begin{bincollection}[mr]
\bauthor{\bsnm{Kulldorff},~\bfnm{Martin}\binits{M.}}
(\byear{1999}).
\btitle{Spatial scan statistics: Models, calculations, and applications}.
In \bbooktitle{Scan Statistics and Applications}
\bpages{303--322}.
\bpublisher{Birkh\"auser}, \baddress{Boston, MA}.
\bid{mr={1697758}}
\bptok{imsref}%
\end{bincollection}
\endbibitem

\bibitem{kuto98}
\begin{bbook}[mr]
\bauthor{\bsnm{Kutoyants},~\bfnm{Yu.~A.}\binits{Y.~A.}}
(\byear{1998}).
\btitle{Statistical Inference for Spatial {P}oisson Processes}.
\bseries{Lecture Notes in Statist.}
\bvolume{134}.
\bpublisher{Springer}, \baddress{New York}.
\bid{mr={1644620}}
\bptok{imsref}%
\end{bbook}
\endbibitem

\bibitem{lastpenr11}
\begin{barticle}[author]
\bauthor{\bsnm{Last},~\bfnm{G.}\binits{G.}} \AND
  \bauthor{\bsnm{Penrose},~\bfnm{M.}\binits{M.}}
(\byear{2011}).
\btitle{Poisson process {F}ock space representation, chaos expansion and
  covariance inequalities}.
\bjournal{Probab. Theory Related Fields}
\bvolume{150}
\bpages{663--690}.
\bptok{imsref}%
\end{barticle}
\endbibitem

\bibitem{laws93a}
\begin{barticle}[author]
\bauthor{\bsnm{Lawson},~\bfnm{A.~B.}\binits{A.~B.}}
(\byear{1993}).
\btitle{On the analysis of mortality events around a prespecified fixed point}.
\bjournal{J. Roy. Statist. Soc. Ser. A}
\bvolume{156}
\bpages{363--377}.
\bptok{imsref}%
\end{barticle}
\endbibitem

\bibitem{lotwsilv82}
\begin{barticle}[mr]
\bauthor{\bsnm{Lotwick},~\bfnm{H.~W.}\binits{H.~W.}} \AND
  \bauthor{\bsnm{Silverman},~\bfnm{B.~W.}\binits{B.~W.}}
(\byear{1982}).
\btitle{Methods for analysing spatial processes of several types of points}.
\bjournal{J. Roy. Statist. Soc. Ser. B}
\bvolume{44}
\bpages{406--413}.
\bid{issn={0035-9246}, mr={0693241}}
\bptok{imsref}%
\end{barticle}
\endbibitem

\bibitem{mollsyvewaag98}
\begin{barticle}[mr]
\bauthor{\bsnm{M{\o}ller},~\bfnm{Jesper}\binits{J.}},
  \bauthor{\bsnm{Syversveen},~\bfnm{Anne~Randi}\binits{A.~R.}} \AND
  \bauthor{\bsnm{Waagepeter\-sen},~\bfnm{Rasmus~Plenge}\binits{R.~P.}}
(\byear{1998}).
\btitle{Log {G}aussian {C}ox processes}.
\bjournal{Scand. J.~Statist.}
\bvolume{25}
\bpages{451--482}.
\bid{doi={10.1111/1467-9469.00115}, issn={0303-6898}, mr={1650019}}
\bptok{imsref}%
\end{barticle}
\endbibitem

\bibitem{mollwaag04}
\begin{bbook}[mr]
\bauthor{\bsnm{M{\o}ller},~\bfnm{Jesper}\binits{J.}} \AND
  \bauthor{\bsnm{Waagepetersen},~\bfnm{Rasmus~Plenge}\binits{R.~P.}}
(\byear{2004}).
\btitle{Statistical Inference and Simulation for Spatial Point Processes}.
\bseries{Monogr. Statist. Appl. Probab.}
\bvolume{100}.
\bpublisher{Chapman \& Hall/CRC, Boca Raton, FL}.
\bid{mr={2004226}}
\bptok{imsref}%
\end{bbook}
\endbibitem

\bibitem{mollwaag07}
\begin{barticle}[mr]
\bauthor{\bsnm{M{\o}ller},~\bfnm{Jesper}\binits{J.}} \AND
  \bauthor{\bsnm{Waagepetersen},~\bfnm{Rasmus~P.}\binits{R.~P.}}
(\byear{2007}).
\btitle{Modern statistics for spatial point processes}.
\bjournal{Scand. J. Statist.}
\bvolume{34}
\bpages{643--684}.
\bid{issn={0303-6898}, mr={2392447}}
\bptnote{check related}%
\bptok{imsref}%
\end{barticle}
\endbibitem

\bibitem{nguyzess79a}
\begin{barticle}[mr]
\bauthor{\bsnm{Nguyen},~\bfnm{Xuan-Xanh}\binits{X.-X.}} \AND
  \bauthor{\bsnm{Zessin},~\bfnm{Hans}\binits{H.}}
(\byear{1979}).
\btitle{Integral and differential characterizations of the {G}ibbs process}.
\bjournal{Math. Nachr.}
\bvolume{88}
\bpages{105--115}.
\bid{doi={10.1002/mana.19790880109}, issn={0025-584X}, mr={0543396}}
\bptok{imsref}%
\end{barticle}
\endbibitem

\bibitem{numa61}
\begin{barticle}[author]
\bauthor{\bsnm{Numata},~\bfnm{M.}\binits{M.}}
(\byear{1961}).
\btitle{Forest vegetation in the vicinity of Choshi---Coastal flora and
  vegetation at {Choshi, Chiba} prefecture, {IV (in Japanese)}}.
\bjournal{Bull. Choshi Mar. Lab.}
\bvolume{3}
\bpages{28--48}.
\bptok{imsref}%
\end{barticle}
\endbibitem

\bibitem{numa64}
\begin{barticle}[author]
\bauthor{\bsnm{Numata},~\bfnm{M.}\binits{M.}}
(\byear{1964}).
\btitle{Forest vegetation, particularly pine stands in the vicinity of
  Choshi---Flora and vegetation in {Choshi, Chiba} prefecture, {VI (in
  Japanese)}}.
\bjournal{Bull. Choshi Mar. Lab.}
\bvolume{6}
\bpages{27--37}.
\bptok{imsref}%
\end{barticle}
\endbibitem

\bibitem{ogattane81}
\begin{barticle}[author]
\bauthor{\bsnm{Ogata},~\bfnm{Y.}\binits{Y.}} \AND
  \bauthor{\bsnm{Tanemura},~\bfnm{M.}\binits{M.}}
(\byear{1981}).
\btitle{Estimation of interaction potentials of spatial point patterns through
  the maximum likelihood procedure}.
\bjournal{Ann. Inst. Statist. Math.}
\bvolume{33}
\bpages{315--338}.
\bptok{imsref}%
\end{barticle}
\endbibitem

\bibitem{ogattane86}
\begin{binproceedings}[author]
\bauthor{\bsnm{Ogata},~\bfnm{Y.}\binits{Y.}} \AND
  \bauthor{\bsnm{Tanemura},~\bfnm{M.}\binits{M.}}
(\byear{1986}).
\btitle{Likelihood estimation of interaction potentials and external fields of
  inhomogeneous spatial point patterns}.
In \bbooktitle{Pacific Statistical Congress}
(\beditor{\bfnm{I.~S.}\binits{I.~S.}~\bsnm{Francis}},
  \beditor{\bfnm{B.~J.~F.}\binits{B.~J.~F.}~\bsnm{Manly}} \AND
  \beditor{\bfnm{F.~C.}\binits{F.~C.}~\bsnm{Lam}}, eds.)
\bpages{150--154}.
\bpublisher{Elsevier}, \baddress{Amsterdam}.
\bptok{imsref}%
\end{binproceedings}
\endbibitem

\bibitem{ohse83}
\begin{barticle}[mr]
\bauthor{\bsnm{Ohser},~\bfnm{J.}\binits{J.}}
(\byear{1983}).
\btitle{On estimators for the reduced second moment measure of point
  processes}.
\bjournal{Math. Operationsforsch. Statist. Ser. Statist.}
\bvolume{14}
\bpages{63--71}.
\bid{issn={0323-3944}, mr={0697340}}
\bptok{imsref}%
\end{barticle}
\endbibitem

\bibitem{ohsestoy81}
\begin{barticle}[mr]
\bauthor{\bsnm{Ohser},~\bfnm{J.}\binits{J.}} \AND
  \bauthor{\bsnm{Stoyan},~\bfnm{D.}\binits{D.}}
(\byear{1981}).
\btitle{On the second-order and orientation analysis of planar stationary point
  processes}.
\bjournal{Biometrical J.}
\bvolume{23}
\bpages{523--533}.
\bid{issn={0323-3847}, mr={0635658}}
\bptok{imsref}%
\end{barticle}
\endbibitem

\bibitem{papa74b}
\begin{barticle}[mr]
\bauthor{\bsnm{Papangelou},~\bfnm{F.}\binits{F.}}
(\byear{1974}).
\btitle{The conditional intensity of general point processes and an application
  to line processes}.
\bjournal{Z. Wahrsch. Verw. Gebiete}
\bvolume{28}
\bpages{207--226}.
\bid{mr={0373000}}
\bptnote{check year}%
\bptok{imsref}%
\end{barticle}
\endbibitem

\bibitem{preg82}
\begin{binproceedings}[author]
\bauthor{\bsnm{Pregibon},~\bfnm{D.}\binits{D.}}
(\byear{1982}).
\btitle{Score tests in {GLIM} with applications}.
In \bbooktitle{{GLIM} 82: Proceedings of the International Conference on
  Generalized Linear Models}.
\bseries{Lecture Notes in Statist.}
\bvolume{14}.
\bpublisher{Springer}, \baddress{New York}.
\bptok{imsref}%
\end{binproceedings}
\endbibitem

\bibitem{rao48}
\begin{barticle}[mr]
\bauthor{\bsnm{Radhakrishna~Rao},~\bfnm{C.}\binits{C.}}
(\byear{1948}).
\btitle{Large sample tests of statistical hypotheses concerning several
  parameters with applications to problems of estimation}.
\bjournal{Proc. Cambridge Philos. Soc.}
\bvolume{44}
\bpages{50--57}.
\bid{mr={0024111}}
\bptok{imsref}%
\end{barticle}
\endbibitem



\bibitem{rathcres94b}
\begin{barticle}[mr]
  \bauthor{\bsnm{Rathbun},~\bfnm{Stephen~L.}\binits{S.~L.}}
\AND
  \bauthor{\bsnm{Cressie},~\bfnm{Noel}\binits{N.}}
(\byear{1994}).
  \btitle{Asymptotic properties of estimators for the parameters of spatial
  inhomogeneous {P}oisson point processes}.
\bjournal{Adv. in Appl. Probab.}
 \bvolume{26}
\bpages{122--154}.
\bid{doi={10.2307/1427583}, issn={0001-8678},
  mr={1260307}}
\bptok{imsref}
\end{barticle}
\endbibitem\


\bibitem{ripl76}
\begin{barticle}[mr]
\bauthor{\bsnm{Ripley},~\bfnm{B.~D.}\binits{B.~D.}}
(\byear{1976}).
\btitle{The second-order analysis of stationary point processes}.
\bjournal{J. Appl. Probab.}
\bvolume{13}
\bpages{255--266}.
\bid{issn={0021-9002}, mr={0402918}}
\bptok{imsref}%
\end{barticle}
\endbibitem

\bibitem{ripl77}
\begin{barticle}[mr]
\bauthor{\bsnm{Ripley},~\bfnm{B.~D.}\binits{B.~D.}}
(\byear{1977}).
\btitle{Modelling spatial patterns (with discussion)}.
\bjournal{J. Roy. Statist. Soc. Ser. B}
\bvolume{39}
\bpages{172--212}.
\bid{issn={0035-9246}, mr={0488279}}
\bptok{imsref}%
\end{barticle}
\endbibitem

\bibitem{ripl88}
\begin{bbook}[mr]
\bauthor{\bsnm{Ripley},~\bfnm{B.~D.}\binits{B.~D.}}
(\byear{1988}).
\btitle{Statistical Inference for Spatial Processes}.
\bpublisher{Cambridge Univ. Press}, \baddress{Cambridge}.
\bid{mr={0971986}}
\bptok{imsref}%
\end{bbook}
\endbibitem

\bibitem{riplkell77}
\begin{barticle}[mr]
\bauthor{\bsnm{Ripley},~\bfnm{B.~D.}\binits{B.~D.}} \AND
  \bauthor{\bsnm{Kelly},~\bfnm{F.~P.}\binits{F.~P.}}
(\byear{1977}).
\btitle{Markov point processes}.
\bjournal{J. Lond. Math. Soc. (2)}
\bvolume{15}
\bpages{188--192}.
\bid{issn={0024-6107}, mr={0436387}}
\bptok{imsref}%
\end{barticle}
\endbibitem

\bibitem{schlbadd00}
\begin{barticle}[mr]
\bauthor{\bsnm{Schladitz},~\bfnm{K.}\binits{K.}} \AND
  \bauthor{\bsnm{Baddeley},~\bfnm{A.~J.}\binits{A.~J.}}
(\byear{2000}).
\btitle{A third order point process characteristic}.
\bjournal{Scand. J. Statist.}
\bvolume{27}
\bpages{657--671}.
\bid{doi={10.1111/1467-9469.00214}, issn={0303-6898}, mr={1804169}}
\bptok{imsref}%
\end{barticle}
\endbibitem

\bibitem{silv96}
\begin{barticle}[mr]
\bauthor{\bsnm{Silvapulle},~\bfnm{Mervyn~J.}\binits{M.~J.}}
(\byear{1996}).
\btitle{A test in the presence of nuisance parameters}.
\bjournal{J. Amer. Statist. Assoc.}
\bvolume{91}
\bpages{1690--1693}.
\bid{issn={0162-1459}, mr={1439111}}
\bptok{imsref}%
\end{barticle}
\endbibitem

\bibitem{stei95}
\begin{barticle}[mr]
\bauthor{\bsnm{Stein},~\bfnm{Michael~L.}\binits{M.~L.}}
(\byear{1995}).
\btitle{An approach to asymptotic inference for spatial point processes}.
\bjournal{Statist. Sinica}
\bvolume{5}
\bpages{221--234}.
\bid{issn={1017-0405}, mr={1329294}}
\bptok{imsref}%
\end{barticle}
\endbibitem

\bibitem{stiletal00}
\begin{barticle}[mr]
\bauthor{\bsnm{Stillinger},~\bfnm{Dorothea~K.}\binits{D.~K.}},
  \bauthor{\bsnm{Stillinger},~\bfnm{Frank~H.}\binits{F.~H.}},
  \bauthor{\bsnm{Torquato},~\bfnm{Salvatore}\binits{S.}},
  \bauthor{\bsnm{Truskett},~\bfnm{Thomas~M.}\binits{T.~M.}} \AND
  \bauthor{\bsnm{Debenedetti},~\bfnm{Pablo~G.}\binits{P.~G.}}
(\byear{2000}).
\btitle{Triangle distribution and equation of state for classical rigid disks}.
\bjournal{J. Statist. Phys.}
\bvolume{100}
\bpages{49--72}.
\bid{doi={10.1023/A:1018675208867}, issn={0022-4715}, mr={1791553}}
\bptok{imsref}%
\end{barticle}
\endbibitem\eject

\bibitem{stoykendmeck87}
\begin{bbook}[mr]
\bauthor{\bsnm{Stoyan},~\bfnm{D.}\binits{D.}},
  \bauthor{\bsnm{Kendall},~\bfnm{W.~S.}\binits{W.~S.}} \AND
  \bauthor{\bsnm{Mecke},~\bfnm{J.}\binits{J.}}
(\byear{1987}).
\btitle{Stochastic Geometry and Its Applications}.
\bpublisher{Wiley}, \baddress{Chichester}.
\bid{mr={0895588}}
\bptok{imsref}%
\end{bbook}
\endbibitem

\bibitem{stoystoy95}
\begin{bbook}[author]
\bauthor{\bsnm{Stoyan},~\bfnm{D.}\binits{D.}} \AND
  \bauthor{\bsnm{Stoyan},~\bfnm{H.}\binits{H.}}
(\byear{1995}).
\btitle{Fractals, Random Shapes and Point Fields}.
\bpublisher{{Wiley}}, \baddress{Chichester}.
\bptok{imsref}%
\end{bbook}
\endbibitem

\bibitem{stra75}
\begin{barticle}[mr]
\bauthor{\bsnm{Strauss},~\bfnm{David~J.}\binits{D.~J.}}
(\byear{1975}).
\btitle{A model for clustering}.
\bjournal{Biometrika}
\bvolume{62}
\bpages{467--475}.
\bid{issn={0006-3444}, mr={0383493}}
\bptok{imsref}%
\end{barticle}
\endbibitem

\bibitem{lies00}
\begin{bbook}[mr]
\bauthor{\bparticle{van} \bsnm{Lieshout},~\bfnm{M.~N.~M.}\binits{M.~N.~M.}}
(\byear{2000}).
\btitle{Markov Point Processes and Their Applications}.
\bpublisher{Imperial College Press}, \baddress{London}.
\bid{doi={10.1142/9781860949760}, mr={1789230}}
\bptok{imsref}%
\end{bbook}
\endbibitem

\bibitem{wald41}
\begin{barticle}[mr]
\bauthor{\bsnm{Wald},~\bfnm{Abraham}\binits{A.}}
(\byear{1941}).
\btitle{Some examples of asymptotically most powerful tests}.
\bjournal{Ann. Math. Statist.}
\bvolume{12}
\bpages{396--408}.
\bid{issn={0003-4851}, mr={0006683}}
\bptok{imsref}%
\end{barticle}
\endbibitem

\bibitem{walletal92}
\begin{barticle}[author]
\bauthor{\bsnm{Waller},~\bfnm{L.}\binits{L.}},
  \bauthor{\bsnm{Turnbull},~\bfnm{B.}\binits{B.}},
  \bauthor{\bsnm{Clark},~\bfnm{L.~C.}\binits{L.~C.}} \AND
  \bauthor{\bsnm{Nasca},~\bfnm{P.}\binits{P.}}
(\byear{1992}).
\btitle{Chronic Disease Surveillance and testing of clustering of disease and
  exposure: Application to leukaemia incidence and {TCE}-contaminated dumpsites
  in upstate {New York}}.
\bjournal{Environmetrics}
\bvolume{3}
\bpages{281--300}.
\bptok{imsref}%
\end{barticle}
\endbibitem

\bibitem{wang85}
\begin{barticle}[mr]
\bauthor{\bsnm{Wang},~\bfnm{P.~C.}\binits{P.~C.}}
(\byear{1985}).
\btitle{Adding a variable in generalized linear models}.
\bjournal{Technometrics}
\bvolume{27}
\bpages{273--276}.
\bid{issn={0040-1706}, mr={0797565}}
\bptok{imsref}%
\end{barticle}
\endbibitem

\bibitem{widorowl70}
\begin{barticle}[author]
\bauthor{\bsnm{Widom},~\bfnm{B.}\binits{B.}} \AND
  \bauthor{\bsnm{Rowlinson},~\bfnm{J.~S.}\binits{J.~S.}}
(\byear{1970}).
\btitle{New model for the study of liquid--vapor phase transitions}.
\bjournal{J.~Chem. Phys.}
\bvolume{52}
\bpages{1670--1684}.
\bptok{imsref}%
\end{barticle}
\endbibitem




\bibitem{wu00}
\begin{barticle}[mr]
  \bauthor{\bsnm{Wu},~\bfnm{Liming}\binits{L.}}
(\byear{2000}).
\btitle{A new
  modified logarithmic {S}obolev inequality for {P}oisson point processes and
  several applications}.
\bjournal{Probab. Theory Related Fields}
\bvolume{118}
\bpages{427--438}.
\bid{doi={10.1007/PL00008749}, issn={0178-8051},
  mr={1800540}}
\bptok{imsref}
\end{barticle}
\endbibitem
\end{thebibliography}
\end{document}